\documentclass[paper]{JHEP3}
\usepackage{cite}
\usepackage{amssymb,amsmath}
\usepackage{graphicx}
\usepackage{booktabs}
\usepackage{subfigure}
\usepackage{enumerate}

\title{
Double real radiation corrections to $t\bar{t}$ production at the LHC: the all-fermion processes}

\author{
Gabriel Abelof, Aude Gehrmann--De Ridder\\
Institute for Theoretical Physics, ETH, CH-8093 Z\"urich,
Switzerland}


\abstract{
We present the double real radiation corrections to the hadronic $t \bar{t}$ production stemming from partonic processes with fermions only. For this purpose, we extend the NNLO antenna subtraction formalism developed originally for the computation of jet observables in $e^+e^-$ annihilation to include the evaluation of hadronic observables involving a massive pair of particles. In all partonic processes, we checked  the validity of our subtraction terms given for leading and subleading colour contributions numerically by showing that the ratio between real radiation matrix elements and subtraction terms approaches unity in all single and double unresolved configurations.
}

\allowdisplaybreaks[4]

\newcommand{\beq}{\begin{equation}}
\newcommand{\eeq}{\end{equation}}
\newcommand{\beqa}{\begin{eqnarray}}
\newcommand{\eeqa}{\end{eqnarray}}
\newcommand{\cm}{{\cal M}^0}
\newcommand{\ds}{{\rm d}\hat{\sigma}}
\newcommand{\dphi}{{\rm d}\Phi}
\newcommand{\wt}{\widetilde}
\newcommand{\re}{{\rm{Re}}}
\newcommand{\norm}{{\cal N}}
\newcommand{\ssoft}[3]{{\cal S}_{#1 #2 #3}}
\newcommand{\soft}[4]{{\cal S}_{#1 #2 #3 #4}}
\newcommand{\order}[1]{{\cal O}(#1)}
\newcommand{\q}[1]{#1_q}
\newcommand{\qb}[1]{#1_{\bar{q}}}
\newcommand{\qp}[1]{#1_{q'}}
\newcommand{\qpb}[1]{#1_{\bar{q}'}}
\newcommand{\Q}[1]{#1_Q}
\newcommand{\Qb}[1]{#1_{\bar{Q}}}
\newcommand{\qi}[1]{\hat{#1}_q}
\newcommand{\qbi}[1]{\hat{#1}_{\bar{q}}}
\newcommand{\qpi}[1]{\hat{#1}_{q'}}
\newcommand{\qpbi}[1]{\hat{#1}_{\bar{q}'}}
\newcommand{\gl}[1]{#1_g}
\newcommand{\gli}[1]{\hat{#1}_g}
\newcommand{\ph}[1]{#1_{\gamma}}
\newcommand{\phin}[1]{\hat{#1}_{\gamma}}
\newcommand{\del}[2]{\delta_{i_{#1},i_{#2}}}
\def\d{\hbox{d}}

\def\JET{J}
\def\e{\epsilon}

\begin{document}

\section{Introduction}\label{sec.intro}

The production of top-antitop pairs is one of the core measurements at LHC. Top quarks \cite{cdftop,d0top} are measured through their decay into a bottom quark and a subsequently decaying $W$-boson, yielding up to six-jet final states for top quark pair production. With a mass $m_{t}=173 \pm 1.3$ GeV, the top quark is the heaviest Standard Model particle produced at colliders and due to its very large mass it decays before it hadronises. By studying its properties in detail, it is hoped to elucidate the origin of particle masses and the mechanism of electroweak symmetry breaking. Since its discovery at the Fermilab Tevatron, a number of these properties (mass, couplings) have been determined to an accuracy of ten to twenty per cent. With the large number of top quark pairs expected to be produced at the LHC, the study of the top quark is becoming precision physics. This large production rate aims to obtain precise measurements of its properties and production cross sections with an expected experimental accuracy of the order of five per cent. In order to extract fundamental parameters like couplings and masses by comparing theoretical predictions and data, it is therefore mandatory to have theoretical predictions with a comparable accuracy.

In order to match the expected experimental precision for these heavy quark production processes at LHC, fixed-order calculations need to be considered at least at the next-to-leading (NLO), if not even at next-to-next-to-leading order (NNLO) in perturbative QCD.  

Most of the calculations involving top particles are performed in the so-called narrow width approximation. The top quark appears as virtual particle in hadron collider processes and due to the small ratio between its width and its mass, it is possible to factor the cross section of processes involving top quarks into the product of the production cross section for on-shell top quarks and the corresponding decay width. In this paper, we shall also follow this approach, in which the top quark pairs are considered to be produced on-shell.

Current theoretical predictions for the top quark pair production cross section include NLO corrections \cite{ttfull1,ttfull2,ttnlo} and next-to-leading-logarithmic resummation (NLL)~\cite{newresum}. More recently even the NNLL resummation effects have been completed in\cite{resumnnll}. These predictions lead to a theoretical uncertainty of the order of ten per cent. The same precision is available for single top quark production \cite{singtop}, top-pair-plus-jets production \cite{ttj,ttjj} and for top-pair-plus-bottom-pair production \cite{ttbb1,ttbb2}.

At present, a full fixed order calculation of the total top-antitop rate at NNLO is still missing. In such a calculation, three essential ingredients enter: double real ${\rm d}\sigma^{RR}$, mixed real-virtual, ${\rm d}\sigma^{RV}$,  and double virtual contributions ${\rm d}\sigma^{VV}$. 

Recent progresses has been accomplished concerning the two-loop contributions. The parts of these two-loop virtual corrections that are built with products of one-loop virtual amplitudes have been already computed in \cite{oneloop2}. Concerning the two-loop virtual corrections which are built with products of two-loop and tree-level amplitudes, the situation is different since the two-loop virtual corrections for the processes $q\bar q \to t\bar{t}$ and $gg\to t\bar{t}$ are not fully available at present. However, a purely numerical evaluation of the quark-initiated process \cite{czakontt} could be partly confirmed by analytical results \cite{ourtt}, which were most recently extended also to the gluon-induced subprocess.

Most recently, progress towards the computation of mixed real-virtual corrections has been achieved, and the infrared structure of these contributions has been studied \cite{newczakon}. Also, following a semi-numerical approach \cite{stripper} partonic processes contributing to double real radiation to top quark pair production have been evaluated \cite{czakonsub}.  

In this paper,  we shall evaluate the double real contributions to hadronic top quark pair production arising from partonic processes involving only fermions. \\

Generally, for hard scattering observables, the inclusive cross section with two incoming hadrons $H_1,H_2$ can be written as  
\beq\label{eq.hadroncross}
{\rm d} \sigma = \sum_{a,b} \int 
\frac{{\rm d} \xi_1}{\xi_1} \frac{{\rm d} \xi_2}{\xi_2}\, f_{a/1}(\xi_1,\mu_F) \,f_{b/2}(\xi_2,\mu_F)
\, {\rm d} \hat{\sigma}_{ab}(\xi_1H_1,\xi_2H_2,\mu_F,\mu_R)\ .
\eeq
$\xi_1$ and $\xi_2$ are the momentum fractions of the partons of species $a$ and $b$ in both incoming hadrons, $f_i$ being the corresponding parton distribution functions. ${\rm d} \hat{\sigma}_{ab}$ denotes the parton-level scattering cross section for incoming partons $a$ and $b$ which depends on the the renormalisation and factorisation scales denoted by $\mu_R$ and $\mu_{F}$ respectively.

The partonic cross section ${\rm d }\hat{\sigma}_{ab}$ has a perturbative expansion in the strong coupling $\alpha_{s}$ which depends on the renormalisation scale $\mu_{R}$. For a hadronic observable, a theoretical prediction is obtained at a given order in $\alpha_{s}$ when all partonic channels contributing at that order to the partonic cross section are summed and convoluted with the appropriate parton distribution functions as in eq.(\ref{eq.hadroncross}).

Beyond leading order, these partonic channels contain both ultraviolet and infrared (soft and collinear) singularities. The ultraviolet poles are removed by renormalisation, while collinear poles originating from the radiation of intial-state partons are cancelled by mass factorisation counterterms. The remainining soft and collinear infrared poles cancel among each other only when all partonic channels are summed over \cite{kln}.

While infrared singularities from purely virtual corrections are obtained immediately after integration over the loop momenta, their extraction is more involved for real emission (or mixed real-virtual) contributions. There, the infrared singularities only become explicit after integrating  the matrix elements over the phase space appropriate to  the differential cross section under consideration. Since these observables depend in general in a non trivial manner on the experimental criteria needed to define them, they can only be calculated numerically. The computation of hadronic observables including higher order corrections therefore requires a systematic procedure to cancel infrared singularities among different partonic channels before any numerical computation of the observable can be performed. Subtraction methods explicitly constructing infrared subtraction terms are well-known solutions to this problem.

The crucial points that all subtraction terms must satisfy are that (a) they approximate the full real radiation matrix elements in all singular limits, (b) they are still sufficiently simple to be integrated analytically, (c) they should account for the limit they are aimed at without introducing spurious infrared singularities in other limits. 

For massless QCD, and for the task of next-to-leading order (NLO) calculations \cite{ks} of hadronic and jet observables, various subtraction methods have been proposed in the literature \cite{cs,Frixione:1995ms,Nagy:1996bz,Frixione:1997np,Somogyi:2006cz}. The dipole formalism of Catani and Seymour \cite{cs} and the FKS \cite{Frixione:1995ms} of Frixione, Kunszt and Signer are the most widely used ones. Both have also been implemented in an automated way in \cite{Gleisberg:2007md,Seymour:2008mu,Hasegawa:2008ae,Hasegawa:2009tx,Frederix:2008hu,Czakon:2009ss}.

Several methods for systematically constructing general subtraction terms have been proposed at NNLO \cite{antennannlo,Weinzierl:2003fx,Kilgore:2004ty,Frixione:2004is,Somogyi:2005xz,Somogyi:2006da,Somogyi:2006db,Somogyi:2008fc,Aglietti:2008fe,Somogyi:2009ri,Bolzoni:2009ye,Bolzoni:2010bt,stripper,czakonsub,Anastasiou:2010pw,Anastasiou:2011qx,Boughezal:2011jf}. Another NNLO subtraction scheme has been proposed in~\cite{Catani:2007vq}. It is not a general subtraction scheme, but it can nevertheless deal with an entire class of processes, those without coloured final states,  in hadron-hadron collisions.  It has been explicitly applied to several observables~\cite{Catani:2007vq,Grazzini:2008tf,Catani:2009sm,Catani:2010en,Ferrera:2011bk,Catani:2011qz}.

In addition, there is the sector decomposition approach which avoids the need for analytical integration, and which has been developed for virtual~\cite{Binoth:2000ps,Binoth:2003ak,Heinrich:2008si,Carter:2010hi} and real radiation~\cite{Heinrich:2002rc,Anastasiou:2003gr, Binoth:2004jv, Heinrich:2006ku,Carter:2010hi} corrections to NNLO, and applied to several observables already~\cite{Anastasiou:2004qd,Anastasiou:2004xq,Anastasiou:2005qj,Melnikov:2006di}.

In order to evaluate double real contributions to $t \bar{t}$ production at hadron colliders, we will follow the NNLO antenna subtraction method. This method has been originally developed for the production of massless partons in electron-positron annihilation~\cite{antennannlo,GehrmannDeRidder:2005hi,GehrmannDeRidder:2005aw}, leading to the successful description of the infrared structure of three-jet events at NNLO \cite{3jet,Weinzierl:2009nz} and the subsequent numerical calculation of the NNLO corrections to event shape distributions~\cite{GehrmannDeRidder:2007bj,GehrmannDeRidder:2007hr,Weinzierl:2008iv,Weinzierl:2009ms}, the moments of event shapes~\cite{GehrmannDeRidder:2009dp,Weinzierl:2009yz} and jet rates~\cite{GehrmannDeRidder:2008ug,Weinzierl:2010cw}.  

For processes with initial-state partons and massless final states, the antenna subtraction formalism has been so far fully worked out to NLO in \cite{k,daleo}. It has been extended to NNLO for processes involving one initial state parton relevant for electron-proton scattering in \cite{Gionata}. Several essential steps towards a full NNLO calculation of jet rates at hadron collider have been achieved in \cite{radja, joao,Monni,RVnew}. 

For QCD observables involving massive particles, subtraction methods have been so far only developed completely up to the NLO level using generalisations of the dipole formalism \cite{cdst1,cdt2,weinzierl}, or of  the antenna formalism~\cite{mathias,us}. An extension of the antenna formalism to deal with the production of a massive pair of particles in association with jets up to NNLO will be presented in this paper.

For these observables, QCD radiation from massive particles can lead to soft divergencies but cannot lead to strict collinear divergencies, since the mass acts as an infrared regulator. As a consequence, less divergent contributions arise and the infrared structure of such processes is expected to be simpler. The kinematics is more involved due to the presence of finite values of the parton masses, though. In addition, in calculations of such observables, logarithms involving the ratio of the scales present in a given reaction occur.  These finite effects are related to the singular behaviour of the matrix elements in the massless limit \cite{cdst1,cdt2,mathias,us}. Depending on the kinematics of the reactions involved, these logarithms, although finite, may or may not be enhanced.

For the kinematical situation under consideration in this paper, i.e. the production of a heavy fermion pair whose mass is of the the same order as as the partonic center-of-mass energy, we shall ignore these finite effects as also done in \cite{czakonsub}, since they are not expected to be large. As a consequence, in our calculation of the double real corrections to $t \bar{t}$ hadronic production, singular infrared behaviours are considered to arise only through the presence of collinear and/or soft radiation of massless particles and only subtraction terms capturing these features will be considered in the context of the present paper.

For hadronic processes with massive fermions and jets in the final state, the antenna subtraction formalism has been extended in \cite{mathias, us} up to NLO. In \cite{us} subtraction terms capturing single unresolved, i.e. soft or collinear, radiation of the real contributions to the hadronic reactions $ pp \to Q\bar{Q}$ and $p p \to Q\bar{Q} +{\rm jet}$ were explicitly constructed for leading and subleading colour contributions. The results obtained in this previous paper provide us with essential ingredients for the construction of  subtraction terms for the hadronic reaction $pp \to Q\bar{Q}$ evaluated at NNLO and considered in this paper.\\

At NNLO, any massless or massive calculation of the partonic contributions to an hadronic observable involving an $m$ particle final state is given by
\beqa
{\rm d}\hat\sigma_{NNLO}&=&\int_{{\rm{d}}\Phi_{m+2}} {\rm{d}}\hat\sigma_{NNLO}^{RR}
+\int_{{\rm{d}}\Phi_{m+1}} {\rm{d}}\hat\sigma_{NNLO}^{RV} 
+\int_{{\rm{d}}\Phi_m}{\rm{d}}\hat\sigma_{NNLO}^{VV},
\label{eq:subnnlo1}
\eeqa
where there are three distinct contributions due to: double real radiation ${\rm{d}}\hat\sigma_{NNLO}^{RR}$, mixed real-virtual radiation ${\rm{d}}\hat\sigma_{NNLO}^{RV}$ and double virtual radiation ${\rm{d}}\hat\sigma_{NNLO}^{VV}$, which are separetely infrared divergent. Employing a subtraction method, eq.(\ref{eq:subnnlo1}) amounts to \cite{antennannlo}
\beqa
{\rm d}\hat\sigma_{NNLO}&=&\int_{{\rm{d}}\Phi_{m+2}}\left({\rm{d}}\hat\sigma_{NNLO}^{RR}-{\rm{d}}\hat\sigma_{NNLO}^S\right)
+\int_{{\rm{d}}\Phi_{m+2}}{\rm{d}}\hat\sigma_{NNLO}^S\nonumber\\
&+&\int_{{\rm{d}}\Phi_{m+1}}\left({\rm{d}}\hat\sigma_{NNLO}^{RV}-{\rm{d}}\hat\sigma_{NNLO}^{V S}\right)
+\int_{{\rm{d}}\Phi_{m+1}}{\rm{d}}\hat\sigma_{NNLO}^{V S}
+\int_{{\rm{d}}\Phi_{m+1}}{\rm{d}}\hat\sigma_{NNLO}^{MF,1}\nonumber\\
&+&\int_{{\rm{d}}\Phi_m}{\rm{d}}\hat\sigma_{NNLO}^{VV}
+\int_{{\rm{d}}\Phi_m}{\rm{d}}\hat\sigma_{NNLO}^{MF,2}\label{eq.sigNNLO},
\eeqa
where ${\rm d} \hat\sigma^{S}_{NNLO}$ denotes the subtraction term for the $(m+2)$-parton final state which behaves like the double real radiation contribution ${\rm d} \hat\sigma^{RR}_{NNLO}$ in all singular limits. Likewise, ${\rm d} \hat\sigma^{VS}_{NNLO}$ is the one-loop virtual subtraction term coinciding with the one-loop $(m+1)$-final state ${\rm d} \hat\sigma^{RV}_{NNLO}$ in all singular limits. The two-loop correction to the $(m+2)$-parton final state is denoted by ${\rm d}\hat\sigma^{VV}_{NNLO}$.  In addition, as there are partons in the initial state, there are also two mass factorisation contributions, ${\rm d}\hat\sigma^{MF,1}_{NNLO}$ and ${\rm d}\hat\sigma^{MF,2}_{NNLO}$, for the $(m+1)$- and $m$-particle final states respectively.

In order to construct a numerical implementation of the NNLO cross section, the various contributions must be reorganised according to the number of final state particles,
\beqa
\ds_{NNLO}&=&\int_{{\rm{d}}\Phi_{m+2}}\left[\ds_{NNLO}^{RR}-\ds_{NNLO}^S\right]
\nonumber \\
&+& \int_{{\rm{d}}\Phi_{m+1}}
\left[
\ds_{NNLO}^{RV}-\ds_{NNLO}^{T}
\right] \nonumber \\
&+&\int_{{\rm{d}}\Phi_{m\phantom{+1}}}\left[
\ds_{NNLO}^{VV}-\ds_{NNLO}^{U}\right],\label{eq.subnnlo}
\eeqa
where the terms in each of the square brackets is finite and well behaved in the infrared singular regions.  

Due to the presence of massive fermions in the final state, fewer subtractions terms are required since the real matrix elements develop singular behaviours in fewer regions of phase space than in the massless case. However, the kinematics is more involved due to the finite value of the parton masses and the integration of the subtraction terms is more difficult. A major step concerning the latter has recently been accomplished in \cite{bernreuther}, where the authors integrated one of the genuine NNLO antenna functions. We shall also use their results in this paper. 

More precisely, the purpose of this paper is to evaluate  the first line of eq.(\ref{eq.subnnlo}), i.e. to evaluate $\ds_{NNLO}^{RR}-\ds_{NNLO}^S$ applied to the production of $t \bar{t}$ at LHC using a generalisation of the antenna formalism at NNLO. Towards a computation of all partonic channels contributing to this process, we shall here restrict ourselves to the partonic processes involving only initial and final state fermions. Those can be obtained by crossing any two massless final state fermions to the initial state in the fictitious process $0 \to Q \bar{Q} q \bar{q} q'\bar{q}'$, where we also allow the possibility that both massless quark-antiquark pairs are of the same flavour. It is worth noting here that the contributions coming from the identical-flavour process $ q \bar{q} \to Q \bar{Q} q \bar{q}$ as well as the contributions from the process $ q q'  \to Q \bar{Q} q q'$ with its identical flavour counterpart are presented here for the first time.\\

The plan of the paper is as follows: In section \ref{sec.formalism} we present a generalisation of the antenna subtraction formalism to evaluate hadronic reactions involving a final-state heavy quark pair at NNLO. There, we present the construction of the double real emission subtraction terms capturing single and double unresolved contributions of real radiation matrix elements related to the all-fermion processes. We particularly focus on the changes introduced by the presence of massive final states in the expressions of the subtraction terms compared to the case in which only massless partons are present. In section \ref{sec.antennae} we give a list of all genuine NNLO four-parton tree-level antennae required. In section \ref{sec.limits}, after having presented all massless and massive double unresolved factors encountered in singular configurations of the double real matrix elements and four-parton antennae, we give the single and double unresolved limits of the massless and massive four-parton tree-level antennae encountered in our calculation. Section \ref{sec.subterms} contains our main results. There, we present the colour decomposition of the real contributions and give their associated subtraction terms for the all-fermion processes involved. In all cases, leading and subleading colour contributions are considered. In section \ref{sec.results}, we test the validity of the subtraction terms. We check that the ratio between the real radiation matrix elements and the corresponding subtraction terms approaches unity in all single and double unresolved limits. Finally, section \ref{sec.conclusions} contains our conclusions.

Two appendices are also enclosed: appendix A gives a list of all the tree-level three-parton antennae required in the context of this paper, while appendix B summarises the phase space mappings required to construct the subtraction terms presented in section \ref{sec.subterms} in all configurations.

\section{Antenna subtraction for double real radiation with massive final state fermions}\label{sec.formalism}
Beyond the leading order and up to the next-to-next-to leading order in perturbation theory, the tree-level matrix elements squared associated to the real emission contributions, when integrated over the phase space, develop singularities when one or two final-state partons  are unresolved, i.e. are soft or collinear. The construction of subtraction terms can face this problem as mentioned in section \ref{sec.intro}.

The aim of this section is to present the general structure of the subtraction terms which capture the single and double unresolved features of the real radiation contributions related to the hadronic production of a pair of heavy quarks $Q\bar{Q}$ in association with $(m-2)$ jets at NNLO in perturbative QCD. For this purpose, we develop  an extended version of the antenna formalism, in order to account for hadronic processes with a massive fermion pair in the final state.

As mentioned in section \ref{sec.intro}, for the kinematical situation under consideration in this paper, i.e. the production of a heavy fermion pair whose mass is of the the same order  as the partonic center-of-mass energy, the single and double unresolved features are related only to collinear and/or soft radiation of up to two massless partons. Mass effects, however, are present in the infrared limits of the matrix elements squared: The presence of massive partons in the final state changes the kinematics and influences the soft behaviour of the real matrix elements in a non-trivial way as explained in sections~\ref{sec.limits} and~\ref{sec.subterms}.  

Quite generally within the antenna formalism, the subtraction terms, which reproduce the singular behaviour of tree-level real matrix elements,
are constructed from products of antenna functions with reduced matrix elements. These antenna functions capture all unresolved radiation emitted between two hard partons, the radiators, which can be either massless or massive. The subtraction terms can be integrated over a phase space which is factorised into an antenna phase space (involving all unresolved partons and the two radiators) multiplied by a reduced phase space, where the momenta of the radiators and the unresolved radiation are replaced by two redefined momenta. These redefined momenta can be in the initial or in the final state depending on where the corresponding radiator momenta are, and they are defined by appropriate mappings. Depending on where the two radiators are located, in the initial or in the final state, we distinguish three types of configurations: final-final, initial-final and initial-initial. They will be denoted with the shorthand notation: (f-f, i-f, i-i). The full subtraction term is then obtained by summing over all antennae required in one configuration and by summing over all configurations needed for the problem under consideration.

Subtraction terms are constructed using the fundamental factorisation properties of QCD amplitudes in their soft and collinear limits. However, in QCD (unlike in QED) these fundamental properties are satisfied by colour-ordered sub-amplitudes rather than by the the full amplitudes. The antenna formalism uses colour ordering properties in an essential way: Real emission matrix elements are decomposed into a linear combination of colour-ordered amplitudes, and subtraction terms are constructed for each of these amplitudes separetely. Furthermore, in colour-ordered amplitudes, only partons which are colour-connected can lead to a singular  behaviour of those. This helps  considerably to determine the type of antenna functions that are to be used in each subtraction term. 

At the next-to-leading order, antenna subtraction terms are constructed solely with tree-level three-parton antenna functions which can be massless \cite{k,antennannlo,daleo} or massive \cite{mathias, us} depending on the process under consideration. These three-parton antenna functions encapsulate all singular limits due to the emission of {\it one} unresolved parton between two colour-connected hard radiators.

At NNLO, the subtraction terms aiming to capture the single and double unresolved features of the double real matrix elements squared are constructed with either products of two three-parton antennae, or with genuine NNLO four-parton antennae. This will depend on the colour connection  between the unresolved partons as explained in the massless case in \cite{antennannlo,joao} or as will be explained below in the massive case. 

Compared to the massless antenna formalism, the presence of massive partons in the final states modifies the subtraction terms in a non-trivial way. Parton masses lead to modified kinematics and have to be taken into account for the phase space factorisations. Furthermore, those also modify the soft behaviour of the real matrix elements and it is precisely in order to properly capture these soft behaviours within this formalism, that antennae with massive radiators have to be considered.

For the NNLO corrections to $Q\bar{Q}$ + jets production in hadronic collisions, we will need all three types of subtraction terms and therefore all three types (f-f, i-f, i-i) of three and four-parton antenna functions, involving either massless or massive radiators. Three-parton antennae (massless and massive) have been derived before. Those of relevance in our context, the $t \bar{t}$ hadronic production from all-fermion process, will be given in appendix A. All required four-parton antenna functions will be presented in section \ref{sec.antennae} and their limits will be given in section~\ref{sec.limits}. The phase space factorisation required to define the genuine NNLO subtraction terms, i.e. those involving four-parton-antennae will be presented below.

\subsection{Double real radiation contributions to heavy quark pair production} 
We are interested in describing the double real radiation corrections to the process 
\begin{equation}
 p p \to Q \bar{Q} + (m-2) {\rm jets}.
 \end{equation}
To setup the normalisation, we first present the leading order (LO) $m$-partonic contribution to this process where the partons can either 
be massless or massive. It reads,
\begin{eqnarray}
\lefteqn{{\rm d} \hat\sigma_{LO}(p_1,p_2) =
{\cal N}_{LO}\,
\sum_{{m-2}}{\rm d}\Phi_{m}(p_{Q},p_{\bar{Q}},p_{5},\ldots,p_{m-2};
p_1,p_2)} \nonumber \\ && \times 
\frac{1}{S_{{m-2}}}\,
|{\cal M}_{m+2}(p_{Q},p_{\bar{Q}},p_{5},\ldots,p_{m-2};p_1,p_2)|^{2}\; 
\JET_{m}^{(m)}(p_{Q},p_{\bar{Q}},p_{5},\ldots,p_{m-2}).
\label{eq.siglo}
\end{eqnarray}
$p_1$ and $p_{2}$ are the momenta of the initial state partons, the massive partons $Q$ and $\bar{Q}$ have momenta $p_{Q}$ and $p_{\bar{Q}}$, while the momenta of the remaining $(m-2)$ massless final state partons are labelled $p_{5} \ldots p_{m-2}$. $S_{m-2}$ is a symmetry factor for identical massless partons in the final state. $\JET_{m}^{(m)}(p_{Q},p_{\bar{Q}},p_{5},\ldots p_{m-2})$ is the jet function. It ensures that out of $(m-2)$ massless partons and a pair of heavy quarks $Q$ and $\bar{Q}$ present in the final state at parton level, an observable with a pair of heavy quark jets in association with $(m-2)$ jets is built. At this order each massless or massive parton forms a jet on its own. The normalization factor ${\cal N}_{LO}$ includes all QCD-independent factors as well as the dependence on the renormalised QCD coupling constant $\alpha_s$. $\sum_{m-2}$ denotes the sum over all configurations with $(m-2)$ massless partons. ${\rm d}\Phi_{m}$ is the phase space for an $m$-parton final state containing $(m-2)$ massless and two massive partons with total four-momentum $p_1^{\mu}+p_2^{\mu}$. In $d=4-2\e$ space-time dimensions, this phase space takes the form:
\begin{eqnarray}
\lefteqn{\d \Phi_m(p_{Q},p_{\bar{Q}},p_{5},\ldots,p_{m-2};p_1,p_2) = 
\frac{\d^{d-1} p_Q}{2E_Q (2\pi)^{d-1}}\; 
\frac{\d^{d-1} p_{\bar{Q}}}{2E_{\bar{Q}} (2\pi)^{d-1}}} \nonumber \\
&&\times  \frac{\d^{d-1} p_5}{2E_5 (2\pi)^{d-1}}\; \ldots \;
\frac{\d^{d-1} p_{m-2}}{2E_{m-2} (2\pi)^{d-1}}\; (2\pi)^{d} \;
\delta^d (p_1+p_2 - p_{Q}-p_{\bar{Q}}-p_5 - \ldots p_{m-2}) \,.\hspace{2mm}
\label{eq:phasem}
\end{eqnarray}

In eq.(\ref{eq.siglo}) $|{\cal M}_{m+2}|^2$ denotes a colour-ordered tree-level matrix element squared with  $m$ final state partons, out of which two are massive and two are initial state partons. These terms only account for the leading colour contributions to the squared matrix elements. On the other hand, subleading colour contributions involve in general interferences between two colour-ordered parton amplitudes. However, to keep the notation simpler we denote these interference contributions also as $|{\cal M}_{m+2}|^2$. 

The NLO real radiation partonic contributions to the production of a heavy quark pair in association with $(m-2)$ jets involve $(m+1)$-final state partons (one more parton than the leading-order case) with two of them being massive. They have been derived in \cite{us} together with their corresponding subtraction terms capturing all single unresolved radiation. 

Here we are interested in the construction of the subtraction terms for the double real radiation ${\rm d}\hat\sigma^{S}_{NNLO}$ which correctly subtract all single and double unresolved singularities contained in the matrix elements with $(m+2)$-final state partons present in the double real emission contributions. These may be written as
 \begin{eqnarray}
\lefteqn{{\rm d}\hat\sigma^{RR}_{NNLO}(p_1,p_2)=
{\cal N}_{NNLO}\,
\sum_{{m+2}}{\rm d}\Phi_{m+2}(p_{Q},p_{\bar{Q}},p_{5},\ldots,\,p_{m};
p_1,p_2) }\nonumber \\ && \times 
\frac{1}{S_{{m+2}}}\,
|{\cal M}_{m+4}(p_{Q},p_{\bar{Q}},p_{5},\ldots,p_{m};p_1,p_2)|^{2}\; 
\JET_{m}^{(m+2)}(p_{Q},p_{\bar{Q}},p_{5},\ldots,p_{m}).\hspace{3mm} \\
&\equiv&
{\cal N}_{NNLO}\,
\sum_{{m+2}}{\rm d}\Phi_{m+2}(p_{3}, \ldots, p_{m+4}; p_1,p_2) \nonumber \\ && \times 
\frac{1}{S_{{m+2}}}\,
|{\cal M}_{m+4}(p_{3},\ldots, p_{m+4},;p_1,p_2)|^{2}\; 
\JET_{m}^{(m+2)}(p_{3},\ldots,p_{m+4}),\hspace{3mm}\label{eq.real}
\end{eqnarray}
where the last line is obtained by relabelling all final state partons. This procedure enables us to follow the massless notation as given in \cite{joao,RVnew} more closely. In this equation, the jet function $\JET_{m}^{(m+2)}$ ensures that out of $m$ massless partons and a $Q\bar{Q}$ pair, an observable with a pair of heavy quark jets in addition to $(m-2)$ jets, is built.

This contribution to the NNLO cross section develops singularities if one or two partons are unresolved (soft or collinear). Depending on the colour  connection between these unresolved partons, the following configurations must be distinghuished \cite{antennannlo,joao}:  

\begin{itemize}
\item[(a)] One unresolved parton but the experimental observable selects only $m$ jets.
\item[(b)] Two colour-connected unresolved partons (colour-connected).
\item[(c)] Two unresolved partons that are not colour-connected but share a common radiator (almost colour-unconnected).
\item[(d)] Two unresolved partons that are well separated from each other in the colour chain (colour-unconnected).
\item[(e)] Compensation terms for the over-subtraction of large angle soft emission
\end{itemize}

This separation among subtraction contributions according to colour connection is valid in final-final, initial-final or initial-initial configurations 
and, in any of them, the subtraction formulae have a characteristic structure in terms of the required antenna functions. This antenna structure has been derived for processes involving only massless partons, for the final-final and initial-final cases in \cite{antennannlo,Gionata} and in \cite{joao} for the initial-initial case. The presence of massive partons in the final state does not modify the general structure of the subtraction terms required to match the unresolved features given above in itemized form.    

Concerning the items related to the colour-connection of the unresolved partons, the configuration $(c)$, the almost colour-unconnected case, and the configuration $(e)$, regarding the treatment of large angle soft radiation, do not occur in the all-fermion processes that contribute to the double real corrections to $t \bar{t}$ production. On one hand, there are not enough final state partons (besides the pair of massive ones) to cover the configuration $(c)$ and, on the other hand, the absence of final state gluons forbids configuration $(e)$. In the following, we shall therefore restrict ourselves to describe how the configurations $(a)$, $(b)$ and $(d)$ are dealt with here while leaving the discussion of configurations $(c)$ and $(e)$ to be treated elsewhere.

\subsection{Subtraction terms for single unresolved radiation ${\rm d}\hat\sigma^{S,a}_{NNLO}$}
The subtraction terms for single unresolved radiation associated to $Q \bar{Q}$ production in association with $(m-2)$- jets can be taken from the expressions  derived in \cite{us}. It is recalled below in all three final-final (f-f), initial-final (i-f)  and initial-initial (i-i) configurations.
\begin{eqnarray}
\lefteqn{{\rm d}\hat\sigma_{NNLO}^{S,a,(ff)}
=  {\cal N}_{NNLO}\,\sum_{\textrm{perms}}{\rm d}\Phi_{m+2}(p_{3},\ldots,p_{m+4};p_1,p_2)
\frac{1}{S_{{m+2}}} }\nonumber \\
&\times& \,  \sum_{j}\;X^0_{ijk}\,
|{\cal M}_{m+3}(\ldots,I,K,\ldots)|^2 J_{m}^{(m+1)}(p_{3},\ldots,p_I,p_K,\ldots,p_{m+4})\;
,\label{eq.sub2aff}  \\
\lefteqn{{\rm d}\hat\sigma_{NNLO}^{S,a,(if)}
={\cal N}_{NNLO}\,\sum_{\textrm{perms}}{\rm d}\Phi_{m+2}(p_3,\ldots,p_{m+4};p_1,p_2)
  \frac{1}{S_{m+2}} }\nonumber\\
&\times&\, \sum_{{i}=1,2}\sum_{j} X^{0}_{i,jk}\,
 |{\cal M}_{m+3}(\ldots,\hat{I},K,\ldots)|^2 J^{(m+1)}_{m}(p_{3},\ldots,p_K,\ldots,p_{m+4})\;,\label{eq.sub2aif}\\
\lefteqn{{\rm d}\hat\sigma_{NNLO}^{S,a,(ii)}
={\cal N}_{NNLO}\, \sum_{\textrm{perms}}{\rm d}\Phi_{m+2}(p_{3},\ldots,p_{m+4};p_1,p_2)
  \frac{1}{S_{m+2}}}\nonumber \\
&\times&\, \sum_{i,k=1,2} \sum_{j} X^{0}_{ik,j}
  |{\cal M}_{m+3}(\ldots,\hat{I},\hat{K},\ldots)|^2
  \,J^{(m+1)}_{m}(\tilde{p}_3,\ldots,\tilde{p}_{m+4}).\label{eq.sub2aii} 
\end{eqnarray}
In these equations, the sum over $j$ is the sum over all unresolved partons in a colour-ordered amplitude between radiators $i$ and $k$ which can be both located in the final state eq.(\ref{eq.sub2aff}), $\hat{i}$ in the initial state and $k$ in the final state eq.(\ref{eq.sub2aif}) or both in the initial state eq.(\ref{eq.sub2aii}). 

In the reduced matrix elements $|{\cal M}|^2$, the redefined momenta are respectively denoted by $(I,K)$ in the final-final, by $(\hat{I},K)$ in the initial-final and by $(\hat{I},\hat{K})$ in the initial-initial configurations, respectively. We shall adopt this notation throughout the paper: Initial state momenta will be denoted with a hat everywhere.

The subtraction terms given in the equations above, involve the phase space for the production of $(m+2)$ partons, $\d \Phi_{m+2}$ with two of them being massive, the three-parton antenna functions appropriate to a given configuration $X^{0}_{ijk}$, $X^{0}_{i,jk}$ and $X^{0}_{ik,j}$, the colour-ordered reduced $(m+3)$-parton amplitude squared $|{\cal M}_{m+3}|^2$ and the jet function $\JET^{(m+1)}_{m}$.

Compared to the NLO expressions given in \cite{us} the NLO jet function $J_{m}^{(m)}$ is now replaced by $J_{m}^{(m+1)}$ in all three configurations. It ensures that out of $(m-1)$ massless partons and a pair of massive final state partons, $(m-2)$ jets and a $Q\bar{Q}$ jet pair is built. The jet function and the reduced $m$-parton amplitude do not depend on the individual momenta ${p}_{i}$, $p_j$ and ${p}_{k}$ but will only depend on the redefined momenta $p_{I}$ and $p_{K}$ which are linear combinations of the original momenta $p_{i},p_{j},p_{k}$ in the appropriate configuration. The three types of  three-parton antenna functions $X^{0}_{i,jk}$ ,$X^{0}_{i,,jk}$ and $X^{0}_{i,k,j}$ can involve either massless or massive hard radiators. They depend only on the original momenta $p_i$, $p_j$ and $p_{k}$ , and in  the presence of massive hard radiators, also on the masses of those final state partons. These antennae have all been derived before and the ones relevant in the context of this paper are given in appendix A for completeness. 

Depending on the configuration considered, the momenta present in the reduced matrix elements and in the jet function are defined with a particular $3\to 2$ mapping. Specially, for the initial-initial mapping, all momenta, massless and massive, in the arguments of the reduced matrix elements $|{\cal M}_{m+3}|^2$ and the jet function $\JET^{(m+1)}_{m} $ have to be redefined. They are denoted with tildes in eq.(\ref{eq.sub2aii}). The two hard radiators are simply rescaled by factors $x_1$ and $x_2$ respectively. In \cite{us} we showed that the original initial-initial mapping derived in \cite{daleo} for massless partons only, can also be used for processes involving massive partons. These mappings are recalled in appendix B.

In the subtraction terms above,  ${\rm d}\hat\sigma_{NNLO}^{S,a}$ coincides with the matrix element (\ref{eq.real}) when $j$ is unresolved.
At NNLO, however, the jet function $J_m^{(m+1)}$ allows one of the $(m+1)$ momenta to become unresolved. As a consequence, the matrix element present in the subtraction terms can have an unresolved parton. In this limit ${\rm d}\hat\sigma_{NNLO}^{S,a}$ does not coincide with the matrix element (\ref{eq.real}) and, as explained in detail in \cite{antennannlo,joao}, these spurious double unresolved configurations  will cancel against  similar unresolved features encountered for the subtraction terms coming from other configurations. In this paper, those other configurations will arise only for the items $(b)$ or $(d)$, associated to two unresolved particles which are colour-connected or colour-disconnected respectively.

The subtraction terms ${\rm d}\hat\sigma^{S,a}_{NNLO}$ defined above, need to be added back in their integrated form to the $(m+1)$ partonic contributions. Poles from those will cancel against explicit poles in the one-loop virtual ${\rm d}{\sigma}^{RV}$ contributions and against those present in the mass factorisation contributions $ {\rm d}\sigma^{MF,1}$ given in eq.(\ref{eq:subnnlo1}).

To define the integrated forms of these subtraction terms ${\rm d}\hat\sigma_{NNLO}^{S,a}$ involving only three-parton antennae, an appropriate phase space factorisation needs to be considered in all three configurations. Those have been derived in the massless case in \cite{antennannlo, daleo} and in the massive case in \cite{us,mathias}.

\subsection{Subtraction terms for colour-connected double unresolved radiation ${\rm d}\hat\sigma^{S,b}_{NNLO}$}
When two unresolved partons $j$ and $k$ are adjacent, we build the subtraction term starting from the four-particle tree-level antennae in all three configurations $X_{ijkl}$, $X_{i,jkl}$ and  $X_{il,jk}$. By construction, they contain all colour-connected double unresolved limits of the $(m+4)$-parton matrix element associated with partons $j$ and $k$ unresolved between radiators $i$ and $l$. However, these antennae can also become singular in single unresolved limits associated with $j$ or $k$ where they do not coincide with limits of the matrix elements. To ensure that these subtraction terms are only active in the double unresolved limits of the real matrix elements, we subtract the appropriate single unresolved limits of the four-particle tree-level  antennae. For this purpose we use products of two tree-level three-particle antennae in the appropriate configurations. As in the single unresolved case, we replace the original hard radiators with new particles, $I$ and $L$. When one of the hard radiators is in the initial state, $p_{\hat{I}} \equiv \hat{\bar{p}}_i = x_i p_i$ and when both are in the initial state,
$p_{\hat{I}} \equiv \hat{\bar{p}}_i = x_i p_i$, $p_{\hat{L}} \equiv \hat{\bar{p}}_l = x_l p_l$ and all other momenta have to be Lorentz boosted.

The colour-connected double subtraction term in all three configurations then reads:
\begin{eqnarray}
\lefteqn{{\rm d}\hat\sigma_{NNLO}^{S,b,(ff)}
=  {\cal N}_{NNLO}\,\sum_{\textrm{perms}}{\rm d}\Phi_{m+2}(p_{3},\ldots,p_{m+4};p_1,p_2)
\frac{1}{S_{{m+2}}}\,}\nonumber \\
&\times&  \sum_{jk}\left( X^0_{ijkl}
- X^0_{ijk} X^0_{IKl} - X^0_{jkl} X^0_{iJL} \right)\nonumber\\
&\times&|{\cal M}_{m+2}(\ldots,I,L,\ldots)|^2\,
J_{m}^{(m)}(\ldots,p_{I},p_{L},\ldots),\;
\label{eq.sub2bff}\\
\lefteqn{{\rm d}\hat\sigma_{NNLO}^{S,b,(if)}
=  {\cal N}_{NNLO}\,\sum_{\textrm{perms}}{\rm d}\Phi_{m+2}(p_{3},\ldots,p_{m+4};p_1,p_2)
\frac{1}{S_{{m+2}}} }\nonumber \\
&\times&\sum_{i=1,2} \sum_{jk}\;\left( X^0_{i,jkl}
- X^0_{i,jk} X^0_{I,Kl} - X^0_{jkl} X^0_{i,JL} \right)\nonumber\\
&\times&|{\cal M}_{m+2}(\ldots,\hat{I},L,\ldots)|^2\,
J_{m}^{(m)}(\ldots,p_{L},\ldots) 
 \;,
\label{eq.sub2bif}\\
\lefteqn{{\rm d}\hat\sigma_{NNLO}^{S,b,(ii)}
=  {\cal N}_{NNLO}\,\sum_{\textrm{perms}}{\rm d}\Phi_{m+2}(p_{3},\ldots,p_{m+4};p_1,p_2)
\frac{1}{S_{{m+2}}} }\nonumber \\
&\times& \,\sum_{il=1,2} \sum_{jk}\;\left( X^0_{il,jk}
- X^0_{l,jk} X^0_{iL,K} - X^0_{i,kj} X^0_{Il,J} \right)\nonumber \\
&\times&
|{\cal M}_{m+2}(\ldots,\hat{I},\hat{L},\ldots)|^2\,
J_{m}^{(m)}(\tilde{p}_3,\ldots,\tilde{p}_{m+4}),\,
\label{eq.sub2bii}
\end{eqnarray}
where the sum runs over all colour-adjacent pairs $j,k$ and implies the appropriate selection of hard momenta $i,l$ which, as usual, have three possible assignments of radiators. In all cases the $(m+2)$-parton matrix element is evaluated with new on-shell momenta given by a momentum mapping appropriate in each configuration. Those $4 \to 2$ mappings have been derived in \cite{daleo,joao} in the massless case. We checked that the same mappings, which are presented in appendix B for completeness, can also be applied in the presence of massive partons. 
 
The ${\rm d}\hat\sigma_{NNLO}^{S,b}$ subtraction terms involve: the phase space for the production of $(m+2)$ partons, $\d \Phi_{m+2}$, with two of them being massive, the colour-ordered reduced $(m+2)$-parton amplitude squared $|{\cal M}_{m+2}|^2$, the products of three-parton antennae and four-parton antennae appropriate to a given configuration and the jet function  $\JET^{(m)}_{m}$. As before, reduced amplitudes and jet functions only depend on redefined momenta in the appropriate configurations, while the tree-level  four-parton antennae depend on the original momenta $i,j,k,l$.

The relevant massless and massive four-parton antenna functions required to derive the subtraction terms for the double real contributions to $t \bar{t}$ production coming from fermionic processes will be given in section~\ref{sec.antennae} while their single and double unresolved limits will be given in section~\ref{sec.limits}.

\subsubsection{Phase space factorisation for two colour-connected unresolved particles}
As mentioned before and explained in \cite{RVnew}, the part of  ${\rm d}\hat\sigma_{NNLO}^{S,b}$ involving products of three-parton antennae needs to be added back in its integrated form at the $(m+1)$-parton level. The part of ${\rm d}\hat\sigma_{NNLO}^{S,b}$ involving four-parton antennae, instead, needs to be added back in its integrated form as an $m$-parton contribution. It can then be combined with the virtual two-loop contributions ${\rm d}\sigma^{VV}$ and the mass factorisation counterterms ${\rm d}\sigma^{MF,2}$ , which have also $m$ final state partons as shown in  eq.(\ref{eq.sigNNLO}), and as explained in \cite{antennannlo,joao,RVnew}.  

In order to be able to integrate these subtraction terms, ${\rm d}\hat\sigma_{NNLO}^{S,b}$ involving four-parton antennae, one therefore needs the $2 \to (m +2)$ parton phase space to factorise into an antenna phase space involving four-partons, the unresolved partons and the radiators, times a reduced $2 \rightarrow m$ phase space. This factorisation will be different depending on whether the hard radiators are in the initial or in the final state. For the cases involving only massless particles the phase space factorisations for final-final, initial-final and initial-initial configurations have been derived in \cite{antennannlo,daleo,joao}.

The presence of massive hard radiators introduces slight differences with respect to the massless case in the factorisation formulae for final-final and initial-final configurations. In the initial-initial configuration, the situation is unchanged as the radiators are massless. In the following, we shall present the factorisation formulae for the phase space in these three configurations while restricting ourselves to the kinematical situations where the two unresolved partons are massless. 

The final-final case has already been derived in \cite{bernreuther} . We recall it here for completeness. We start from the $ 2\to (m+2)$ parton phase space ${\rm d} \Phi_{m+2}$ as given in eq.(\ref{eq:phasem}) involving two massive and $m$ massless partons whose momenta are denoted by 
$p_3,\ldots,p_{m+4}$. It can be written as  
\begin{eqnarray}
\label{eq.psx4}
{\rm d} \Phi_{m+2}(p_3,\ldots,p_{m+4};p_1,p_2)&=& 
{\rm d} \Phi_{m}(p_{3},\ldots,p_I,p_L,\ldots,p_{m+2};p_1,p_2)\nonumber\\
&\times&{\rm d} \Phi_{X_{ijkl}} (p_i,p_j,p_k,p_l;p_I+p_L).\label{eq.psfi4}
\end{eqnarray}
In this case, $p_i$ and $p_{l}$ are the momenta of the massive radiators $i,l$  between which the colour-connected and massless unresolved partons $j,k$ with momenta $p_j$ and $p_k$ are emitted. The antenna phase space  ${\rm d}\Phi_{X_{ijkl}} (p_i,p_j,p_k,p_l;p_I+p_L)$ is closely related to the 4-particle massive phase space ${\rm d} \Phi_{4}(p_i,p_j,p_k,p_l;p_I+p_L)$ as follows
\begin{equation}
{\rm d} \Phi_{4}(p_i,p_j,p_k,p_l;p_I +p_L )=P_{2}(q^2, m_{Q}) 
\times{\rm d} \Phi_{X_{ijkl}} (p_i,p_j,p_k,p_l;p_I+p_L),\\
\end{equation}
where $q^2= (p_I+p_L)^2$ and $P_{2}(q^2,m_{Q})$ denotes the inclusive phase space measure for two particles of equal masses $m_{Q}$ in $d=4-2 \e$ dimensions.\\

For the subtraction terms built with massive initial-final four-parton antenna functions (with a massless initial state hard radiator and a massive final state one) we want our $2\rightarrow (m+2)$ phase space to factorise into a $2\rightarrow m$ reduced phase space (with massive and massles final state particles) times a $2\rightarrow 3$ antenna phase space with at least one massive final state particle. To show how the factorisation is derived, we start again from the same $2\rightarrow (m+2)$ particle phase space and consider the following situation: the massless initial state hard radiator has momentum $p_1$, the massive final state radiator has momentum $p_l$, the unresolved and colour-connected particles have momenta  $p_j$ and $p_k$. We insert
\beq
1=\int {\rm d}^d q\:\delta(q+p_1-p_j-p_k-p_l),
\eeq
and
\beq
1=\frac{Q^2+m_{jkl}^2}{2\pi}\int\frac{{\rm d} x}{x}\int [{\rm d}\wt{p_{jkl}}](2\pi)^d\:\delta(q+xp_1-\wt{p_{jkl}})
\eeq
with
\beq
Q^2=-q^2\hspace{20mm}m_{jkl}^2=m_j^2+m_k^2+m_l^2
\eeq
and
\beq\label{eq.x}
 x=\frac{Q^2+m_{jkl}^2}{2p_1\cdot q}=\frac{s_{1j}+s_{1k}+s_{1l}-s_{jk}-s_{jl}-s_{kl}}{s_{1j}+s_{1k}+s_{1l}},
\eeq
in eq.(\ref{eq.psfi4}), the expression of the $2\rightarrow (m+2)$ phase space. By doing so, we find that it can be conveniently written as follows
\beqa
\dphi_{m+2}(p_3,\ldots,p_{m+4};p_1,p_2)=\dphi_m (p_1,\ldots,\wt{p_{jkl}},\ldots,p_{m+2};xp_1,p_2)\nonumber\\
\times\frac{Q^2+m_{jkl}^2}{2\pi}\dphi_3 (p_j,p_k,p_l;p_1,q)\frac{{\rm d} x}{x}.\label{eq.psif4}
\eeqa
The four-parton intial-final phase space ${\rm d} \Phi_{X_{i,jkl}}$ involving one massless and one massive radiators in the intial and final state respectively, given for $i=1$ by
\begin{eqnarray}
{\rm d} \Phi_{X_{i,jkl}}=\frac{Q^2+m_{jkl}^2}{2\pi}\dphi_3 (p_j,p_k,p_l;p_1,q),
\end{eqnarray}
can be regarded as the new initial-final massive antenna phase space. It is worth noting that, in the case where $j,k$ are massless and only $l$ is massive, we have $m_{jkl}=m_{l}=m_{L}$. Moreover, another important observation is that, when written in terms of $s_{ab}=2p_a\cdot p_b$, the rescalling variable $x$ defined in eq.(\ref{eq.x}) has the same form as in the massless case \cite{Gionata}. Since this same variable $x$ is used in the definition of the phase space mappings for initial-final double unresolved configurations (see appendix B.2.2), the aforementioned observation suggests that the phase space mappings should be the same in the massless and massive cases provided that all variables are expressed in terms of invariants as in eq.(\ref{eq.x}). This is indeed the case. \\

For the initial-initial phase space factorisation, we can use the phase space factorisation derived in \cite{daleo,joao} as in the massless case. The hard radiators situated in the initial state are obviously massless, and massless (and massive) final state partons have to be redefined using a Lorentz boost. In this case the phase space can be factorised as
\begin{eqnarray}
\label{eq.psii4}
{\rm d}\Phi_{m+2}(p_3,\ldots,p_{m+4};p_1,p_2)&=&
{\rm d}\Phi_{m}(\tilde{p}_3,\ldots,\tilde{p}_{m+4};x_1p_1,x_2p_2)
\nonumber\\
&&\times\delta(x_1-\hat{x}_1)\,\delta(x_2-\hat{x}_2)\,[{\rm d}p_j][{\rm d}p_k]\, {\rm d}x_1\, {\rm d}x_2\,,\label{eq.psii4}
\end{eqnarray}
where the tilde over the final state momenta indicates that they are boosted.\\

Using the factorised forms of the phase space in the three configurations as defined above in eqs. (\ref{eq.psfi4}), (\ref{eq.psif4}) and (\ref{eq.psii4}) one can explicitly rewrite each of the four-particle subtraction terms present in 
${\rm d}\sigma_{NNLO}^{S,b}$ in the forms,
 \begin{eqnarray}
&&|{\cal M}_{m+2}|^2\,
J_{m}^{(m)}\; 
{\rm d}\Phi_{m}
\int {\rm d} \Phi_{X_{ijkl}}\;X^0_{ijkl},\\
&&|{\cal M}_{m+2}|^2\,
J_{m}^{(m)}\; 
{\rm d}\Phi_{m}
\int \frac{Q^2+m_{jkl}^2}{2\pi} {\rm d}\Phi_{3}(p_j,p_k,p_l;p,q)\;X^0_{i,jkl}\frac{{\rm d}x}{x},\\
&&|{\cal M}_{m+2}|^2\,
J_{m}^{(m)}\; 
{\rm d}\Phi_{m}
\int [{\rm d} p_j][{\rm d} p_k] \delta(x_1-\hat{x}_1)\,\delta(x_2-\hat{x}_2)\;X^0_{il,jk}  {\rm d}x_1\,  {\rm d}x_2.
\end{eqnarray}

Using this, we then carry out the integration of the corresponding four-parton antenna functions. The integrated forms of the four-parton antennae are defined by
\begin{eqnarray}
\label{eq.x4intff}
&&{\cal X}^0_{ijkl} = \frac{1}{[C(\epsilon)]^2}
\int {\rm d} \Phi_{X_{ijkl}}\;X^0_{ijkl},\\
\label{eq.x4intif}
&&{\cal X}^0_{i,jkl}(x_i)=\frac{1}{[C(\epsilon)]^2}\int {\rm d}\Phi_3 \frac{Q^2+m_{jkl}^2}{2\pi} X^0_{i,jkl},\\
\label{eq.x4intii}
&&{\cal X}^0_{il,jk}(x_i,x_l)=\frac{1}{[C(\epsilon)]^2}\int [{\rm d} p_j][{\rm d} p_k]\;x_i\;x_l\; \delta(x_i-\hat{x}_i)\,\delta(x_l-\hat{x}_l)\,X^0_{il,jk},\,
\end{eqnarray}
where 
\beq\label{eq.ceps}
C(\epsilon)=(4\pi)^{\epsilon}\frac{e^{-\epsilon \gamma}}{8\pi^2},
\eeq
and where the antennae and antenna phase spaces including possibly the masses of the radiators have been derived above in eqs. (\ref{eq.psfi4}), (\ref{eq.psif4}) and (\ref{eq.psii4}).

These integrations are performed analytically in $d$ dimensions to make the infrared singularities explicit.

\subsection{Subtraction terms for colour-unconnected unresolved radiation ${\rm d}\hat\sigma^{S,d}_{NNLO}$}
We finally arrive at the last configuration considered in this section and relevant for our purpose in this paper, namely the configuration $(d)$ involving the situation where the two unresolved partons are pair-wise colour-unconnected. For colour-ordered amplitudes of the type  ${\cal M}(\hdots,i,j,k,\hdots,n,o,p,\hdots)$, where partons $j$ and $o$ are unresolved, the $(m+4)$-parton matrix element factorises into the product of two uncorrelated single unresolved factors with hard partons $I,K$ and $N,P$ respectively multiplied by a reduced $(m+2)$-parton matrix element. The structure is the same as in the massless case \cite{joao}. The difference arises only from the use of  three-parton antennae with massive radiators and massive phase spaces.

The subtraction terms for the colour-unconnected configuration read,
\begin{eqnarray}
\lefteqn{{\rm d}\hat\sigma_{NNLO}^{S,d,(ff)}
= - {\cal N}_{NNLO}\,\sum_{\textrm{perms}}{\rm d}\Phi_{m+2}(p_{3},\ldots,p_{m+4};p_1,p_2)
\frac{1}{S_{{m+2}}}} \nonumber \\
&\times& \,\Bigg [ \sum_{j,o}\;X^0_{ijk}\;X^0_{nop}\,
|{\cal M}_{m+2}(\ldots,I,K,\ldots,N,P,\ldots)|^2\,
\nonumber \\ &&\hspace{3cm}\times
J_{m}^{(m)}(p_3,\ldots,p_I,p_K,\ldots,p_N,p_P,\ldots,p_{m+4})\;\Bigg
]\;,\nonumber\\  
\label{eq.sub2dff}\\
\lefteqn{ {\rm d}\hat\sigma_{NNLO}^{S,d,(if)}
= - {\cal N}_{NNLO}\,\sum_{\textrm{perms}}{\rm d}\Phi_{m+2}(p_{3},\ldots,p_{m+4};p_1,p_2)
\frac{1}{S_{{m+2}}} }\nonumber \\
&\times &\Bigg [ \sum_{i=1,2}
\sum_{j,o}\;X^0_{i,jk}\;X^0_{nop}\,
|{\cal M}_{m+2}(\ldots,\hat{I},K,\ldots,N,P,\ldots)|^2\,
\nonumber \\ &&\hspace{3cm}\times
J_{m}^{(m)}(p_3,\ldots,p_K,\ldots,p_N,p_P,\ldots,p_{m+4})\;\Bigg
],\nonumber\\
\label{eq.sub2dif}\\
\lefteqn{ {\rm d}\hat\sigma_{NNLO}^{S,d,(ii)}
= - {\cal N}_{NNLO}\,\sum_{\textrm{perms}}{\rm d}\Phi_{m+2}(p_{3},\ldots,p_{m+4};p_1,p_2)
\frac{1}{S_{{m+2}}}}\nonumber \\
&\times&\Bigg [ \sum_{i,n=1,2}\sum_{j,o}\;X^0_{i,jk}\;X^0_{n,op}\, \,|{\cal M}_{m+2}(\ldots,\hat{I},K,\ldots,\hat{N},P,\ldots)|^2\,
\nonumber \\ &&\hspace{3cm}\times
J_{m}^{(m)}( {p}_3,\ldots,p_K,\ldots,p_P,\ldots, {p}_{m+4})\;
\phantom{\Bigg]}
\nonumber\\
&&+\sum_{k,n=1,2}\sum_{j,o}\;X^0_{k,ji}\;X^0_{n,op}\, \,|{\cal M}_{m+2}(\ldots,I,\hat{K},\ldots,\hat{N},P,\ldots)|^2\,
\nonumber \\ &&\hspace{3cm}\times
J_{m}^{(m)}({p}_3,\ldots,p_I,\ldots,p_P,\ldots,{p}_{m+4})\;
\phantom{\Bigg]}
\nonumber\\
&&+\, \sum_{i,k=1,2}\sum_{j,o}\;X^0_{ik,j}\;X^0_{nop}\,
|{\cal M}_{m+2}
(\ldots,\hat{I},\hat{K},\ldots,N,P,\ldots)|^2\,
\nonumber \\ &&\hspace{3cm}\times
J_{m}^{(m)}(\tilde{p}_3,\ldots,p_N,p_P,\ldots,\tilde{p}_{m+4})\;\Bigg
]\;,
\label{eq.sub2dii}
\end{eqnarray}
where the summation over $o$ is such that it only includes two antenna configurations with no common momenta. The nature of the radiator pairs $(i,k)$ and $(n,p)$ defines the antennae to be used. As it was already mentioned before, this term compensates spurious double unresolved features arising in subtraction terms for configurations $(a)$ and for terms in $(b)$ proportional to products of two three-parton antenna functions. 

To obtain the integrated form of this counterterm we exploit the factorisation of the $(m+2)$-parton phase space into an $m$-parton phase space multiplied by two independent phase space factors for each of the two antennae. The integrated form is thus simply the product of two integrated three-parton antennae. Finally, as discussed in \cite{RVnew} this subtraction term has to be added back in integrated form at the $m$-parton level. \\

As it was already the case at the NLO level~\cite{us}, the subtraction terms derived above in all configurations entering at NNLO, are strictly valid for the subtraction of infrared singularities of colour-ordered matrix elements squared. Interferences between partial amplitudes with different colour orderings appearing in the subleading colour pieces can also develop infrared singularities which need to be subtracted. As we will see in section~\ref{subsec.subleading}, the subtraction of infrared singularities in those terms does require special care. However, to keep our equations as brief and clear as possible, we still write our subtraction terms in all configurations (f-f,i-f and i-i) and in all cases  
$(a)$, $(b)$ or $(d)$ as above.

\section{Antenna functions}\label{sec.antennae}
In this section, we present the expressions of the four-parton antenna functions which are explicitly needed in the subtraction terms for the double real radiation corrections to hadronic heavy quark pair production due to the partonic processes obtained by crossing any pair of massless fermions in the process $0 \rightarrow Q\bar{Q}q \bar{q} q'\bar{q}'$. The two massless quark-antiquark pairs can (but do not necessarily have to) be of the same flavour.

As explained in the previous sections, antenna functions are the key ingredients needed to build subtraction terms in the antenna formalism.
In general, those functions are denoted with the character $X$. Each antenna is determined not only by its particle content (hard radiators and unresolved particles) but also by the pair of hard particles that it collapses to in its singular limits. Antennae that collapse onto a quark-antiquark pair are $X=A$ for $qg\bar{q}$ and $qgg\bar{q}$, $X=B$ for $q\bar{q}'q'\bar{q}$ and $X=C$ for $qq\bar{q}\bar{q}$. Antennae that collapse onto a (anti) quark and a gluon are $X=D$ for $qgg$ and $qggg$, and $X=E$ for $qq'\bar{q}'$ and $qq'\bar{q}'g$. Finally gluon-gluon antennae are $X=F$ for $ggg$ and $gggg$, $X=G$ for $gq\bar{q}$ and $gq\bar{q}g$, and $X=H$ for $q\bar{q}q'\bar{q}'$.

All antenna functions are derived from physical colour-ordered matrix elements squared. Each type of antenna is obtained from a different physical process: quark-antiquark antennae are related to processes of the form $\gamma^* \rightarrow q\bar{q} +$(partons) \cite{GehrmannDeRidder:2004tv}, quark gluon antennae, to $\tilde\chi \rightarrow \tilde{g}+$(partons) \cite{GehrmannDeRidder:2005hi}, and gluon-gluon antennae are derived from $H\rightarrow$(partons) \cite{GehrmannDeRidder:2005aw}. The tree-level antenna functions are obtained, both in the massless or massive case, by normalising the colour-ordered three-(or four)-parton tree-level squared matrix elements to the squared matrix element for the basic two-parton process, omitting couplings and colour factors:
\beqa
&&X_{ijk}^0 = S_{ijk,IK}\, \frac{|{\cal M}^0_{ijk}|^2}{|{\cal M}^0_{IK}|^2}\\
&&X_{ijkl}^0 = S_{ijkl,IL}\, \frac{|{\cal M}^0_{ijkl}|^2}{|{\cal M}^0_{IL}|^2}\label{eq.4partonantenna}.
\eeqa
$S$ denotes the symmetry factor associated with the antenna, which accounts both for potential identical particle symmetries and for the presence of more than one antenna in the basic two-parton process. 

The three and four-parton initial-final and initial-initial antenna functions are, in principle, defined by crossing one or two massless partons from the final to the initial state in the corresponding final-final antennae. The three-parton initial-final and initial-initial antennae are denoted respectively by $X_{i,jk}^0$ and $X_{ik,j}^0$ while the corresponding four-parton antennae will  be denoted by $X_{i,jkl}^0$ and $X_{ik,jl}^0$ \footnote{It is woth noting that, in the initial-final case this crossing may not be unambiguous and may result in the necessity of a further splitting of the antenna as noted in \cite{daleo,us}. However, those antennae are not needed in the present context of this paper.}.

While at NLO only three-parton tree-level antennae are used, at NNLO also four-parton tree-level and three-parton one-loop antenna functions need to be considered. The latter, however, are not needed for the double real radiation contributions, and they will not be discussed here. All the three-parton antenna functions used in the subtraction terms that will be presented in section \ref{sec.subterms} are of types A and E in different configurations with massive and/or massless partons. These antennae have been computed, integrated and extensively studied in \cite{antennannlo,daleo,mathias,us}, and we shall therefore not discuss them further here. They are listed in appendix \ref{sec.3partonantennae} together with their soft and collinear limits.

We shall here focus only on the four-parton antennae which are genuine NNLO objects capturing double unresolved radiation features of real matrix element squared. In the context of this paper, the double unresolved limits which need to be subtracted with four-parton antennae are double soft and triple collinear limits involving only fermions in initial and final states. All these limits can be subtracted with B-type and C-type antenna functions. The former are needed in their final-final, initial-final and initial-initial forms, while only initial-initial massless forms of the latter are needed. The B and C-type antennae only exist at the four-parton level. As the A-type antenna functions, $B_4^0$ and $C^0_4$ antennae collapse onto a hard quark-antiquark pair in their double unresolved limits, but they have a secondary $q\bar{q}$ pair in the final state instead of gluons. More precisely, in the final-final case, $B_4^0$ and $C_4^0$ antennae are derived from $\gamma^* \rightarrow q\bar{q}q' \bar{q'}$ \cite{antennannlo}. While $B_4^0$ is obtained from a colour-stripped amplitude squared, $C_4^0$ is a pure interference term. Explicitly, they are given by
\beqa
&& B_{4}^0 (\q{1},\qpb{4},\qp{3},\qb{2})\left|{\cal M}^0_{q\bar q}\right|^2 = \left|\cm_4 (\q{1},\qpb{4},\qp{3},\qb{2})\right|^2,\\
&& C_{4}^0 (\q{1},\q{3},\qb{4},\qb{2})\left|{\cal M}^0_{q\bar q}\right|^2= - \re \left(\cm_4(\q{1},\qpb{4},\qp{3},\qb{2})\cm_4(\q{1},\qpb{2},\qp{3},\qb{4})\right),
\eeqa
where the amplitude $\cm_4 (\q{i},\qpb{l},\qp{k},\qb{j})$ contains those Feynman diagrams contributing to the colour stripped amplitude for $\gamma^*\rightarrow \q{i}\qb{j}\qp{k}\qpb{l}$ in which the pair $\q{i}\qb{j}$ is the one that couples to the virtual photon. Explicit expressions for the final-final massless antennae $B_4^0$ and $C_4^0$ are given in \cite{antennannlo}.

As mentioned before, the initial-final and intial-initial antennae are obtained by crossing partons in the final-final functions. Their infrared limits are in principle known as they can also be inferred from the final-final case, although they have not been documented so far. Furthermore, concerning the four-parton antennae involving massive partons, only the final-final massive $B_4^0$ is known \cite{bernreuther}. 

In addition to the massive flavour conserving A-type (involving three-parton final states) or B-type antennae (involving four-parton final states) which can be derived from the decay of a virtual photon, the construction of our subtraction terms will also require the so-called massive flavour violating antennae. These involve radiators of different flavours, one of them being massive and the other massless. For the three-parton case, flavour violating antennae have been derived and integrated in \cite{us}. The required massive flavour violating antenna with four final state partons is new and will be given below.

In the remainder of this section we give the forms of the four-parton antennae used in our calculation. Their infrared limits will be presented in section \ref{sec.limits}. The conventions that we will follow  for the labelling of the partons in our antenna functions are the following: Massless partons will be indexed with $q$ while massive ones with $Q$ and their mass with $m_Q$.  The first and the last particles in the argument of a given antenna are the hard radiators, while the partons placed in between the radiators are the unresolved particles. For conciseness, the ${\cal O}(\epsilon)$ in the expression of the unintegrated antennae will be omitted.\\

\parindent 0em

{\bf{B-type antennae}}\\
The subtraction of the double unresolved singularities in the processes that we are presently considering requires B-type antenna functions in their massive final-final, flavour violating initial-final, and massless initial-initial forms. They are given below.\\

{\it Final-final}\\
\parindent 1.5em

The only final-final $B^0_4$ antenna function which is needed in our subtraction terms is the one involving a pair of massive radiators ($Q\bar{Q}$) and a secondary massless quark-antiquark pair ($q\bar{q}$). It is denoted by $B_4^0(\Q{1},\qb{4},\q{3},\Qb{2})$, and it has been computed and integrated in \cite{bernreuther}. This antenna function is used to subtract the infrared limits in which a soft quark-antiquark pair is emitted between a massive quark and a massive antiquark in the colour chain. Its explicit expression reads,
\beqa
B_4^0(\Q{1},\qb{4},\q{3},\Qb{2})&=&\frac{1}{\left(E_{cm}^2+2m_Q^2\right)}\bigg\{ \frac{1}{s_{34}s_{134}^2}\left( s_{12}s_{13}+s_{12}s_{14}+s_{13}s_{23}+s_{14}s_{24}\right)\nonumber\\
&&+\frac{1}{s_{34}s_{234}^2}\left( s_{12}s_{23}+s_{12}s_{24}+s_{13}s_{23}+s_{14}s_{24}\right)\nonumber\\
&&+\frac{1}{s_{34}^2 s_{134}^2}\left( 2s_{12}s_{13}s_{14}+s_{13}s_{14}s_{24}+s_{13}s_{14}s_{23}-s_{13}^2 s_{24}-s_{14}^2 s_{23}\right)\nonumber\\
&&+\frac{1}{s_{34}^2 s_{234}^2}\left( 2s_{12}s_{23}s_{24}+s_{13}s_{23}s_{24}+s_{14}s_{23}s_{24}-s_{13} s_{24}^2-s_{14} s_{23}^2\right)\nonumber\\
&&+\frac{1}{s_{34}s_{134}s_{234}}\left( 2s_{12}^2+s_{12}s_{23}+s_{12}s_{24}+s_{12}s_{13}+s_{12}s_{14}\right)\nonumber\\
&&+\frac{1}{s_{34}^2s_{134}s_{234}}\big( -s_{13}s_{24}^2-s_{14}s_{23}^2-s_{13}^2 s_{24}-s_{14}^2 s_{23}\nonumber\\
&& +s_{13}s_{14}s_{23}+s_{13}s_{14}s_{24}+s_{13}s_{23}s_{24}+s_{14}s_{23}s_{24} \nonumber\\
&& -2s_{12}s_{13}s_{24}-2s_{12}s_{14}s_{23}\big)+\frac{2s_{12}}{s_{134}s_{234}}\nonumber\\
&& +m_Q^2\bigg( \frac{8s_{13}s_{14}}{s_{34}^2 s_{134}^2}+\frac{8s_{23}s_{24}}{s_{34}^2 s_{234}^2}-\frac{4}{s_{134}^2}-\frac{4}{s_{234}^2}\nonumber\\
&&-\frac{2}{s_{34}s_{134}^2}(s_{12}+s_{23}+s_{24})-\frac{2}{s_{34}s_{234}^2}(s_{12}+s_{13}+s_{14})\nonumber\\
&&+\frac{2}{s_{34}s_{134}s_{234}}(4s_{12}-s_{13}-s_{14}-s_{23}-s_{24})\nonumber\\
&& -\frac{8}{s_{34}^2 s_{134}s_{234}}(s_{14}s_{23}+s_{13}s_{24})\bigg)\nonumber\\
&&-m_Q^4\bigg(\frac{8}{s_{34}s_{134}^2}+\frac{8}{s_{34}s_{234}^2}\bigg)\bigg\}+\order{\epsilon},\label{eq.B04ffm}
\eeqa
with the normalisation given by the tree-level two-parton matrix element (with couplings and colour factors omitted)
\beq
\left| \cm_2(\gamma^*\rightarrow Q\bar{Q})\right|^2=4\left[ (1-\epsilon)E_{cm}^2+2m_Q^2\right].
\eeq

\parindent 0em
{\it Initial-final}\\
\parindent 1.5em

To account for the singularities that arise when a massless quark-antiquark pair becomes unresolved between a massive final state fermion and a massless incoming one, we need a massive initial-final  flavour-violating B-type antenna. The hard radiators are a massive final state quark $Q$, and a massless initial state $q$, and the expression of this antenna is 
\beqa
B_4^0(\Q{1},\qb{4},\q{3},\qi{2})&=&-\frac{1}{\left(Q^2+m_Q^2\right)}\bigg\{ -\frac{1}{s_{34}s_{134}^2}\left( s_{12}s_{13}+s_{12}s_{14}+s_{13}s_{23}+s_{14}s_{24}\right)\nonumber\\
&&+\frac{1}{s_{34}s_{234}^2}\left( s_{12}s_{23}+s_{12}s_{24}+s_{13}s_{23}+s_{14}s_{24}\right)\nonumber\\
&&-\frac{1}{s_{34}^2 s_{134}^2}\left( 2s_{12}s_{13}s_{14}+s_{13}s_{14}s_{24}+s_{13}s_{14}s_{23}-s_{13}^2 s_{24}-s_{14}^2 s_{23}\right)\nonumber\\
&&+\frac{1}{s_{34}^2 s_{234}^2}\left( -2s_{12}s_{23}s_{24}+s_{13}s_{23}s_{24}+s_{14}s_{23}s_{24}-s_{13} s_{24}^2-s_{14} s_{23}^2\right)\nonumber\\
&&+\frac{1}{s_{34}s_{134}s_{234}}\left( 2s_{12}^2+s_{12}s_{23}+s_{12}s_{24}-s_{12}s_{13}-s_{12}s_{14}\right)\nonumber\\
&&+\frac{1}{s_{34}^2s_{134}s_{234}}\big( -s_{13}s_{24}^2-s_{14}s_{23}^2+s_{13}^2 s_{24}+s_{14}^2 s_{23}\nonumber\\
&& -s_{13}s_{14}s_{23}-s_{13}s_{14}s_{24}+s_{13}s_{23}s_{24}+s_{14}s_{23}s_{24} \nonumber\\
&& -2s_{12}s_{13}s_{24}-2s_{12}s_{14}s_{23}\big)-\frac{2s_{12}}{s_{134}s_{234}}\nonumber\\
&& +2m_Q^2\bigg( \frac{1}{s_{134}^2}+\frac{1}{s_{134}s_{234}}-\frac{1}{s_{34}s_{134}}+\frac{s_{12}-s_{234}}{s_{34}s_{134}^2} \bigg)\bigg\}+\order{\epsilon},\label{eq.B04fl}
\eeqa
where $Q^2=-(p_1-p_2+p_3+p_4)^2$, $s_{134}=s_{13}+s_{14}+s_{34}$, and $s_{234}=-s_{23}-s_{24}+s_{34}$. In this case the normalisation is
\beq
\left| \cm_2(\gamma^*q\rightarrow Q)\right|^2=4(1-\epsilon)(Q^2+m_Q^2).
\eeq\\

\parindent 0em
{\it Initial-initial}\\
\parindent 1.5em

Two initial-initial forms of B-type antennae are needed in our subtraction terms. The first of them is $B_4^0(\qbi{1},\qpb{4},\qp{3},\qi{2})$ and it used to capture the limits where a final state quark-antiquark pair becomes unresolved between the incoming quark and antiquark, which act as the hard radiators. As will be shown in section \ref{sec.limits}, this antenna contains the $\qp{3},\qpb{4}$ double soft limit as well as the triple collinear limits $\qbi{1}||\qpb{4}||\qp{3}$ and $\qpb{4}||\qp{3}||\qi{2}$. The expression of this first initial-initial antenna function is,
\beqa
B_4^0(\qbi{1},\qpb{4},\qp{3},\qi{2})&=&-\frac{1}{Q^2}\bigg\{ -\frac{1}{s_{34}s_{134}^2}\left( s_{12}s_{13}+s_{12}s_{14}+s_{13}s_{23}+s_{14}s_{24}\right)\nonumber\\
&&-\frac{1}{s_{34}s_{234}^2}\left( s_{12}s_{23}+s_{12}s_{24}+s_{13}s_{23}+s_{14}s_{24}\right)\nonumber\\
&&+\frac{1}{s_{34}^2 s_{134}^2}\left( 2s_{12}s_{13}s_{14}-s_{13}s_{14}s_{24}-s_{13}s_{14}s_{23}+s_{13}^2 s_{24}+s_{14}^2 s_{23}\right)\nonumber\\
&&+\frac{1}{s_{34}^2 s_{234}^2}\left( 2s_{12}s_{23}s_{24}-s_{13}s_{23}s_{24}-s_{14}s_{23}s_{24}+s_{13} s_{24}^2+s_{14} s_{23}^2\right)\nonumber\\
&&+\frac{1}{s_{34}s_{134}s_{234}}\left( 2s_{12}^2-s_{12}s_{23}-s_{12}s_{24}-s_{12}s_{13}-s_{12}s_{14}\right)\nonumber\\
&&+\frac{1}{s_{34}^2s_{134}s_{234}}\big( s_{13}s_{24}^2+s_{14}s_{23}^2+s_{13}^2 s_{24}+s_{14}^2 s_{23}\nonumber\\
&& -s_{13}s_{14}s_{23}-s_{13}s_{14}s_{24}-s_{13}s_{23}s_{24}-s_{14}s_{23}s_{24} \nonumber\\
&& -2s_{12}s_{13}s_{24}-2s_{12}s_{14}s_{23}\big)+\frac{2s_{12}}{s_{134}s_{234}}\bigg\}+\order{\epsilon},\label{eq.$B_4^0$ii}
\eeqa
with $Q^2=-(p_1+p_2-p_3-p_4)^2$, $s_{134}=-s_{13}-s_{14}+s_{34}$, and $s_{234}=-s_{23}-s_{24}+s_{34}$. This antenna function was computed and integrated in \cite{radja}. It is normalised to 
\beq\label{eq.initialinitialnorm}
\left| \cm_2(q\bar{q}\rightarrow \gamma^*)\right|^2=4(1-\epsilon)Q^2.
\eeq

The second B-type initial-initial antenna function that we use in our subtraction terms is $B_4^0(\q{1},\qpi{4},\qp{3},\qi{2})$ and it only contains the $\q{1}||\qpi{4}||\qp{3}$ triple collinear limit. The unintegrated form of this antenna is
\beqa
B_4^0(\q{1},\qpi{4},\qp{3},\qi{2})&=&-\frac{1}{Q^2}\bigg\{ \frac{1}{s_{34}s_{134}^2}\left( s_{12}s_{13}-s_{12}s_{14}+s_{13}s_{23}+s_{14}s_{24}\right)\nonumber\\
&&+\frac{1}{s_{34}s_{234}^2}\left( -s_{12}s_{23}+s_{12}s_{24}+s_{13}s_{23}-s_{14}s_{24}\right)\nonumber\\
&&+\frac{1}{s_{34}^2 s_{134}^2}\left( 2s_{12}s_{13}s_{14}-s_{13}s_{14}s_{24}+s_{13}s_{14}s_{23}-s_{13}^2 s_{24}+s_{14}^2 s_{23}\right)\nonumber\\
&&+\frac{1}{s_{34}^2 s_{234}^2}\left( 2s_{12}s_{23}s_{24}-s_{13}s_{23}s_{24}+s_{14}s_{23}s_{24}-s_{13} s_{24}^2+s_{14} s_{23}^2\right)\nonumber\\
&&-\frac{1}{s_{34}s_{134}s_{234}}\left( 2s_{12}^2+s_{12}s_{23}-s_{12}s_{24}-s_{12}s_{13}+s_{12}s_{14}\right)\nonumber\\
&&+\frac{1}{s_{34}^2s_{134}s_{234}}\big( -s_{13}s_{24}^2+s_{14}s_{23}^2-s_{13}^2 s_{24}+s_{14}^2 s_{23}\nonumber\\
&& +s_{13}s_{14}s_{23}-s_{13}s_{14}s_{24}-s_{13}s_{23}s_{24}+s_{14}s_{23}s_{24} \nonumber\\
&& +2s_{12}s_{13}s_{24}+2s_{12}s_{14}s_{23}\big)-\frac{2s_{12}}{s_{134}s_{234}}\bigg\}+\order{\epsilon}\label{eq.$B_4^0$ii2}
\eeqa
with $Q^2=-(p_1-p_2+p_3-p_4)^2$, $s_{134}=s_{13}-s_{14}-s_{34}$, and $s_{234}=-s_{23}+s_{24}-s_{34}$. Its integrated form has been given in \cite{radja}, and the normalisation is the same one given in eq.(\ref{eq.initialinitialnorm}), but with the $Q^2$ corresponding to this case. \\

\parindent 0em

{\bf C-type antennae} \\
To subtract triple collinear limits in amplitudes with two $q\bar{q}$ pairs of the same flavour, two different  initial-initial C-type antenna functions involving two massless final and two initial state partons are needed.

\parindent 1.5em

The first of these initial-initial C-type antennae is
\beqa
C_4^0(\qbi{1},\q{3},\qb{2},\qi{4})&=&\frac{1}{Q^2}\bigg\{ -\frac{s_{12}s_{13}s_{14}}{s_{23}s_{34}s_{123}s_{134}}+\frac{1}{2s_{23}s_{34}s_{134}s_{234}}\left( s_{12}s_{13}s_{24} + s_{13}s_{14}s_{24} \right)\nonumber\\
&&+\frac{1}{2s_{23}s_{34}s_{123}s_{234}}\left( s_{12}s_{13}s_{24} + s_{13}s_{14}s_{24} \right)\nonumber\\
&&-\frac{s_{13}s_{24}^2}{s_{23}s_{34}s_{234}^2}+\frac{s_{12}s_{13}}{s_{23}s_{123}s_{134}}-\frac{s_{13}s_{14}}{s_{34}s_{123}s_{134}}\nonumber\\
&&+\frac{1}{2s_{23}s_{123}s_{234}}\left( s_{12}s_{14}-s_{12}s_{34}-s_{12}^2+s_{13}s_{24} \right)\nonumber\\
&&+\frac{1}{2s_{34}s_{134}s_{234}}\left( s_{12}s_{14}-s_{14}s_{23}-s_{14}^2+s_{13}s_{24} \right)\nonumber\\
&&-\frac{1}{2s_{23}s_{134}s_{234}}\left( -s_{12}s_{14}+s_{12}s_{34}+s_{12}^2+s_{13}s_{24} \right)\nonumber\\
&&-\frac{1}{2s_{34}s_{123}s_{234}}\left( -s_{12}s_{14}+s_{14}s_{23}+s_{14}^2+s_{13}s_{24} \right)\nonumber\\
&&-\frac{s_{13}}{s_{123}s_{134}}+\frac{1}{s_{23}s_{234}^2}\left( s_{12}s_{24}-s_{14}s_{24} \right)-\frac{1}{s_{34}s_{234}^2}\left( s_{12}s_{24}-s_{14}s_{24} \right)\nonumber\\
&&+\frac{1}{2s_{123}s_{234}}\left( s_{12}+s_{14} \right)+\frac{1}{2s_{134}s_{234}}\left( s_{12}+s_{14} \right)\bigg\}+\order{\epsilon},\label{eq.B4C41}
\eeqa
which is used to subtract the triple collinear limit $\q{3}||\qb{2}||\qi{4}$. The second one is given by
\beqa
C_4^0(\qbi{1},\qbi{3},\qb{2},\qb{4})&=&\frac{1}{Q^2}\bigg\{ -\frac{s_{12}s_{13}s_{14}}{s_{23}s_{34}s_{123}s_{134}}-\frac{1}{2s_{23}s_{34}s_{134}s_{234}}\left( s_{12}s_{13}s_{24} - s_{13}s_{14}s_{24} \right)\nonumber\\
&&-\frac{1}{2s_{23}s_{34}s_{123}s_{234}}\left(- s_{12}s_{13}s_{24} + s_{13}s_{14}s_{24} \right)\nonumber\\
&&-\frac{s_{13}s_{24}^2}{s_{23}s_{34}s_{234}^2}-\frac{s_{12}s_{13}}{s_{23}s_{123}s_{134}}-\frac{s_{13}s_{14}}{s_{34}s_{123}s_{134}}\nonumber\\
&&+\frac{1}{2s_{23}s_{123}s_{234}}\left( -s_{12}s_{14}-s_{12}s_{34}-s_{12}^2+s_{13}s_{24} \right)\nonumber\\
&&+\frac{1}{2s_{34}s_{134}s_{234}}\left( -s_{12}s_{14}-s_{14}s_{23}-s_{14}^2+s_{13}s_{24} \right)\nonumber\\
&&+\frac{1}{2s_{23}s_{134}s_{234}}\left( s_{12}s_{14}+s_{12}s_{34}+s_{12}^2+s_{13}s_{24} \right)\nonumber\\
&&+\frac{1}{2s_{34}s_{123}s_{234}}\left( s_{12}s_{14}+s_{14}s_{23}+s_{14}^2+s_{13}s_{24} \right)\nonumber\\
&&-\frac{s_{13}}{s_{123}s_{134}}+\frac{1}{s_{23}s_{234}^2}\left( s_{12}s_{24}+s_{14}s_{24} \right)+\frac{1}{s_{34}s_{234}^2}\left( s_{12}s_{24}+s_{14}s_{24} \right)\nonumber\\
&&+\frac{1}{2s_{123}s_{234}}\left( -s_{12}+s_{14} \right)+\frac{1}{2s_{134}s_{234}}\left( s_{12}-s_{14} \right)\bigg\}+\order{\epsilon},\label{eq.B4C42}
\eeqa
and it is used to subtract the triple collinear limit $\qbi{3}||\qb{2}||\qb{4}$.

The normalisation for the C-type initial-initial antennae is again given in eq.(\ref{eq.initialinitialnorm}) with $Q^2$ given by the square of the sum of incoming parton momenta in each case.

\section{Infrared limits}\label{sec.limits}
The factorisation properties of QCD tree-level squared amplitudes have been extensively studied in \cite{campbellglover,catanigrazzini1,catanigrazzini2,deflorian}. While at NLO only single soft and collinear singularities may arise, at NNLO there are several double unresolved regions of phase space in which the amplitudes can develop singularities when integrated over.

\subsection{Single unresolved factors}\label{subsec.factors} 
In their infrared limits, colour-ordered matrix elements factorise into universal unresolved factors and reduced matrix elements. Depending on whether the partons involved in each limit are massive or massless, and depending also on whether one or two partons become unresolved, the universal factors will be different. We shall start by presenting the single unresolved massless and massive factors.

Retaining only those infrared limits which lead to poles in $\epsilon$ when integrated over the phase space, the only single unresolved limits that colour-ordered amplitudes involving only fermions have are single collinear massless limits arising when a quark-antiquark pair becomes collinear and clusters to form a gluon. If both collinear particles are in the final state, the parent gluon will also be in the final state, and it will be in the initial state in the cases where one of the collinear partons is in the initial state. Thus, the reduced colour-ordered matrix element associated with this limit involves a gluon in the final or in the initial state. Therefore, although only quarks are involved in the matrix elements for the processes that we are considering in this paper, unresolved gluon limits have to be considered as well. While the collinear limits of this gluon only give poles in $\epsilon$ if the other collinear particle is massless (a massless fermion in this case), in soft limits mass terms will appear in the soft eikonal factors if the gluon is emitted from a massive hard radiator.

\subsubsection{Single unresolved massless factors}
When two massless partons become collinear, a colour-ordered sub-amplitude factorises into a reduced matrix element times a specific Altarelli-Parisi splitting function corresponding to the particular parton-parton splitting. These functions depend on $z$, the momentum fraction carried by the unresolved parton. Depending whether the unresolved parton is collinear to an initial or to a final state parton, the definition of $z$ is different. For two final state particles $i$ and $j$ of momenta $p_i$ and $p_j$ becoming collinear, we have, in the limit,
\beq\label{eq:zdefinal}
p_i\to z p_{ij},\quad p_j\to (1-z) p_{ij} ,\quad s_{ik}\to z s_{ijk},
\quad s_{jk}\to (1-z) s_{ijk}\,,
\eeq
whereas for a final state particle $j$ of momentum $p_j$ becoming collinear with an initial state parton $i$ of momentum $p_i$ we have
\beq\label{eq:zdefinitial}
p_j\to z p_{i},\quad p_{ij}\to (1-z) p_{i} ,
\quad s_{ik}\to \frac{s_{ijk}}{1-z},\quad s_{jk}\to \frac{z s_{ijk}}{1-z}\;.
\eeq

The splitting functions denoted by $P_{ij \rightarrow (ij)}(z)$ corresponding to the collinear limit of two final state partons $i$ and $j$, given in the conventional dimensional regularisation scheme \cite{CDR} with all particles treated in  $d=4-2\epsilon$ dimensions, are \cite{AP}:
\beqa
&& P_{qg\rightarrow Q}(z)=\frac{1+(1-z)^2-\epsilon z^2}{z}\label{eq.splitting1}\\
&& P_{q\bar{q}\rightarrow G}(z)=\frac{z^2+(1-z)^2-\epsilon}{1-\epsilon}\label{eq.splitting2}\\
&& P_{gg\rightarrow G}(z)=2\left[\frac{z}{1-z}+\frac{1-z}{z}+z(1-z)\right].\label{eq.splitting3}
\eeqa

When the collinearity arises between an initial  $i$ and a final state parton $j$, the splitting functions denoted by $P_{ij \leftarrow (ij)}(z)$ are given by \cite{AP}:
\beqa
&& P_{gq\leftarrow Q}(z)=\frac{1+z^2-\epsilon(1-z)^2}{(1-\epsilon)(1-z)^2}
=\frac{1}{1-z}\frac{1}{1-\epsilon}P_{qg\rightarrow Q}(1-z)\label{eq.splitting4}\\
&& P_{qg\leftarrow Q}(z)=\frac{1+(1-z)^2-\epsilon z^2}{z(1-z)}
=\frac{1}{1-z}P_{qg\rightarrow Q}(z)\label{eq.splitting5}\\
&& P_{q\bar{q}\leftarrow G}(z)=\frac{z^2+(1-z)^2-\epsilon}{1-z}
=\frac{1-\epsilon}{1-z}P_{q\bar{q}\rightarrow G}(z)\label{eq.splitting6}\\
&& P_{gg\leftarrow G}(z)=\frac{2(1-z+z^2)^2}{z(1-z)^2}
=\frac{1}{1-z}P_{gg\rightarrow G}(z).\label{eq.splitting7}
\eeqa
The additional factors $(1-\epsilon)$ and $1/(1-\epsilon)$  account for the different number of polarizations of quark and gluons in the cases in which the particle entering the hard processes changes its type.

In all the splitting functions defined above, the label $q$ can stand for a massless quark or an antiquark since charge conjugation implies that  $P_{qg\rightarrow Q}=P_{\bar{q}g\rightarrow\bar{Q}}$ and $P_{qg\leftarrow Q}=P_{\bar{q}g\leftarrow\bar{Q}}$ . The labels $Q$ and $G$ denote the parent parton of the two collinear partons, which is massless. \\

When a gluon $j$ emitted between two massless hard radiators $i$ and $k$ becomes soft, the eikonal factor that factorises off the colour-ordered squared matrix element is
\beq \label{eq.eikonal}
S_{ijk}=\frac{2s_{ik}}{s_{ij}s_{jk}}.
\eeq
This limit is obtained by letting $p_{j} \to \lambda p_{j}$ with $\lambda \to 0$ or, alternatively, by keeping only terms with any product of the inverse of the two invariants $s_{ij}$ and $s_{jk}$ in the colour-ordered matrix element squared.\\

\subsubsection{Single unresolved massive factors}
In the present context, the only single unresolved massive factor that needs to be considered is the massive soft factor, which is a generalisation of the massless eikonal factor defined above. The generalised soft eikonal factor $S_{ijk}(m_i,m_k)$ for a massless
gluon $j$ emitted between two massive partons $i$ and $k$ depends on the invariants $s_{lm}=2p_l\cdot p_m$ built with the partons $i$, $j$
and $k$ but also on the masses $m_i$ and $m_k$ of $i$ and $k$. It is given by \cite{cdst1,mathias}
\beq\label{eq.eikonalmassive}
S_{ijk}(m_i,m_k)=\frac{2s_{ik}}{s_{ij}s_{jk}}-\frac{2m_i^2}{s_{ij}^2}-
\frac{2m_k^2}{s_{jk}^2}.
\eeq
This limit is again obtained by letting $p_{j} \to \lambda p_{j}$ with $\lambda \to 0$ or alternatively by keeping in the matrix element squared only terms with the inverse of any product of the invariants $s_{ij}$ and $s_{jk}$. The above expression for the massive soft factor 
agrees with eq.(\ref{eq.eikonal}) if the masses are set to zero.

\subsection{Double unresolved factors}
In general, when two particles are unresolved in a tree-level process,  a variety of different configurations can arise:
\begin{itemize}
\item[{(1)}]  two soft particles,
\item[{(2)}]  two pairs of collinear particles,
\item[{(3)}]  three collinear particles,
\item[{(4)}]  one soft and two collinear.
\end{itemize}

As we saw in section \ref{sec.formalism}, the resulting double unresolved configurations of the colour-ordered matrix element squared need to be separated into three categories depending on the colour connections of the unresolved partons and the hard radiators associated with them. In the following, we shall describe only the double unresolved factors encountered in the context of this paper.

\subsubsection{Colour-unconnected pairs of unresolved particles} 
For unresolved particles that are disjoint in the colour chain, which arise in the item $(d)$ as presented in section \ref{sec.formalism}, the colour-ordered matrix element can be factorised into the product of two disjoint single unresolved factors multiplied by a reduced matrix element with two partons less than the original colour-ordered matrix element squared. 

For example, in the case of two pairs of colour-connected particles $(a,b)$ and $(c,d)$ which are all located in the final state, 
the colour-ordered matrix elements given by $|\cm_n(\ldots,a,b,\ldots,c,d,\ldots)|^2$ undergo the following factorisation in the double collinear limit:
\beq
|\cm_n(\ldots,a,b,\ldots,c,d,\ldots)|^2 \rightarrow P_{ab\to P}(z_1,s_{ab})\,P_{cd\to Q}(z_2,s_{cd})|\cm_{n-2}(\ldots,P,\ldots,Q,\ldots)|^2,
\eeq
where partons $a$ and $b$ form $P$, while $c$ and $d$ cluster to form $Q$, so that $P$ and $Q$ are themselves colour-unconnected. The collinear splitting functions appearing in this equation are the (colourless) Altarelli-Parisi splitting functions given by eqs. (\ref{eq.splitting1}-\ref{eq.splitting7}).

In the processes that we are considering in this paper, the only double unresolved colour-unconnected limits that can occur are double quark-antiquark collinear limits, with each single collinear pair given by a final state (anti) quark and an initial state (anti) quark. The unresolved factor associated to this limit is a product of two splitting functions of the type given in eqs. (\ref{eq.splitting1}-\ref{eq.splitting7}). Note that none of the four-parton antennae required for our subtraction term captures these double collinear singularities. These are accounted for by subtraction terms involving only the product of two three-parton antennae.\\

\subsubsection{Colour-connected pairs of unresolved particles} 
The colour-connected double unresolved limits that appear in the partonic processes considered here are
\begin{enumerate}
\item [A)] Triple collinear limits of three massless fermions where one massless fermion plays the role of the hard radiator and is in the initial state. 
\item [B)] Double soft limits of a massless final state $q\bar{q}$ pair emitted between massive or massless radiators. 
\end{enumerate}
As it was mentioned above, these colour-connected double unresolved limits are captured by four-parton antenna functions and arise in subtraction terms ${\rm d}\sigma_{NNLO}^{S,b}$ as presented in section \ref{sec.formalism}. In the following, we shall give the massive and massles double unresolved factors associated with these two types of limits.\\

\parindent 0em

A) {\bf Massless triple collinear factors}\\
In those regions of phase space where three colour-connected massless partons $(a,b,c)$ become collinear, a generic colour-ordered amplitude squared denoted by $|\cm_n(\ldots,a,b,c,\ldots)|^2$ factorises as:
\beq
|\cm_{n}(\ldots,a,b,c,\ldots)|^2 \rightarrow P_{abc \rightarrow P}|\cm_{n-2}(\ldots,P,\ldots)|^2.
\eeq
where the three colour-connected final state particles $(a,b,c)$ cluster to form a single parent particle $P$. The triple collinear splitting function for partons $a$, $b$ and $c$ clustering to form the parent parton $P$ is generically denoted by,
\beq
P_{abc \rightarrow P}(w,x,y,s_{ab},s_{ac},s_{bc},s_{abc}),
\eeq
where $w$, $x$ and $y$ are the momentum fractions of the clustered partons,
\beq\label{momfrac}
p_a=wp_P, \qquad p_b=xp_P, \qquad p_c=yp_P, \hspace{10mm} \mbox{with }\hspace{2mm} w+x+y=1.
\eeq
In addition to its dependence on the momentum fractions carried by the clustering partons, the splitting function also depends on the
invariant masses of parton-parton pairs and the invariant mass of the whole cluster. The explicit forms of the triple collinear splitting functions  $P_{abc \rightarrow P}$ are obtained by retaining terms in the colour-ordered matrix element squared that possess two of the `small' denominators $s_{ab}$,
$s_{ac}$, $s_{bc}$ and $s_{abc}$.

\parindent 1.5em

The triple collinear limits that have to be considered here are those involving three massless fermions, one of which is in the initial state. The splitting functions for these types of triple collinear limits can be obtained from the analoguous limit where the three collinear particles are in the final state \cite{deflorian,campbellglover}.

We therefore start by considering the case where a quark-antiquark pair $q'\bar{q}'$ and a quark $q$ of different flavour cluster together to form a parent quark $\tilde{q}$ of the same flavour as $q$. In such a configuration a squared sub-amplitude in which the three collinear fermions are colour-connected factorises as:
\beq
|\cm_n(...\q{i},\qpb{j},\qp{k},...)|^2 \stackrel{^{\q{i}||\qpb{j}||\qp{k}}}{\longrightarrow} P_{q\bar{q}'q'\rightarrow \tilde{q}}^{{\rm non-ident.}}(w,x,y,s_{q\bar{q}'},s_{q'\bar{q}'},s_{q\bar{q}'q'}) |\cm_{n-2}(...,\q{l},...)|^2,
\eeq
with $p_l=p_i+p_j+p_k$. The corresponding splitting function is 
\beqa
&&\hspace{-5mm}P_{q\bar{q}'q'\rightarrow \tilde{q}}^{{\rm non-ident.}}(w,x,y,s_{q\bar{q}'},s_{q'\bar{q}'},s_{q\bar{q}'q'})=-\frac{1}{s_{q\bar{q}'q'}^2}\left( 1-\epsilon+\frac{2s_{q\bar{q}'}}{s_{q'\bar{q}'}}\right)\nonumber\\
&&\hspace{10mm}-\frac{2(x s_{q\bar{q}'q'}-(1-w))s_{q\bar{q}'})^2}{s_{q'\bar{q}'}^2 s_{q\bar{q}'q'}^2 (1-w)^2}+\frac{1}{s_{q'\bar{q}'} s_{q\bar{q}'q'}}\left( \frac{1+x^2+(x+w)^2}{1-w} -\epsilon (1-w) \right)\nonumber\\ \label{eq.triplecoll}
\eeqa
where $w$, $x$, and $y$ are the momentum fractions of $q$, $\bar{q}'$ and $q'$ respectively, and $w+x+y=1$.

As we will see in section \ref{sec.subterms}, the squared amplitudes for processes with identical quark flavours contain interferences of partial amplitudes with different flavour assignments for the massless antiquarks. Although they are finite in all other single and double unresolved limits, these interference terms can still be singular in triple collinear configurations. In these cases, the factorisation reads
\beqa
&&\re \left( \cm_n(...,\q{i},\qpb{j},\qp{k},m,...) \cm_n(...,\qp{i},\qpb{j},\q{k},n,...)^{\dagger}\right)\stackrel{^{\q{i}||\qb{j}||\q{k}}}{\longrightarrow}\nonumber \\
&&\hspace{10mm}-\frac{1}{2}P_{q\bar{q}'q'\rightarrow \tilde{q}}^{{\rm ident.}}(w,x,y,s_{q\bar{q}'},s_{q'\bar{q}'},s_{q\bar{q}'q'})\re \left( \cm_{n-2}(...,\q{l},m,...)\cm_{n-2}(...,\q{l},n,...)^{\dagger}\right).\nonumber\\ \label{eq.facttriplecollid}
\eeqa 
The corresponding splitting function, given by
\beqa
&&\hspace{-5mm}P_{q\bar{q}'q'\rightarrow \tilde{q}}^{{\rm ident.}}(w,x,y,s_{q\bar{q}'},s_{q'\bar{q}'},s_{q\bar{q}'q'})=\nonumber\\
&&\hspace{5mm}-\frac{1}{s_{q\bar{q}'q'}^2}\left( 2+\epsilon+\frac{2s_{q\bar{q}'}}{s_{q'\bar{q}'}}\right)-\frac{1}{2s_{q\bar{q}'} s_{q'\bar{q}'}}\left( \frac{x(1+x^2)}{(1-y)(1-w)}-\epsilon x\left( \frac{2(1-y)}{1-w}+1+\epsilon \right)\right)\nonumber\\ 
&&\hspace{5mm}\frac{1}{s_{q'\bar{q}'} s_{q\bar{q}'q'}}\left( \frac{1+x^2}{1-y}+\frac{2x}{1-w}-\epsilon\left( \frac{(1-w)^2}{1-y}+1+x+\frac{2x}{1-w}+\epsilon(1-w)\right)\right)\nonumber\\
&&\hspace{5mm}+(s_{q\bar{q}'} \leftrightarrow s_{q'\bar{q}'},y\leftrightarrow w),\label{eq.triplecollid}
\eeqa
is manifestly symetric under the interchange $q\leftrightarrow q'$ and $y\leftrightarrow w$.

The triple collinear splitting functions in eqs. (\ref{eq.triplecoll}) and (\ref{eq.triplecollid}) correspond to configurations in which all three collinear particles are outgoing. However, having only two massless final state particles, the double real radiation matrix elements for top-antitop production can only have triple collinear limits involving one of the incoming partons and both massless outgoing particles. The corresponding initial-final splitting functions are related to their final-final counterparts by \cite{deflorian}
\beqa
&&\hspace{-5mm}P_{q\bar{q}'q'\leftarrow \tilde{q}}(z_1,z_2,z_3,s_{\hat{q}\bar{q}'},s_{q'\bar{q}'},s_{\hat{q}\bar{q}'q'})=\nonumber\\
&&\hspace{30mm}P_{q\bar{q}'q'\rightarrow \tilde{q}}(1/z_3,-z_1/z_3,-z_2/z_3,-s_{q\bar{q}'},s_{q'\bar{q}'},s_{q'\bar{q}'}-s_{qq'}-s_{q\bar{q}'})\nonumber\\
&&\hspace{-5mm}P_{\bar{q}qq'\leftarrow \tilde{q}'}(z_1, z_2,z_3,s_{q\hat{q}'},s_{q'\hat{q}'},s_{q\hat{q}'q'})=\nonumber\\
&&\hspace{30mm}P_{q\bar{q}'q'\rightarrow \tilde{q}}(-z_1/z_3,1/z_3,-z_2/z_3,-s_{q\bar{q}'},-s_{q'\bar{q}'},s_{qq'}-s_{q'\bar{q}'}-s_{q\bar{q}'}),\nonumber\\ \label{eq.triplecollif}
\eeqa
where $z_1$ and $z_2$ are the momentum fractions of the final state particles participating in the limit, and $z_3=1-z_1-z_2$. eq.(\ref{eq.triplecollif}) is valid for both `non-identical' and `identical' triple collinear splitting functions.\\

\parindent 0em

B) {\bf Massive and massless double soft factors}\\
When a massless quark-antiquark pair becomes soft between two hard radiators, the amplitude squared factorises into a double soft factor
and a reduced matrix element squared with the quark-antiquark pair removed from it. As it was the case for the single soft gluon eikonal factor,  the soft factor for a soft $q\bar{q}$ pair also depends on the masses of the hard radiators if these are massive. Again, the massless factor can be obtained from the massive one by setting the masses to zero.

\parindent 1.5em
   
When the quark-antiquark pair $(c,d)$ becomes soft between the hard radiators $(a,b)$ of masses $m_{a}$ and $m_{b}$, the massive double soft factor is given by, 
\beqa\label{eq:softm}
\soft{a}{c}{d}{b}(m_a,m_b)&=&\frac{2(s_{ab}s_{cd}-s_{ac}s_{bd}-s_{bc}s_{ad})}{s_{cd}^2(s_{ac}+s_{ad})(s_{bc}+s_{bd})}+\frac{2}{s_{cd}^2}\left[ \frac{s_{ac}s_{ad}}{(s_{ac}+s_{ad})^2} +  \frac{s_{bc}s_{bd}}{(s_{bc}+s_{bd})^2}\right]\nonumber\\
&&-\frac{2m_{a}^2}{s_{cd}(s_{ac}+s_{ad})^2}-\frac{2m_{b}^2}{s_{cd}(s_{bc}+s_{bd})^2}.\label{eq.seik}
\eeqa
This factor is obtained by setting $p_c \to \lambda p_c,\:\:p_{d} \to \lambda p_d$ with $\lambda \to 0$ in the matrix elements squared and was also derived  in \cite{bernreuther}. This limit is only present in the processes $q\bar{q}\rightarrow Q\bar{Q}q'\bar{q}'$ and $q\bar{q}\rightarrow Q\bar{Q}q\bar{q}$. It occurs in those diagrams in which the final state quark-antiquark pair splits from a gluon propagator that becomes soft. As it will be seen in section \ref{sec.subterms}, we encounter all possible combinations of hard radiators (two massive, one massive and one massless, and two massless), and therefore eikonal factors with two masses, with one mass, and with no masses are needed. They can all be obtained from eq.(\ref{eq:softm}).

Furthermore, a derivation of this double soft factor using the soft current formalism will be given in section \ref{sec.subterms}.

\subsection {Single and double unresolved limits of four-parton antennae}\label{subsec.limitsA}
In this section we present the single and double unresolved limits of all four-parton antenna functions defined in section \ref{sec.antennae}. Those are the ones required to construct the subtraction terms for the double real contributions to heavy quark pair production in hadronic collisions due to processes involving only fermions. In their infrared limits, these antennae yield the universal single and double 
unresolved factors defined above. Additionally, some of these four-parton antennae yield angular correlation terms in single collinear limits. These angular terms arise when a gluon splits into a quark-antiquark pair or into two gluons. The treatment of these angular dependent terms  will be discussed in section \ref{subsec.ang}. \\

\parindent 0em
{\it Final-final}\\
\parindent 1.5em

The massive final-final B-type antenna given in eq.(\ref{eq.B04ffm}) is only singular when the massless $q\bar{q}$ is soft or when this quark-antiquark pair is collinear. The behaviour of this antenna in each of these limits is given by
\beqa
&&B_4^0(\Q{1},\qb{4},\q{3},\Qb{2})\stackrel{^{\q{3},\qb{4}\rightarrow0}}{\longrightarrow}\soft{1}{3}{4}{2}(m_Q,m_Q),\\
&&B_4^0(\Q{1},\qb{4},\q{3},\Qb{2})\stackrel{^{\q{3}||\qb{4}}}{\longrightarrow}\frac{1}{s_{34}}P_{q\bar{q}\rightarrow G}(z)A_3^0(\Q{1},\gl{(34)},\Qb{2})+{\text{ang.}},
\eeqa
where (ang.) indicates the presence of angular correlation terms.\\

\parindent 0em
{\it Initial-final}\\
\parindent 1.5em

The flavour violating B-type antenna function given in eq.(\ref{eq.B04fl}) contains a triple collinear singularity, in addition to the double soft and single collinear limits. It reads,
\beqa
&&B_4^0(\Q{1},\qpb{4},\qp{3},\qi{2})\stackrel{^{\q{3},\qb{4}\rightarrow0}}{\longrightarrow}\soft{1}{3}{4}{2}(m_Q,0)\\
&&B_4^0(\Q{1},\qpb{4},\qp{3},\qi{2})\stackrel{^{\q{3}||\qb{4}||\qi{2}}}{\longrightarrow} P_{q\bar{q}'q'\leftarrow \tilde{q}}^{{\rm non-ident.}}(z_1,z_2,z_3,s_{24},s_{34},s_{234}) \\
&&B_4^0(\Q{1},\qpb{4},\qp{3},\qi{2})\stackrel{^{\q{3}||\qb{4}}}{\longrightarrow}\frac{1}{s_{34}}P_{q\bar{q}\rightarrow G}(z)A_3^0(\Q{1},\gl{(34)},\qi{2})+{\text{ang.}}
\eeqa
where (ang.) means that angular terms are  present here. \\

\parindent 0em
{\it Initial-initial}\\
\parindent 1.5em

The initial-initial B-type antenna with the massless hard radiators in the initial state has a double soft singularity, as well as two triple collinear and a single collinear limit:
\begin{eqnarray}
&&B_4^0(\qbi{1},\qpb{4},\qp{3},\qi{2})\stackrel{^{\q{3},\qb{4}\rightarrow0}}{\longrightarrow}\soft{1}{3}{4}{2}(0,0)\\
&&B_4^0(\qbi{1},\qpb{4},\qp{3},\qi{2})\stackrel{^{\q{3}||\qb{4}||\qbi{1}}}{\longrightarrow}P_{q\bar{q}'q'\leftarrow \tilde{q}}^{{\rm non-ident.}}(z_1,z_2,z_3,s_{14},s_{34},s_{134})\\
&&B_4^0(\qbi{1},\qpb{4},\qp{3},\qi{2})\stackrel{^{\q{3}||\qb{4}||\qi{2}}}{\longrightarrow}P_{q\bar{q}'q'\leftarrow \tilde{q}}^{{\rm non-ident.}}(z_1,z_2,z_3,s_{24},s_{34},s_{234})\\
&&B_4^0(\qbi{1},\qpb{4},\qp{3},\qi{2})\stackrel{^{\q{3}||\qb{4}}}{\longrightarrow}\frac{1}{s_{34}}P_{q\bar{q}\rightarrow G}(z)A_3^0(\qbi{1},\gl{(34)},\qi{2})+{\text{ang.}}.
\end{eqnarray}
Again we see that the splitting of a gluon into a collinear quark-antiquark pair yields angular terms.

The other initial-initial B-type antenna function does not have a double soft singularity, since one of the unresolved particles is crossed to the initial state. It only contains a triple collinear as well as a single collinear limit:
\begin{eqnarray}
&&B_4^0(\q{1},\qpi{4},\qp{3},\qi{2})\stackrel{^{\q{1}||\qp{3}||\qpi{4}}}{\longrightarrow}P_{\bar{q}qq'\leftarrow \tilde{q}'}^{{\rm non-ident.}}(z_1, z_2,z_3,s_{14},s_{34},s_{134})\\
&&B_4^0(\q{1},\qpi{4},\qp{3},\qi{2})\stackrel{^{\qp{3}||\qpi{4}}}{\longrightarrow}\frac{1}{s_{34}}P_{gq\leftarrow Q}(z)+{\text{ang.}}
\end{eqnarray}

As mentioned in section \ref{sec.antennae}, the C-type antennae given in eqs. (\ref{eq.B4C41}, \ref{eq.B4C42}) are used to subtract triple collinear limits in processes with two pairs of identical flavour massless quarks. These are actually the only singularities that these antenna functions contain. For the present calculation, we only need to consider initial-initial C-type antennae, whose limiting behaviour is,
\beq
C_4^0(\qbi{1},\q{3},\qb{2},\qi{4})\stackrel{^{\qb{2}||\q{3}||\qi{4}}}{\longrightarrow}\frac{1}{2}P_{q\bar{q}'q'\leftarrow \tilde{q}}^{{\rm ident.}}(z_1,z_2,z_3,s_{24},s_{34},s_{234}),
\eeq
and
\beq
C_4^0(\qbi{1},\qbi{3},\qb{2},\qb{4})\stackrel{^{\qb{2}||\qb{4}||\qbi{3}}}{\longrightarrow}\frac{1}{2}P_{\bar{q}qq'\leftarrow \tilde{q}'}^{{\rm ident.}}(z_1, z_2,z_3,s_{23},s_{24},s_{234}).
\eeq

\subsection{Angular terms}\label{subsec.ang} 
As seen above, in single collinear limits where a gluon splits into a quark-antiquark pair (or into two gluons), antenna functions do not just yield a product of a spin averaged splitting function (\ref{eq.splitting1}-\ref{eq.splitting7}) and a spin averaged reduced matrix element: angular correlation terms are also present. This is a general feature of QCD amplitudes squared and it holds both for four-parton antenna functions and for the real radiation matrix elements. In these types of single collinear limits, arising from gluon splittings, neither the real radiation matrix elements squared nor the four-parton antenna functions yield the unpolarised splitting functions multiplied by spin-averaged reduced matrix elements. Instead, amplitudes squared (or, similarly, four-parton antennae) factorise into spin-dependent tensorial splitting functions contracted with a tensorial reduced matrix element \cite{cs,catanigrazzini2}:
\beq\label{eq.collccunp}
|\cm_n(...,\qb{a};;\q{b},...)|^2\stackrel{^{a||b}}{\longrightarrow}\frac{1}{s_{ab}}P_{q\bar{q}\rightarrow G}^{\mu\nu}(z,k_{\perp}) |\cm_{n-1}(...,c_g,...)|_{\mu\nu}^2.
\eeq
In a slight abuse of notation, $|\cm_{n-1}(...,c_g,...)|_{\mu\nu}^2$ denotes the reduced matrix element squared without the average over the polarisations of gluon $c$. The spin dependent splitting function is given by
\beq\label{eq.splitunp}
P_{q\bar{q}\rightarrow G}^{\mu\nu}(z,k_{\perp})=-g^{\mu\nu}+4z(1-z)\frac{k_{\perp}^{\mu}k_{\perp}^{\nu}}{k_{\perp}^2}
\eeq 
and the collinear limit is parametrised as
\beqa
&& p_{a}^{\mu}=zp^{\mu}+k_{\perp}^{\mu}-\frac{k_{\perp}^2}{z}\frac{n^{\mu}}{2p\cdot n}\nonumber\\
&& p_{b}^{\mu}=(1-z)p^{\mu}-k_{\perp}^{\mu}-\frac{k_{\perp}^2}{1-z}\frac{n^{\mu}}{2p\cdot n}.
\eeqa
$n$ is a reference light-like four-vector needed to uniquely determine the transverse component $k_{\perp}$, which satisfies $k_{\perp}\cdot p$=0 and $k_{\perp}\cdot n$=0. The collinear limit is approached as $k_{\perp}\rightarrow 0$, and, as it can be seen from eqs. (\ref{eq.collccunp}) and (\ref{eq.splitunp}), the angular correlation terms are proportional to
\beq\label{eq.azimuthal}
\frac{1}{s_{ab}}\frac{k_{\perp}^{\mu}k_{\perp}^{\nu}}{k_{\perp}^2}|\cm_{n-1}(...,c_g,...)|_{\mu\nu}^2.
\eeq

Since in the antenna formalism single unresolved limits of the amplitudes squared are subtracted with spin-averaged three-parton antenna functions multiplied with reduced matrix elements which do not contain the azimuthal terms produced by the spin correlations of the matrix element, it is clear that the subtraction terms do not have the same pointwise singular behaviour as the amplitudes squared. 

In the same line of thought, a similar subtlety is found in the subtraction terms for two colour-connected unresolved partons. As explained in section \ref{sec.formalism} these subtraction terms (${\rm d}\sigma^{S,b}_{NNLO}$) are constructed as a difference of four-parton antenna functions and products of two three-parton antennae multiplied with reduced matrix elements squared. The former are intended to subtract all singularities in the double unresolved region, while the latter ensure that the whole subtraction term is free of singularities in all single unresolved regions. If the four-parton antenna in eqs. (\ref{eq.sub2bff}-\ref{eq.sub2bii}) has single unresolved limits where angular terms are present, the product of the two three-parton antenna functions in the same equation will no longer subtract its single collinear behaviour locally. In both cases (real radiation matrix elements and unintegrated subtraction terms), however, angular correlation terms are averaged out once the integration over the azimuthal angle between the momenta $p_i$ ($i=a,b$)  and $p$ (the collinear direction), has been carried out for all collinear limits. Moreover, it can be shown \cite{weinzierl2j,3jet,joao} that in gluon splittings the functional dependence of the angular terms in eq.(\ref{eq.azimuthal}) on the azimuthal angle $\phi$ is 
\beq\label{eq.azimuthal}
\frac{1}{s_{ab}}\frac{k_{\perp}^{\mu}k_{\perp}^{\nu}}{k_{\perp}^2}|\cm_{n-1}(...,c_g,...)|_{\mu\nu}^2 \sim \cos(2\phi + \alpha).
\eeq
This suggests that spin correlations can be averaged out by combining two phase space points whose azimuthal angles are
\beq
\phi\;\;\text{and}\;\;\phi+\frac{\pi}{2}.
\eeq
It is only after this combination of phase space points that the subtraction terms are a good approximation of the double real radiation corrections in all its single and double unresolved limits. This procedure, which has proven to be extremely successful for the double real radiation corrections to gluon scattering \cite{joao}, is the one that we shall follow in section \ref{sec.results} to show that our subtraction terms correctly converge to the real radiation matrix elements squared in all their unresolved limits.

It is also important to note that angular terms in the four-parton antenna functions vanish once they are integrated over the corresponding antenna phase space. This fact implies that the integrated antenna functions do not have any azimuthal correlations. This is crucial for the cancellation of infrared poles when the subtraction terms involving these antennae have to be added back in integrated forms at the $m$-parton level.

\section{Top quark pair production at the LHC}\label{sec.subterms}
In this section we shall present the colour-ordered double real emission contributions to $t\bar{t}$ production at the LHC due to processes involving only fermions. Together with these, we shall give their corresponding antenna subtraction terms, which capture all single and double unresolved limits of the respective real radiation matrix elements squared.

\subsection{Structure of double real radiation contributions}
In a previous paper \cite{us} we have presented the colour-ordered form of the real radiation corrections as well as the corresponding NLO subtraction terms for all the partonic processes contributing to $t\bar{t}+$jet production at hadron colliders. Both the NLO real radiation corrections and the subtraction terms for $t\bar{t}$+jet are essential ingredients for the study of the double real radiation corrections to heavy quark pair production at NNLO: on one hand, the matrix elements involved and their colour decomposition are the same, and, on the other hand, the aforementioned NLO subtraction terms already capture all the single unresolved infrared behaviour of the matrix elements, which is needed at NNLO in the $\ds^{S,a}_{NNLO}$ piece of the subtraction term. Indeed, at partonic level, the forms derived in \cite{us} for the real radiation matrix elements together with their related subtraction terms for single unresolved limits can be taken over to the present paper without any changes. While the real radiation matrix elements are exactly the same, the NLO subtraction terms for $t\bar{t}+$jet production have to be supplemented with additional genuine NNLO pieces in order to obtain subtraction terms that account for the double unresolved features of these six-particle amplitudes squared.

We shall start by discussing the contributions of purely fermionic processes to the double real radiation corrections to $t\bar{t}$ production at the LHC, and by recalling from \cite{us} some of the properties of the corresponding real radiation matrix elements.

\subsubsection{Generalities}
To facilitate the reading of our expressions, we shall closely follow the notation in \cite{us} for matrix elements and subtractions terms. The main points of our conventions are the following: The matrix elements denoted with ${\cal M}$ represent colour-ordered sub-amplitudes in which the coupling constants and colour factors are omitted. Furthermore, to explicitly visualise the colour connection between particles in these colour-ordered amplitudes, a double semicolon is used in the labeling of the partons present in a given matrix element. This double semicolon is used for separating chains of colour-connected partons. Partons within a pair of double semicolons belong to a same colour chain, and adjacent partons within a colour chain are colour-connected. An antiquark (or an initial state quark) at the end of a colour chain and a like flavour quark (or initial state antiquark) at the beginning of a different colour chain are also colour-connected since the two chains merge in the collinear limit, where the $q\bar{q} $ is clusters into a gluon. Since in the antenna framework a parton can only be unresolved with respect to its colour-connected neighbours, this notation helps to identify the unresolved limits present in a given colour-ordered amplitude, and therefore helps to construct the corresponding subtraction terms. 

Notationwise, we also denote gluons which are photon-like and only couple to quark lines, with the index $\gamma$ instead of $g$. In sub-amplitudes where all gluons are photon-like no semicolons are used, since the concept of colour connection in not meaningful in those configurations. A hat over the label of a certain parton indicates that it is an initial state particle (for example, $\qi{1}$ is an initial state quark with momentum $p_1$).

Concerning the notation in the subtraction terms, the conventions for the reduced matrix elements are the same as those for the real radiation matrix elements discussed above. The remapped final-state momenta are denoted with tildes and the remapped momenta of initial state hard radiators are denoted by a bar and a hat, as used in other papers \cite{joao,RVnew}. In the four-parton antenna functions the hard radiators are ``on the edges'' and the uresolved particles are ``in the middle''. \\

Following the general factorisation formula given in eq.(\ref{eq.hadroncross}), while omitting the renormalisation and factorisation scale dependences, the contribution of the partonic process $q\bar{q}\rightarrow Q\bar{Q}$ to the total leading order cross section for $t\bar{t}$ production in hadronic collisions is
\beq
{\rm d}\sigma_{LO}=\int \frac{{\rm d}\xi_1}{\xi_1}\frac{{\rm d}\xi_2}{\xi_2}  f_q(\xi_1) f_{\bar{q}}(\xi_2)\ds_{q\bar{q}\rightarrow Q\bar{Q}},
\eeq
where the leading order partonic differential cross section is given by
\beq
\ds_{q\bar{q}\rightarrow Q\bar{Q}}=\norm_{LO}\dphi_2(p_1,p_2;p_3,p_4)|\cm_4(\Q{1},\Qb{2},\qbi{3},\qi{4})|^2 J_2^{(2)}(p_1,p_2).
\eeq
The normalisation factor $\norm_{LO}$ reads
\beq\label{eq.norm}
\norm_{LO}=\frac{1}{2s}\times \left( \frac{\alpha_s}{2\pi} \right)^2  \frac{\bar{C}(\epsilon)^2}{C(\epsilon)^2}\times (N_c^2-1)\times \frac{1}{4N_c^2},
\eeq
where $s$ is the hadronic center of mass energy, $(N_c^2-1)$ comes from the colour sum, the factor $1/4N_c^2$ accounts for the averaging over  the spin and colour of the incoming $q\bar{q}$ pair, and $(1/2s)$ is the hadron-hadron flux factor. The coupling is defined as usual: $\alpha_s=g^2/4\pi$, $C(\epsilon)$ is given in eq.(\ref{eq.ceps}), and $\bar{C}(\epsilon)=(4\pi)^{\epsilon}e^{-\epsilon \gamma}$. This way of expressing the coupling factors, with each power of $\alpha_s$ accompanied by a power of $\bar{C}(\epsilon)$, keeps the coupling dimensionless in dimensional regularisation.

The colour and coupling-stripped amplitude $\cm_4(\Q{1},\Qb{2},\qbi{3},\qi{4})$ is related to the full amplitude $M_4^0(\Q{1},\Qb{2},\qbi{3},\qi{4})$ through
\beq
M_4^0(\Q{1},\Qb{2},\qbi{3},\qi{4})=g^2\left( \del{1}{4}\del{3}{2}-\frac{1}{N_c}\del{1}{2}\del{3}{4}\right)\cm_4(\Q{1},\Qb{2},\qbi{3},\qi{4}).
\eeq

The jet function denoted by $J_{2}^{(2)}$ ensures that the top and the antitop are in two separate jets.

The contribution of purely fermionic $\order{\alpha_s^4}$ processes to the double real radiation cross section for massive quark pair production in hadronic collisions is
\beqa
{\rm d}\sigma&=&\int \frac{{\rm d}\xi_1}{\xi_1}\frac{{\rm d}\xi_2}{\xi_2}\bigg\{ \sum_q \left[f_q(\xi_1)f_{\bar{q}}(\xi_2)\ds_{q\bar{q}\rightarrow Q\bar{Q}q\bar{q}}+f_q(\xi_1)f_q(\xi_2)\ds_{qq\rightarrow Q\bar{Q}qq}\right.\nonumber\\
&&\left. \hspace{25mm}+f_{\bar{q}}(\xi_1)f_{\bar{q}}(\xi_2)\ds_{\bar{q}\bar{q}\rightarrow Q\bar{Q}\bar{q}\bar{q}}\right]\nonumber\\
&&\hspace{17mm}+\sum_{q\neq q'}\left[ f_q(\xi_1)f_q(\xi_2)\ds_{q\bar{q}\rightarrow Q\bar{Q}q'\bar{q}'}+f_q(\xi_1)f_{q'}(\xi_2)\ds_{qq'\rightarrow Q\bar{Q}qq'} \right.\nonumber\\
&&\left. \hspace{25mm}+f_{\bar{q}}(\xi_q)f_{\bar{q}'}(\xi_2)\ds_{\bar{q}\bar{q}'\rightarrow Q\bar{Q}\bar{q}\bar{q}'}+f_q(\xi_1)f_{\bar{q}'}(\xi_2)\ds_{q\bar{q}'\rightarrow Q\bar{Q}q\bar{q}'}\right]\bigg\},\label{eq.hadronic}
\eeqa
with the partonic cross sections given by
\beqa
&&\ds_{q\bar{q}\rightarrow Q\bar{Q}q'\bar{q}'}={\rm d}\Phi_4(k_Q,k_{\bar{Q}},k_{q'},k_{\bar{q}'};p_q,p_{\bar{q}})|M^0_{q\bar{q}\rightarrow Q\bar{Q}q'\bar{q}'}|^2 J_2^{(4)}(k_Q,k_{\bar{Q}},k_{q'},k_{\bar{q}'})\label{eq.partonic1}\\
&&\ds_{q\bar{q}\rightarrow Q\bar{Q}q\bar{q}}={\rm d}\Phi_4(k_Q,k_{\bar{Q}},k_{q},k_{\bar{q}};p_q,p_{\bar{q}})|M^0_{q\bar{q}\rightarrow Q\bar{Q}q\bar{q}}|^2 J_2^{(4)}(k_Q,k_{\bar{Q}},k_{q},k_{\bar{q}})\label{eq.partonic2}\\
&&\ds_{qq'\rightarrow Q\bar{Q}qq'}={\rm d}\Phi_4(k_Q,k_{\bar{Q}},k_{q},k_{q'};p_q,p_{q'})|M^0_{qq'\rightarrow Q\bar{Q}qq'}|^2 J_2^{(4)}(k_Q,k_{\bar{Q}},k_q,k_{q'})\label{eq.partonic3}\\
&&\ds_{qq\rightarrow Q\bar{Q}qq}={\rm d}\Phi_4(k_Q,k_{\bar{Q}},k_{q},k_{q};p_q,p_{q})|M^0_{qq\rightarrow Q\bar{Q}qq}|^2 J_2^{(4)}(k_Q,k_{\bar{Q}},k_q,k_{q}\label{eq.partonic4}).
\eeqa
Charge conjugation relates all other partonic cross sections in eq.(\ref{eq.hadronic}) to one of those given in eqs.(\ref{eq.partonic1}-\ref{eq.partonic4}). For example $\ds_{\bar{q}\bar{q}\rightarrow Q\bar{Q}\bar{q}\bar{q}}$ is identical to $\ds_{qq\rightarrow Q\bar{Q}qq}$, etc. 

The jet functions appearing in eqs.(\ref{eq.partonic1}-\ref{eq.partonic4}) are all of the type $J_{2}^{(4)}$ and correspond to the selection criteria of a $2$-jet event: out of four partons, from which two are a $Q \bar{Q}$ pair, an event with two jets is built. Each of these two jets has the heavy quark $Q$ or the heavy antiquark $\bar{Q}$ in it.

\subsubsection{Colour structure}\label{subsec:colouramplitudes}
In order to obtain the subtraction terms, the colour decomposition of the real matrix elements present in the partonic cross sections in eqs.(\ref{eq.partonic1}-\ref{eq.partonic4}) has to be performed. However, since colour ordering does not distinguish between initial and final state partons, it is sufficient to consider the colour decomposition for the unphysical processes $0\rightarrow Q\bar{Q}q\bar{q} q' \bar{q'} $ and obtain the matrix elements needed for $t \bar{t}$ production by appropriate crossings. We shall follow this strategy below.

The colour decomposition of the amplitude for $0\rightarrow Q\bar{Q} q \bar{q} q' \bar{q'} $ has been derived in \cite{us}. For completeness we recall it here. Since there are no gluons, the colour structure of the amplitude is rather simple: no adjoint representation indices are present, i.e. no $T^a$ matrices appear, and all colour factors simply consist of products of fundamental representation deltas. Taking all particles as outgoing, we can write the amplitude for the fictitious process $0\rightarrow Q\bar{Q}q\bar{q}q'\bar{q}'$ as 
\beqa
M_6^0(\Q{1},\Qb{2},\q{3},\qb{4},\qp{5},\qpb{6})&=&g^4\bigg[  \del{1}{4}\del{3}{6}\del{5}{2}\cm_6(\Q{1},\qb{4};;\q{3},\qpb{6};;\qp{5},\Qb{2})\nonumber\\
&&\:\:\:+\del{1}{6}\del{3}{2}\del{5}{4}\cm_6(\Q{1},\qpb{6};;\q{3},\Qb{2};;\qp{5},\qb{4})\nonumber\\
&&-\frac{1}{N_c}\del{1}{4}\del{3}{2}\del{5}{6}\cm_6(\Q{1},\qb{4};;\q{3},\Qb{2};;\qp{5},\qpb{6})\nonumber\\
&&-\frac{1}{N_c}\del{1}{6}\del{3}{4}\del{5}{2}\cm_6(\Q{1},\qpb{6};;\q{3},\qb{4};;\qp{5},\Qb{2})\nonumber\\
&&-\frac{1}{N_c}\del{1}{2}\del{3}{6}\del{5}{4}\cm_6(\Q{1},\Qb{2};;\q{3},\qpb{6};;\qp{5},\qb{4})\nonumber\\
&&+\frac{1}{N_c^2}\del{1}{2}\del{3}{4}\del{5}{6}\cm_6(\Q{1},\Qb{2};;\q{3},\qb{4};;\qp{5},\qpb{6})\bigg],\label{eq.meni}
\eeqa

The subleading colour amplitudes in eq.(\ref{eq.meni}) always contain a colour chain made of a quark-antiquark pair of equal flavour. This fact manifests itself in the corresponding colour coefficients, which have a factor $\delta_{i_{q},j_{\bar{q}}}$ (with $i$ and $j$ of the same flavour). The presence of this factor is related to the fact that this $q\bar{q}$ pair splits from a photon-like propagator and as we shall shortly see, this has important consequences in the way in which the colour-ordered sub-amplitudes factorise in their double soft limits.

For those processes involving two pairs of identical flavour quark-antiquark pairs, the colour decomposition can be obtained from eq.(\ref{eq.meni}) by using
\beq\label{eq.mei}
M_6^0(\Q{1},\Qb{2},\q{3},\qb{4},\q{5},\qb{6})=M_6^0(\Q{1},\Qb{2},\q{3},\qb{4},\qp{5},\qpb{6})-M_6^0(\Q{1},\Qb{2},\q{3},\qb{6},\qp{5},\qpb{4}),
\eeq
where the relative minus sign is due to the fact that both amplitudes differ by the interchange of $p_4$ and $p_6$. 

The amplitudes squared for all partonic processes under consideration can be written in terms of colour-ordered amplitudes by squaring either eq.(\ref{eq.meni}) or eq.(\ref{eq.mei}) and crossing the corresponding (anti) quarks to the initial state. When squaring, interferences of different colour-ordered amplitudes appear. Most of the time, they can, however, be eliminated by using the following identities
\beqa
\lefteqn{\cm_6(\Q{1},\Qb{2};;\q{3},\qb{4};;\qp{5},\qpb{6})}\nonumber\\
&&=\cm_6(\Q{1},\qb{4};;\q{3},\qpb{6};;\qp{5},\Qb{2})+\cm_6(\Q{1},\qpb{6};;\q{3},\Qb{2};;\qp{5},\qb{4})\nonumber\\
&&=\frac{1}{2}\big(\cm_6(\Q{1},\qb{4};;\q{3},\Qb{2};;\qp{5},\qpb{6})+\cm_6(\Q{1},\qpb{6};;\q{3},\qb{4};;\qp{5},\Qb{2})\nonumber\\
&&\hspace{10mm}+\cm_6(\Q{1},\Qb{2};;\q{3},\qpb{6};;\qp{5},\qb{4})\big)\label{eq.deci}
\eeqa 
which relate the leading, subleading, and subsubleading colour partial amplitudes.

\subsection{Soft limits of subleading colour contributions}\label{subsec.subleading}
The procedure to subtract the limits in which a quark-antiquark pair becomes soft is rather straightforward for those colour-ordered amplitudes where the neighbours of the $q\bar{q}$ pair in the colour chain are easy to identify. When the soft pair $\qb{c},\q{d}$ is colour-connected to the hard particles $a$ and $b$, the corresponding sub-amplitude factorises as
\beq
|\cm_{n}(...,a,\qb{c};;\q{d},b,...)|^2 \stackrel{^{p_c,p_d\rightarrow 0}}{\longrightarrow}\soft{a}{c}{d}{b}(m_a,m_b)|\cm_{n-2}(...,a,b,...)|^2,
\eeq
where the double soft factor $\soft{a}{c}{d}{b}$ has been given in eq.(\ref{eq.seik}). It is clear that a suitable subtraction term is
\beq\label{eq.factcc}
X_4^0(a,\qb{c},\q{d},b)|\cm_{n-2}(...,\tilde{a},\tilde{b},...)|^2,
\eeq
since
\beq
X_4^0(a,\qb{c},\q{d},b) \stackrel{^{p_c,p_d\rightarrow 0}}{\longrightarrow}\soft{a}{c}{d}{b}(m_a,m_b),
\eeq
and the momentum mapping is such that the reduced matrix element in eq.(\ref{eq.factcc}) gives $|\cm_{n-2}(...,a,b,...)|^2$ when the $q\bar{q}$ pair becomes soft. The type of four-parton antenna function needed depends on the identity of the hard radiators $a$ and $b$, and the type of phase space mapping depends on whether these radiators are in the initial or in the final state.

The same strategy cannot be followed with those subleading colour amplitudes of the form $\cm_n(...,a;;\q{d},\qb{c})$, whose colour factors contain $\del{d}{c}$. In these cases the gluon propagator from which the soft quark and antiquark split is photon-like, and the hard radiators cannot be simply identified as the pair's adjacent particles in the colour chain, because the concept does not even apply.

In the next section we shall explain how we can build antenna subtraction terms for these subleading colour amplitudes by recalling the factorisation properties of colour-ordered amplitudes in their soft limits. Since, from a physical point of view, the emission of a soft $q\bar{q}$ pair is closely related to the emission of a single soft gluon, we shall start by addressing the problem for this latter case. Note that our results will be obtained in the most general massive case, from which the massless counterparts can be inferred by setting all masses to zero.

\subsubsection{Emission of a single photon-like soft gluon}
When a soft gluon is emitted between the hard particles $a$ and $b$, a colour-ordered amplitude factorises as
\beq\label{eq.factampg}
\cm_{n}(...,a,g,b,...)| \stackrel{^{p_g \rightarrow 0}}{\longrightarrow} \epsilon_{\mu}(p_g,\lambda)\left[ J_a^{\mu}(p_g)-J_b^{\mu}(p_g) \right] \cm_{n-1}(...,a,b,...)
\eeq
where $ \epsilon_{\mu}(p_g,\lambda)$ is the soft gluon's polarisation vector, and the soft currents are given by
\beq\label{eq.currentg}
J_i^{\mu}(p_g)=\frac{p_i^{\mu}}{\sqrt{2}p_i\cdot p_g}.
\eeq
After squaring eq.(\ref{eq.factampg}) we obtain the well-known formula
\beq\label{eq.wellknownfact}
|\cm_{n}(...,a,g,b,...)|^2 \stackrel{^{p_g \rightarrow 0}}{\longrightarrow}\ssoft{a}{g}{b}(m_a,m_b)|\cm_{n-1}(...,a,b,...)|^2,
\eeq
with
\beqa
\ssoft{a}{g}{b}(m_a,m_b)&=&\left( \sum_{\lambda = \pm} \epsilon_{\mu}(p_g,\lambda) \epsilon_{\nu}(p_g,-\lambda)\right)\left[ J_a^{\mu}(p_g)-J_b^{\mu}(p_g) \right]\left[ J_a^{\nu}(p_g)-J_b^{\nu}(p_g) \right]\nonumber\\
&=&- \left[ J_a(p_g)-J_b(p_g) \right]^2.\label{eq.seikonalg}
\eeqa
By evaluating the second line in eq.(\ref{eq.seikonalg}) explicitly with the current given in eq.(\ref{eq.currentg}), the form in eq.(\ref{eq.eikonalmassive}) for the soft eikonal factor given is obtained. To get this factor, we furthermore use  
the fact that colour-ordered amplitudes are gauge invariant which allows us to replace 
\beq
\sum_{\lambda = \pm} \epsilon_{\mu}(p_g,\lambda) \epsilon_{\nu}(p_g,-\lambda) \rightarrow -g_{\mu\nu}
\eeq
in eq.(\ref{eq.seikonalg}).

It should be stressed once more that in this case, the hard radiators $a$ and $b$ are the neighbouring particles of gluon $g$ in the colour chain, and they can be both fermions or gluons. This statement is no longer true for subleading colour amplitudes in which the soft gluon $g$ is photon-like. In such cases, the abelian nature of the soft particle implies, on one hand, that it does not couple to the hard gluons which the amplitude might have, and it also implies, on the other hand, that all the fermions involved in the process can act as hard radiators. Therefore, the factorisation of a subleading colour amplitude in the limit where a photon-like gluon becomes soft reads
\beq\label{eq.factabeliang}
\cm_n(...,a;;g_{\gamma})\stackrel{^{p_g\rightarrow 0}}{\longrightarrow}\epsilon_{\mu}(p_g,\lambda) \Bigg( \sum_{i\in\{q\}}J_i^{\mu}(p_g)-\sum_{j\in\{\bar{q}\}}J_j^{\mu}(p_g) \Bigg) \cm_{n-1}(...,a),
\eeq
where $\{q\}$ stands for all  final state quarks and initial state antiquarks in the process, and $\{ \bar{q} \}$ stands for all final state antiquarks and initial state quarks. Squaring eq.(\ref{eq.factabeliang}) and rearranging the result into a more convenient form, we obtain the following factorisation for the squared subleading colour amplitude:
\beqa
&&\hspace{-5mm}|\cm_n(...,a;;g_{\gamma})|^2 \stackrel{^{p_g\rightarrow 0}}{\longrightarrow} \Bigg( \sum_{\substack{i\in\{q\} \\ j\in\{\bar{q}\}}}\ssoft{i}{g}{j}(m_i,m_j)-\frac{1}{2}\sum_{\substack{(i,j) \in\{q\}\\ i\neq j }}\ssoft{i}{g}{j}(m_i,m_j)\nonumber\\
&&\hspace{55mm}-\frac{1}{2}\sum_{\substack{(i,j) \in\{\bar{q}\}\\ i\neq j}}\ssoft{i}{g}{j}(m_i,m_j) \Bigg)|\cm_{n-1}(...,a)|^2.\label{eq.factabeliang2}
\eeqa
This shows that, unlike the case presented in eq.(\ref{eq.wellknownfact}), the soft factor corresponding to a soft photon-like gluon is a combination of several of soft eikonal factors given in eq.(\ref{eq.eikonalmassive}).

It can be immediately seen from eq.(\ref{eq.factabeliang2}) that in order to subtract the soft limits of parton $g$ in a squared sub-amplitude of the form $|\cm_n(...,a;;g)|^2$, we can use the following combination of antenna functions and reduced matrix elements 
\footnote{Note that this result was already obtained in \cite{us} where it was formulated slightly differently and concerned the soft behaviour of interference terms.}:
\beqa
&& \hspace{8.5mm}\sum_{\substack{i\in\{q\} \\ j\in\{\bar{q}\}}} X_3^0(i,g,j) |\cm_{n-1}(...,\tilde{a})_{(i,j)}|^2\nonumber\\
&& -\frac{1}{2}\sum_{\substack{(i,j) \in\{q\}\\ i\neq j }} X_3^0(i,g,j) |\cm_{n-1}(...,\tilde{a})_{(i,j)}|^2\nonumber\\
&& -\frac{1}{2}\sum_{\substack{(i,j) \in\{\bar{q}\}\\ i\neq j}} X_3^0(i,g,j)|\cm_{n-1}(...,\tilde{a})_{(i,j)}|^2\label{eq.subtermabeliang}
\eeqa
The subscripts $(i,j)$ in the reduced matrix elements mean that the phase space mapping for each term is different, since the hard radiators in each term are different. Since in the limit $p_g\rightarrow 0$ the antenna functions $X_3^0(i,g,j)$ give the soft eikonal factors $\ssoft{i}{g}{j}(m_i,m_j)$ and the phase space mappings reduce to the Born kinematics, it is clear that eq.(\ref{eq.subtermabeliang}) constitutes a suitable subtraction term for eq.(\ref{eq.factabeliang2}). 

Two crucial properties of the subtraction term in eq.(\ref{eq.subtermabeliang}) should be pointed out. The first of these is that it subtracts not only the soft limits of the photon-like gluon, but also all the single collinear limits in which this gluon participates. The second property is that it does not introduce any spurious singularities that are not present in the real radiation matrix elements.

\subsubsection{Emission of a soft $q\bar{q}$ pair splitting from a photon-like gluon propagator}\label{sec.subleading}
When a soft quark-antiquark pair is emitted between partons $a$ and $b$ in the colour chain, the colour sub-amplitudes factorise as
\beq\label{eq.factqqbar}
\cm_{n}(...,a,\qb{c};;\q{d},b,...)| \stackrel{^{p_c,p_d\rightarrow 0}}{\longrightarrow} [\bar{u}_{s_d}(p_d)\hspace{0.3mm}\gamma_{\mu} \hspace{0.3mm}v_{s_c}(p_c)] \left( J_a^{\mu}(p_c,p_d)-J_b^{\mu}(p_c,p_d) \right) \cm_{n-2}(...,a,b,...),
\eeq
where the soft currents are given by
\beq\label{eq.dsoftcurrent}
J_i^{\mu}(p_j,p_k)=\frac{p_i^{\mu}}{s_{jk}(s_{ij}+s_{ik})}.
\eeq
Squaring eq.(\ref{eq.factqqbar}) and summing over the spins of the soft quark and antiquark gives
\beq
|\cm_{n}(...,a,\qb{c};;\q{d},b,...)|^2 \stackrel{^{p_c,p_d\rightarrow 0}}{\longrightarrow}\soft{a}{c}{d}{b}(m_a,m_b)|\cm_{n-2}(...,a,b,...)|^2
\eeq
with the double soft factor  $\soft{a}{c}{d}{b}(m_a,m_b)$ encountered in eq.(\ref{eq.seik}). In terms of the currents given above, it can be rewritten as
\beq\label{eq.softfactorderivation}
\soft{a}{c}{d}{b}(m_a,m_b)={\rm tr}(\not{p}_d\gamma_{\mu}\hspace{-1mm}\not{p}_c\gamma_{\nu})\left( J_a^{\mu}(p_c,p_d)-J_b^{\mu}(p_c,p_d) \right)\left( J_a^{\nu}(p_c,p_d)-J_b^{\nu}(p_c,p_d) \right)^{\dagger}
\eeq
By explicitly evaluating the trace in eq.(\ref{eq.softfactorderivation}) and using the definition of the currents in eq.(\ref{eq.dsoftcurrent}) the expresssion given in eq.(\ref{eq.seik}) is obtained.

When a $q\bar{q}$ pair becomes soft in a sub-amplitude whose colour factor contains $\del{d}{c}$, the factorisation given in eq.(\ref{eq.factqqbar}) does not hold. This is due to the fact that the presence of $\del{d}{c}$ factor indicates that the gluon propagator from which the soft pair splits is photon-like. As discussed in the previous section, in this case the soft pair can be emitted from all the other fermions present in the process. All of these can act as hard radiators. Thus, in the limit $p_c,p_d \rightarrow 0$ these type of sub-amplitudes factorise as  
\beq\label{eq.factabelian}
\cm_n(...,a;;\q{d},\qb{c})\stackrel{^{p_c,p_d\rightarrow 0}}{\longrightarrow}[\bar{u}_{s_d}(p_d)\hspace{0.3mm}\gamma_{\mu} \hspace{0.3mm}v_{p_c}(p_c)] \Bigg( \sum_{i\in\{q\}}J_i^{\mu}(p_c,p_d)-\sum_{j\in\{\bar{q}\}}J_j^{\mu}(p_c,p_d) \Bigg) \cm_{n-2}(...,a).
\eeq
Squaring eq.(\ref{eq.factabelian}) we find
\beqa
&&\hspace{-5mm}|\cm_n(...,a;;\q{d},\qb{c})|^2 \stackrel{^{p_c,p_d\rightarrow 0}}{\longrightarrow} \Bigg( \sum_{\substack{i\in\{q\} \\ j\in\{\bar{q}\}}}\soft{i}{c}{d}{j}(m_i,m_j)-\frac{1}{2}\sum_{\substack{(i,j) \in\{q\}\\ i\neq j }}\soft{i}{c}{d}{j}(m_i,m_j)\nonumber\\
&&\hspace{55mm}-\frac{1}{2}\sum_{\substack{(i,j) \in\{\bar{q}\}\\ i\neq j}}\soft{i}{c}{d}{j}(m_i,m_j) \Bigg)|\cm_{n-2}(...,a)|^2,
\eeqa
which suggests that a suitable subtraction term for these limits can be symbolically written as
\beqa
\sum_{\substack{i\in\{q\} \\ j\in\{\bar{q}\}}} (X_4^0(i,c,d,j)-X_3^0(i,c,d)X_3^0(\wt{ic},\wt{cd},j)-X_3^0(c,d,j)X_3^0(i,\wt{cd},\wt{dj}) |\cm_{n-2}(...,\tilde{a})_{(i,j)}|^2\nonumber\\
-\frac{1}{2}\sum_{\substack{(i,j) \in\{q\}\\ i\neq j }} (X_4^0(i,c,d,j)-X_3^0(i,c,d)X_3^0(\wt{ic},\wt{cd},j)-X_3^0(c,d,j)X_3^0(i,\wt{cd},\wt{dj}) |\cm_{n-2}(...,\tilde{a})_{(i,j)}|^2\label{eq.subtermabelian}\nonumber\\
-\frac{1}{2}\sum_{\substack{(i,j) \in\{\bar{q}\}\\ i\neq j}} (X_4^0(i,c,d,j)-X_3^0(i,c,d)X_3^0(\wt{ic},\wt{cd},j)-X_3^0(c,d,j)X_3^0(i,\wt{cd},\wt{dj})|\cm_{n-2}(...,\tilde{a})_{(i,j)}|^2.\nonumber\\
\eeqa
where the products of the three-parton antennae $X^{0}_{3}$ remove the single unresolved behaviours present in the four-parton antennae $X_{4}^{0}$. Crucially, we have checked that this combination of antenna functions not only explicitly subtracts double soft limits in sub-amplitudes of the form $|\cm_n(...,a;;\q{d},\qb{c})|^2$, but also reproduces the triple collinear limits that involve $\q{d},\qb{c}$ in these sub-amplitudes without introducing any spurious singularities.

\subsection{Fermionic processes for $Q \bar{Q}$ production at LHC}
In the following, we shall explicitly write the real radiation matrix elements for each process in terms of colour-ordered amplitudes squared, and give the corresponding double real subtraction terms capturing single and double unresolved radiation features. The colour decompositions as well as the subtraction terms of single unresolved limits have already been derived in \cite{us}. The double unresolved parts are new.

\subsubsection{The $q\bar{q}\rightarrow Q\bar{Q}q'\bar{q}'$ process}
By squaring eq.(\ref{eq.meni}), using eq.(\ref{eq.deci}) to remove interferences, and crossing $\q{3}$ and $\qb{4}$ to the initial state, we can write the amplitude squared for $q\bar{q}\rightarrow Q\bar{Q}q'\bar{q}'$ as
\beqa
\lefteqn{|M_6^0(\Q{1},\Qb{2},\qbi{3},\qi{4},\qp{5},\qpb{6})|^2=g^8(N_c^2-1)}\nonumber\\
&&\bigg\{ N_c \bigg( |\cm_6(\Q{1},\qi{4};;\qbi{3},\qpb{6};;\qp{5},\Qb{2})|^2+|\cm_6(\Q{1},\qpb{6};;\qbi{3},\Qb{2};;\qp{5},\qi{4})|^2\bigg) \nonumber\\
&&+\frac{1}{N_c}\bigg( |\cm_6(\Q{1},\qi{4};;\qbi{3},\Qb{2};;\qp{5},\qpb{6})|^2+|\cm_6(\Q{1},\qpb{6};;\qbi{3},\qi{4};;\qp{5},\Qb{2})|^2\nonumber\\
&&\:\:\:\:\:\:\:\:\: +|\cm_6(\Q{1},\Qb{2};;\qbi{3},\qpb{6};;\qp{5},\qi{4})|^2-3|\cm_6(\Q{1},\Qb{2};;\qbi{3},\qi{4};;\qp{5},\qpb{6})|^2\bigg)\bigg\}.\label{eq:me2ni}
\eeqa

This amplitude is singular in the $\qp{5}||\qpb{6}$ single collinear limit, as well as in the $\qp{5}$,$\qpb{6}$ double soft limit and the $\qi{4}||\qp{5}||\qpb{6}$ and $\qbi{3}||\qp{5}||\qpb{6}$ triple collinear limits. To subtract these divergencies from the differential partonic cross section we use the following subtraction term
\beqa
\lefteqn{\ds^S_{q\bar{q}\rightarrow Q\bar{Q}q'\bar{q}'}=\norm_{LO} N_F \left( \frac{\alpha_s}{2\pi} \right)^2\frac{\bar{C}(\epsilon)^2}{C(\epsilon)^2} \dphi_4(p_1,p_2,p_5,p_6;p_3,p_4)}\nonumber\\
&&\times\bigg\{ N_c\bigg[ \frac{1}{2} E_3^0(\Q{1},\qp{5},\qpb{6})\bigg( |\cm_5(\Q{(\wt{15})},\qi{4};;\qbi{3},\gl{(\wt{56})},\Qb{2})|^2\nonumber\\
&&\:\:\:\:\:\:\:\:\:\:\:\:\:\:\:\:\:\:\:\:\:\:\:\:\:\:\:\:\: +|\cm_5(\Q{(\wt{15})},\gl{(\wt{56})},\qi{4};;\qbi{3},\Qb{2})|^2 \bigg) J_2^{(3)}(\wt{p_{15}},p_2,\wt{p_{56}})\nonumber\\
&&\:\:\:\:\:\:\:\:\:\:\:+\frac{1}{2} E_3^0(\Qb{2},\qp{5},\qpb{6})\bigg( |\cm_5(\Q{1},\qi{4};;\qbi{3},\gl{(\wt{56})},\Qb{(\wt{25})})|^2\nonumber\\
&&\:\:\:\:\:\:\:\:\:\:\:\:\:\:\:\:\:\:\:\:\:\:\:\:\:\:\:\:\:+|\cm_5(\Q{1},\gl{(\wt{56})},\qi{4};;\qbi{3},\Qb{(\wt{25})})|^2\bigg)J_2^{(3)}(p_1,\wt{p_{25}},\wt{p_{56}})\bigg]\nonumber\\
&&\:\:\:\:\:\:\:\:\:\:\:+\bigg( B_4^0(\Q{1},\qpb{6},\qp{5},\qi{4}) - \frac{1}{2}E_3^0(\Q{1},\qp{5},\qpb{6})A_3^0(\Q{(\wt{15})},\gl{(\wt{56})},\qi{4})\nonumber\\
&&\:\:\:\:\:\:\:\:\:\:\:\:\:\:\:\:\:\: -\frac{1}{2} E_3^0(\Qb{2},\qp{5},\qpb{6})A_3^0(\Q{1},\gl{(\wt{56})},\qi{4})\bigg) |\cm_4(\Q{(\wt{156})},\Qb{2},\qbi{3},\qi{\bar{4}})|^2 J_2^{(2)}(\wt{p_{156}},p_2)\nonumber\\
&&\:\:\:\:\:\:\:\:\:\:\:+\bigg( B_4^0(\Qb{2},\qp{5},\qpb{6},\qbi{3}) - \frac{1}{2}E_3^0(\Q{1},\qp{5},\qpb{6})A_3^0(\Qb{2},\gl{(\wt{56})},\qbi{3})\nonumber\\
&&\:\:\:\:\:\:\:\:\:\:\:\:\:\:\:\:\:\: -\frac{1}{2} E_3^0(\Qb{2},\qp{5},\qpb{6})A_3^0(\Qb{(\wt{25})},\gl{(\wt{56})},\qbi{3})\bigg) |\cm_4(\Q{1},\Qb{(\wt{256})},\qbi{\bar{3}},\qi{4})|^2 J_2^{(2)}(p_1,\wt{p_{256}})\nonumber\\
&&\:\:+\frac{1}{N_c}\bigg[\frac{1}{2} E_3^0(\Q{1},\qp{5},\qpb{6})\bigg( |\cm_5(\Q{(\wt{15})},\Qb{2};;\qbi{3},\gl{(\wt{56})},\qi{4})|^2\nonumber\\
&&\:\:\:\:\:\:\:\:\:\:\:\:\:\:\:\:\:\:\:\:\:\:\:\:\:\:\:\:\:+|\cm_5(\Q{(\wt{15})},\gl{(\wt{56})},\Qb{2};;\qbi{3},\qi{4})|^2\nonumber\\
&&\:\:\:\:\:\:\:\:\:\:\:\:\:\:\:\:\:\:\:\:\:\:\:\:\:\:\:\:\:-2|\cm_5(\Q{(\wt{15})},\Qb{2},\qbi{3},\qi{4},\ph{(\wt{56})})|^2\bigg)   J_2^{(3)}(\wt{p_{15}},p_2,\wt{p_{56}})\nonumber\\
&&\:\:\:\:\:\:\:\:\:\:\:+\frac{1}{2}E_3^0(\Qb{2},\qp{5},\qpb{6})\bigg( |\cm_5(\Q{1},\Qb{(\wt{25})};;\qbi{3},\gl{(\wt{56})},\qi{4})|^2\nonumber\\
&&\:\:\:\:\:\:\:\:\:\:\:\:\:\:\:\:\:\:\:\:\:\:\:\:\:\:\:\:\:+|\cm_5(\Q{1},\gl{(\wt{56})},\Qb{(\wt{25})};;\qbi{3},\qi{4})|^2\nonumber\\
&&\:\:\:\:\:\:\:\:\:\:\:\:\:\:\:\:\:\:\:\:\:\:\:\:\:\:\:\:\:-2|\cm_5(\Q{1},\Qb{(\wt{25})},\qbi{3},\qi{4},\ph{(\wt{56})})|^2\bigg)J_2^{(3)}(p_1,\wt{p_{25}},\wt{p_{56}})\nonumber\\
&&\:\:\:\:\:\:\:\:\:\:\:-\bigg( B_4^0(\Q{1},\qpb{6},\qp{5},\Qb{2}) - \frac{1}{2}E_3^0(\Q{1},\qp{5},\qpb{6})A_3^0(\Q{(\wt{15})},\gl{(\wt{56})},\Qb{2})\nonumber\\
&&\:\:\:\:\:\:\:\:\:\:\:\:\:\:\:\:\:\: -\frac{1}{2} E_3^0(\Qb{2},\qp{5},\qpb{6})A_3^0(\Q{1},\gl{(\wt{56})},\Qb{(\wt{25})})\bigg) |\cm_4(\Q{(\wt{156})},\Qb{(\wt{256})},\qbi{3},\qi{4})|^2 J_2^{(2)}(\wt{p_{156}},\wt{p_{256}})\nonumber\\
&&\:\:\:\:\:\:\:\:\:\:\:-\bigg( B_4^0(\qbi{3},\qpb{6},\qp{5},\qi{4}) - \frac{1}{2}E_3^0(\Q{1},\qp{5},\qpb{6})A_3^0(\qbi{3},\gl{(\wt{56})},\qi{4})\nonumber\\
&&\:\:\:\:\:\:\:\:\:\:\:\:\:\:\:\:\:\: -\frac{1}{2} E_3^0(\Qb{2},\qp{5},\qpb{6})A_3^0(\qbi{3},\gl{(\wt{56})},\qi{4})\bigg) |\cm_4(\Q{\tilde{1}},\Qb{\tilde{2}},\qbi{\bar{3}},\qi{\bar{4}})|^2 J_2^{(2)}(\tilde{p_1},\tilde{p_2})\nonumber\\
&&\:\:\:\:\:\:\:\:\:\:\:-2\bigg( B_4^0(\Q{1},\qpb{6},\qp{5},\qi{4}) - \frac{1}{2}E_3^0(\Q{1},\qp{5},\qpb{6})A_3^0(\Q{(\wt{15})},\gl{(\wt{56})},\qi{4})\nonumber\\
&&\:\:\:\:\:\:\:\:\:\:\:\:\:\:\:\:\:\: -\frac{1}{2} E_3^0(\Qb{2},\qp{5},\qpb{6})A_3^0(\Q{1},\gl{(\wt{56})},\qi{4})\bigg) |\cm_4(\Q{(\wt{156})},\Qb{2},\qbi{3},\qi{\bar{4}})|^2 J_2^{(2)}(\wt{p_{156}},p_2)\nonumber\\
&&\:\:\:\:\:\:\:\:\:\:\:-2\bigg( B_4^0(\Qb{2},\qp{5},\qpb{6},\qbi{3})  - \frac{1}{2}E_3^0(\Q{1},\qp{5},\qpb{6})A_3^0(\Qb{2},\gl{(\wt{56})},\qbi{3})\nonumber\\
&&\:\:\:\:\:\:\:\:\:\:\:\:\:\:\:\:\:\: -\frac{1}{2} E_3^0(\Qb{2},\qp{5},\qpb{6})A_3^0(\Qb{(\wt{25})},\gl{(\wt{56})},\qbi{3})\bigg) |\cm_4(\Q{1},\Qb{(\wt{256})},\qbi{\bar{3}},\qi{4})|^2 J_2^{(2)}(p_1,\wt{p_{256}})\nonumber\\
&&\:\:\:\:\:\:\:\:\:\:\:+2\bigg( B_4^0(\Q{1},\qpb{6},\qp{5},\qbi{3}) - \frac{1}{2}E_3^0(\Q{1},\qp{5},\qpb{6})A_3^0(\Q{(\wt{15})},\gl{(\wt{56})},\qbi{3})\nonumber\\
&&\:\:\:\:\:\:\:\:\:\:\:\:\:\:\:\:\:\: -\frac{1}{2} E_3^0(\Qb{2},\qp{5},\qpb{6})A_3^0(\Q{1},\gl{(\wt{56})},\qbi{3})\bigg) |\cm_4(\Q{(\wt{156})},\Qb{2},\qbi{\bar{3}},\qi{4})|^2 J_2^{(2)}(\wt{p_{156}},p_2)\nonumber\\
&&\:\:\:\:\:\:\:\:\:\:\:+2\bigg( B_4^0(\Qb{2},\qp{5},\qpb{6},\qi{4})  - \frac{1}{2}E_3^0(\Q{1},\qp{5},\qpb{6})A_3^0(\Qb{2},\gl{(\wt{56})},\qi{4})\nonumber\\
&&\:\:\:\:\:\:\:\:\:\:\:\:\:\:\:\:\:\: -\frac{1}{2} E_3^0(\Qb{2},\qp{5},\qpb{6})A_3^0(\Qb{(\wt{25})},\gl{(\wt{56})},\qi{4})\bigg) |\cm_4(\Q{1},\Qb{(\wt{256})},\qbi{3},\qi{\bar{4}})|^2 J_2^{(2)}(p_1,\wt{p_{256}})\bigg]\bigg\},\nonumber\\ \label{eq.sub1}
\eeqa
where the pieces containing the $B_4^0$ antennae subtract the double soft and triple collinear limits, while those multiplied by $E_3^0$ antenna functions subtract the single collinear limit $\qp{5}||\qpb{6}$.

To subtract the double unresolved limits of the subleading colour sub-amplitudes squared $|\cm_6(\Q{1},\qi{4};;\qbi{3},\Qb{2};;\qp{5},\qpb{6})|^2$ and $|\cm_6(\Q{1},\Qb{2};;\qbi{3},\qi{4};;\qp{5},\qpb{6})|^2$ we followed the procedure described in section \ref{sec.subleading} and used eq.(\ref{eq.subtermabelian}). The generic four-parton antennae $X^{0}_{4}$ appearing in eq.(\ref{eq.subtermabelian}) are here given by $B_{4}^0$ type antennae, while the generic $X_{3}^{0}X^{0}_{3}$, also present in eq.(\ref{eq.subtermabelian}), are given above by products of $E_{3}^{0}$ and $A^{0}_{3}$-type  antennae.

\subsubsection{The $q\bar{q}\rightarrow Q\bar{Q}q\bar{q}$ process: identical quark flavour contributions}
The amplitude squared for the identical flavour process $q\bar{q}\rightarrow Q\bar{Q}q\bar{q}$ can be written according to eq.(\ref{eq.mei}) as
\beqa
&&\hspace{-3mm}|M_6^0(\Q{1},\Qb{2},\qbi{3},\qi{4},\q{5},\qb{6})|^2=|M_6^0(\Q{1},\Qb{2},\qbi{3},\qi{4},\qp{5},\qpb{6})|^2+|M_6^0(\Q{1},\Qb{2},\qbi{3},\qpi{4},\qp{5},\qb{6})|^2\nonumber\\
&&\hspace{20mm}+2\re \left( M_6^0(\Q{1},\Qb{2},\qbi{3},\qi{4},\qp{5},\qpb{6})M_6^0(\Q{1},\Qb{2},\qbi{3},\qpi{4},\qp{5},\qb{6})\dagger \right)\label{eq.qqbarqqbarid}.
\eeqa
The Feynman diagrams contributing to this process can be divided into two categories: on one hand we have diagrams that are also present in the non-identical flavour case, which are included in the amplitude $M_6^0(\Q{1},\Qb{2},\qbi{3},\qi{4},\qp{5},\qpb{6})$, and on the other hand we have diagrams in which the initial state quark and antiquark go through to the final state. These latter kinds of diagrams are specific to the identical flavour case and are included in the amplitude $M_6^0(\Q{1}, \Qb{2}, \qbi{3}, \qpi{4}, \qp{5}, \qb{6})$. The amplitude squared $|M_6^0(\Q{1},\Qb{2},\qbi{3},\qi{4},\q{5},\qb{6})|^2$ has  therefore a piece that is the same as the non-identical flavour contribution ($|M_6^0(\Q{1}, \Qb{2}, \qbi{3},\allowbreak{} \qi{4}, \qp{5}, \qpb{6})|^2$), and a remainder piece which we call ``identical-flavour-only". The singularities of the former have been already subtracted with (\ref{eq.sub1}) when we also included the identical quark flavour possibility in the sum over flavours, so we shall now focus on the identical-flavour-only piece
\beqa
&&|M_6^0(\Q{1},\Qb{2},\qbi{3},\qi{4},\q{5},\qb{6})|^2_{\rm{IO}}=|M_6^0(\Q{1},\Qb{2},\qbi{3},\qpi{4},\qp{5},\qb{6})|^2\nonumber\\
&&\hspace{20mm}+2\re \left( M_6^0(\Q{1},\Qb{2},\qbi{3},\qi{4},\qp{5},\qpb{6})M_6^0(\Q{1},\Qb{2},\qbi{3},\qpi{4},\qp{5},\qb{6})\dagger \right).
\eeqa
In terms of colour-ordered amplitudes, this matrix element squared is given by 
\beqa
\lefteqn{|M_6^0(\Q{1},\Qb{2},\qbi{3},\qi{4},\q{5},\qb{6})|^2_{\rm{IO}}=g^8(N_c^2-1)}\nonumber\\
&&\times\bigg\{ N_c \bigg(|\cm_6(\Q{1},\qb{6};;\qbi{3},\qpi{4};;\qp{5},\Qb{2})|^2 + |\cm_6(\Q{1},\qpi{4};;\qbi{3},\Qb{2};;\qp{5},\qb{6})|^2  \bigg) \nonumber\\
&&\:\:\:\:\:\:\:\:\:\:\:- 2\re(\cm_6(\Q{1},\qpi{4};;\qbi{3},\Qb{2};;\qp{5},\qb{6})\cm_6(\Q{1},\qi{4};;\qbi{3},\Qb{2};;\qp{5},\qpb{6})^{\dagger})\nonumber\\
&&\:\:\:\:\:\:\:\:\:\:\:- 2\re(\cm_6(\Q{1},\qpi{4};;\qbi{3},\qb{6};;\qp{5},\Qb{2})\cm_6(\Q{1},\qi{4};;\qbi{3},\qpb{6};;\qp{5},\Qb{2})^{\dagger})\nonumber\\
&&\:\:\:\:\:\:\:\:\:\:\:- 2\re(\cm_6(\Q{1},\qb{6};;\qbi{3},\Qb{2};;\qp{5},\qpi{4})\cm_6(\Q{1},\qpb{6};;\qbi{3},\Qb{2};;\qp{5},\qi{4})^{\dagger})\nonumber\\
&&\:\:\:\:\:\:\:\:\:\:\:- 2\re(\cm_6(\Q{1},\qb{6};;\qbi{3},\qpi{4};;\qp{5},\Qb{2})\cm_6(\Q{1},\qpb{6};;\qbi{3},\qi{4};;\qp{5},\Qb{2})^{\dagger})\nonumber\\
&&\:\:\:\:\:\:\:\:\:\:\:+ 2\re(\cm_6(\Q{1},\Qb{2};;\qbi{3},\qb{6};;\qp{5},\qpi{4})\cm_6(\Q{1},\Qb{2};;\qbi{3},\qi{4};;\qp{5},\qpb{6})^{\dagger})\nonumber\\
&&\:\:+\frac{1}{N_c}\bigg(|\cm_6(\Q{1},\qb{6};;\qbi{3},\Qb{2};;\qp{5},\qpi{4})|^2+|\cm_6(\Q{1},\qpi{4};;\qbi{3},\qb{6};;\qp{5},\Qb{2})|^2\nonumber\\
&&\:\:\:\:\:\:\:\:\:\:\: +|\cm_6(\Q{1},\Qb{2};;\qbi{3},\qpi{4};;\qp{5},\qb{6})|^2-3|\cm_6(\Q{1},\Qb{2};;\qbi{3},\qb{6};;\qp{5},\qpi{4})|^2\bigg)\nonumber\\
&&\:\:+\frac{1}{N_c^2}\bigg(6\re(\cm_6(\Q{1},\Qb{2};;\qbi{3},\qb{6};;\qp{5},\qpi{4})\cm_6(\Q{1},\Qb{2};;\qbi{3},\qi{4};;\qp{5},\qpb{6})^{\dagger})\nonumber\\
&&\:\:\:\:\:\:\:\:\:\:\: -2\re(\cm_6(\Q{1},\Qb{2};;\qbi{3},\qb{6};;\qp{5},\qpi{4})\cm_6(\Q{1},\Qb{2};;\qbi{3},\qpb{6};;\qp{5},\qi{4})^{\dagger})\nonumber\\
&&\:\:\:\:\:\:\:\:\:\:\: -2\re(\cm_6(\Q{1},\Qb{2};;\qbi{3},\qpi{4};;\qp{5},\qb{6})\cm_6(\Q{1},\Qb{2};;\qbi{3},\qi{4};;\qp{5},\qpb{6})^{\dagger})\bigg)\bigg\}.\label{eq:me2i1}
\eeqa

The amplitude squared for the identical-flavour-only contribution is singular in the initial-final collinear limits $\qpi{4}||\qp{5}$ and $\qbi{3}||\qb{6}$, in the double collinear limit $\qpi{4}||\qp{5} + \qbi{3}||\qb{6}$, as well as in the triple collinear limits $\qbi{3}||\qp{5}||\qb{6}$ and $\qpi{4}||\qp{5}||\qb{6}$. The interference terms in eq.(\ref{eq:me2i1}), which are interferences between partial amplitudes coming from $M_6^0(\Q{1},\Qb{2},\qbi{3},\allowbreak{} \qi{4},\qp{5},\qpb{6})$ and $M_6^0(\Q{1},\Qb{2},\qbi{3},\qpi{4},\qp{5},\qb{6})$, are only divergent in the triple collinear limits. The full subtraction term for this contribution is
\beqa
\lefteqn{\ds^S_{q\bar{q}\rightarrow Q\bar{Q}q\bar{q},{\rm{IO}}}=\norm_{LO} \left( \frac{\alpha_s}{2\pi} \right)^2 \frac{\bar{C}(\epsilon)^2}{C(\epsilon)^2} \dphi_4(p_1,p_2,p_5,p_6;p_3,p_4)}\nonumber\\
&&\times\bigg\{ N_c \bigg[ \frac{1}{2}E_3^0(\Q{1},\qp{5},\qpi{4}) \bigg( |\cm_5(\Q{(\wt{15})},\gli{\bar{4}},\qb{6};;\qbi{3},\Qb{2})|^2\nonumber\\
&&\:\:\:\:\:\:\:\:\:\:\:\:\:\:\:\:\:\:\:\:\:\:\:\:\:\:\:\:\:+ |\cm_5(\Q{(\wt{15})},\qb{6};;\qbi{3},\gli{\bar{4}},\Qb{2})|^2 \bigg) J_2^{(3)}(\wt{p_{15}},p_2,p_6)\nonumber\\
&&\:\:\:\:\:\:\:\:\:\:\: +\frac{1}{2}E_3^0(\Qb{2},\qp{5},\qpi{4}) \bigg( |\cm_5(\Q{1},\gli{\bar{4}},\qb{6};;\qbi{3},\Qb{(\wt{25})})|^2\nonumber\\
&&\:\:\:\:\:\:\:\:\:\:\:\:\:\:\:\:\:\:\:\:\:\:\:\:\:\:\:\:\:+ |\cm_5(\Q{1},\qb{6};;\qbi{3},\gli{\bar{4}},\Qb{(\wt{25})})|^2 \bigg) J_2^{(3)}(p_1,\wt{p_{25}},p_6)\nonumber\\
&&\:\:\:\:\:\:\:\:\:\:\: +\frac{1}{2}E_3^0(\Q{1},\qb{6},\qbi{3}) \bigg ( |\cm_5(\Q{(\wt{16})},\gli{\bar{3}},\qpi{4};;\qp{5},\Qb{2})|^2\nonumber\\
&&\:\:\:\:\:\:\:\:\:\:\:\:\:\:\:\:\:\:\:\:\:\:\:\:\:\:\:\:\:+|\cm_5(\Q{(\wt{16})},\qpi{4};;\qp{5},\gli{\bar{3}},\Qb{2})|^2 \bigg) J_2^{(3)}(\wt{p_{16}},p_2,p_5)\nonumber\\
&&\:\:\:\:\:\:\:\:\:\:\: +\frac{1}{2}E_3^0(\Qb{2},\qb{6},\qbi{3}) \bigg ( |\cm_5(\Q{1},\gli{\bar{3}},\qpi{4};;\qp{5},\Qb{(\wt{26})})|^2\nonumber\\
&&\:\:\:\:\:\:\:\:\:\:\:\:\:\:\:\:\:\:\:\:\:\:\:\:\:\:\:\:\:+|\cm_5(\Q{1},\qpi{4};;\qp{5},\gli{\bar{3}},\Qb{(\wt{26})})|^2 \bigg) J_2^{(3)}(p_1,\wt{p_{26}},p_5)\nonumber\\
&&\:\:\:\:\:\:\:\:\:\:\: +\bigg( B_4^0(\qb{6},\qp{5},\qpi{4},\qbi{3})-\frac{1}{2}E_3^0(\Q{1},\qp{5},\qpi{4})A_3^0(\qb{6},\gli{\bar{4}},\qbi{3})\nonumber\\
&&\:\:\:\:\:\:\:\:\:\:\:\:\:\:\:\:\:\:\:\:\:\:\:\:\:\:\:\:\:\:\:\:\:\:-\frac{1}{2}E_3^0(\Qb{2},\qp{5},\qpi{4})A_3^0(\qb{6},\gli{\bar{4}},\qbi{3})\bigg) |\cm_4(\Q{\tilde{1}},\Qb{\tilde{2}},\qbi{\bar{3}},\qi{\bar{4}})|^2 J_2^{(2)}(\tilde{p_1},\tilde{p_2})\nonumber\\
&&\:\:\:\:\:\:\:\:\:\:\: +\bigg( B_4^0(\qp{5},\qb{6},\qbi{3},\qpi{4})-\frac{1}{2}E_3^0(\Q{1},\qb{6},\qbi{3})A_3^0(\qp{5},\gli{\bar{3}},\qpi{4})\nonumber\\
&&\:\:\:\:\:\:\:\:\:\:\:\:\:\:\:\:\:\:\:\:\:\:\:\:\:\:\:\:\:\:\:\:\:\:-\frac{1}{2}E_3^0(\Qb{2},\qb{6},\qbi{3})A_3^0(\qp{5},\gli{\bar{3}},\qpi{4})\bigg) |\cm_4(\Q{\tilde{1}},\Qb{\tilde{2}},\qpbi{\bar{3}},\qpi{\bar{4}})|^2 J_2^{(2)}(\tilde{p_1},\tilde{p_2})\nonumber\\
&&\:\:\:\:\:\:\:\:\:\:\: -\frac{1}{2}E_3^0(\Q{1},\qp{5},\qpi{4})E_3^0(\Qb{2},\qb{6},\qbi{3})\bigg( |\cm_4(\Q{(\wt{15})},\gli{\bar{3}},\gli{\bar{4}},\Qb{(\wt{26})})|^2 \nonumber\\
&&\:\:\:\:\:\:\:\:\:\:\:\:\:\:\:\:\:\:\:\:\:\:\:\:\:\:\:\:\:\:\:\:\:\:\:\:\:\:\:\:\:\:\:\:+   |\cm_4(\Q{(\wt{15})},\gli{\bar{4}},\gli{\bar{3}},\Qb{(\wt{26})})|^2 \bigg)J_2^{(2)}(\tilde{p_{15}},\tilde{p_{26}})\nonumber\\
&&\:\:\:\:\:\:\:\:\:\:\: -\frac{1}{2}E_3^0(\Q{1},\qb{6},\qbi{3})E_3^0(\Qb{2},\qp{5},\qpi{4})\bigg( |\cm_4(\Q{(\wt{16})},\gli{\bar{3}},\gli{\bar{4}},\Qb{(\wt{25})})|^2\nonumber\\
&&\:\:\:\:\:\:\:\:\:\:\:\:\:\:\:\:\:\:\:\:\:\:\:\:\:\:\:\:\:\:\:\:\:\:\:\:\:\:\:\:\:\:\:\:+|\cm_4(\Q{(\wt{16})},\gli{\bar{4}},\gli{\bar{3}},\Qb{(\wt{25})})|^2 \bigg) J_2^{(2)}(\tilde{p_{16}},\tilde{p_{25}})\bigg]\nonumber\\
&&\:\:\:\:\:\:\:\:\:\:\:+2C_4^0(\qbi{3},\q{5},\qb{6},\qi{4})|\cm_4(\Q{\tilde{1}},\Qb{\tilde{2}},\qbi{\bar{3}},\qi{\bar{4}})|^2 J_2^{(2)}(\tilde{p_1},\tilde{p_2})\nonumber\\
&&\:\:\:\:\:\:\:\:\:\:\:+2C_4^0(\qi{4},\qb{6},\q{5},\qbi{3})|\cm_4(\Q{\tilde{1}},\Qb{\tilde{2}},\qbi{\bar{3}},\qi{\bar{4}})|^2 J_2^{(2)}(\tilde{p_1},\tilde{p_2})\nonumber\\
&&\:\:+\frac{1}{N_c}\bigg[ \frac{1}{2}E_3^0(\Q{1},\qp{5},\qpi{4}) \bigg( |\cm_5(\Q{(\wt{15})},\gli{\bar{4}},\Qb{2};;\qbi{3},\qb{6})|^2\nonumber\\
&&\:\:\:\:\:\:\:\:\:\:\:\:\:\:\:\:\:\:\:\:\:\:\:\:\:\:\:\:\: +  |\cm_5(\Q{(\wt{15})},\Qb{2};;\qbi{3},\gli{\bar{4}},\qb{6})|^2\nonumber\\
&&\:\:\:\:\:\:\:\:\:\:\:\:\:\:\:\:\:\:\:\:\:\:\:\:\:\:\:\:\:  -2 |\cm_5(\Q{(\wt{15})},\Qb{2},\qbi{3},\qb{6},\phin{\bar{4}})|^2 \bigg) J_2^{(3)}(\wt{p_{15}},p_2,p_6)\nonumber\\
&&\:\:\:\:\:\:\:\:\:\:\: +\frac{1}{2}E_3^0(\Qb{2},\qp{5},\qpi{4}) \bigg( |\cm_5(\Q{1},\gli{\bar{4}},\Qb{(\wt{25})};;\qbi{3},\qb{6})|^2\nonumber\\
&&\:\:\:\:\:\:\:\:\:\:\:\:\:\:\:\:\:\:\:\:\:\:\:\:\:\:\:\:\:+|\cm_5(\Q{1},\Qb{(\wt{25})};;\qbi{3},\gli{\bar{4}},\qb{6})|^2\nonumber\\
&&\:\:\:\:\:\:\:\:\:\:\:\:\:\:\:\:\:\:\:\:\:\:\:\:\:\:\:\:\:-2|\cm_5(\Q{1},\Qb{(\wt{25})},\qbi{3},\qb{6},\phin{\bar{4}})|^2 \bigg) J_2^{(3)}(p_1,\wt{p_{25}},p_6)\nonumber\\
&&\:\:\:\:\:\:\:\:\:\:\: +\frac{1}{2}E_3^0(\Q{1},\qb{6},\qbi{3}) \bigg ( |\cm_5(\Q{(\wt{16})},\gli{\bar{3}},\Qb{2};;\qp{5},\qpi{4})|^2\nonumber\\
&&\:\:\:\:\:\:\:\:\:\:\:\:\:\:\:\:\:\:\:\:\:\:\:\:\:\:\:\:\:+ |\cm_5(\Q{(\wt{16})},\Qb{2};;\qp{5},\gli{\bar{3}},\qpi{4})|^2\nonumber\\
&&\:\:\:\:\:\:\:\:\:\:\:\:\:\:\:\:\:\:\:\:\:\:\:\:\:\:\:\:\: -2 |\cm_5(\Q{(\wt{16})},\Qb{2},\qp{5},\qpi{4},\phin{\bar{3}})|^2 \bigg) J_2^{(3)}(\wt{p_{16}},p_2,p_5)\nonumber\\
&&\:\:\:\:\:\:\:\:\:\:\: +\frac{1}{2}E_3^0(\Qb{2},\qb{6},\qbi{3}) \bigg ( |\cm_5(\Q{1},\gli{\bar{3}},\Qb{(\wt{26})};;\qp{5},\qpi{4})|^2\nonumber\\
&&\:\:\:\:\:\:\:\:\:\:\:\:\:\:\:\:\:\:\:\:\:\:\:\:\:\:\:\:\:+ |\cm_5(\Q{1},\Qb{(\wt{26})};;\qp{5},\gli{\bar{3}},\qpi{4})|^2\nonumber\\
&&\:\:\:\:\:\:\:\:\:\:\:\:\:\:\:\:\:\:\:\:\:\:\:\:\:\:\:\:\:-2|\cm_5(\Q{1},\Qb{(\wt{26})},\qp{5},\qpi{4},\phin{\bar{3}})|^2 \bigg) J_2^{(3)}(p_1,\wt{p_{26}},p_5)\nonumber\\
&&\:\:\:\:\:\:\:\:\:\:\: -\bigg( B_4^0(\qb{6},\qp{5},\qpi{4},\qbi{3})-\frac{1}{2}E_3^0(\Q{1},\qp{5},\qpi{4})A_3^0(\qb{6},\gli{\bar{4}},\qbi{3})\nonumber\\
&&\:\:\:\:\:\:\:\:\:\:\:\:\:\:\:\:\:\:\:\:\:\:\:\:\:\:\:\:\:\:\:\:\:\:-\frac{1}{2}E_3^0(\Qb{2},\qp{5},\qpi{4})A_3^0(\qb{6},\gli{\bar{4}},\qbi{3})\bigg) |\cm_4(\Q{\tilde{1}},\Qb{\tilde{2}},\qbi{\bar{3}},\qi{\bar{4}})|^2 J_2^{(2)}(\tilde{p_1},\tilde{p_2})\nonumber\\
&&\:\:\:\:\:\:\:\:\:\:\: -\bigg( B_4^0(\qp{5},\qb{6},\qbi{3},\qpi{4})-\frac{1}{2}E_3^0(\Q{1},\qb{6},\qbi{3})A_3^0(\qp{5},\gli{\bar{3}},\qpi{4})\nonumber\\
&&\:\:\:\:\:\:\:\:\:\:\:\:\:\:\:\:\:\:\:\:\:\:\:\:\:\:\:\:\:\:\:\:\:\:-\frac{1}{2}E_3^0(\Qb{2},\qb{6},\qbi{3})A_3^0(\qp{5},\gli{\bar{3}},\qbi{4})\bigg) |\cm_4(\Q{\tilde{1}},\Qb{\tilde{2}},\qpbi{\bar{3}},\qpi{\bar{4}})|^2 J_2^{(2)}(\tilde{p_1},\tilde{p_2})\nonumber\\
&&\:\:\:\:\:\:\:\:\:\:\: +\frac{1}{2}E_3^0(\Q{1},\qp{5},\qpi{4})E_3^0(\Qb{2},\qb{6},\qbi{3}) |\cm_4(\Q{(\wt{15})},\phin{\bar{3}},\phin{\bar{4}},\Qb{(\wt{26})})|^2 J_2^{(2)}(\tilde{p_{15}},\tilde{p_{26}})\nonumber\\
&&\:\:\:\:\:\:\:\:\:\:\: +\frac{1}{2}E_3^0(\Q{1},\qb{6},\qbi{3})E_3^0(\Qb{2},\qp{5},\qpi{4})|\cm_4(\Q{(\wt{16})},\phin{\bar{3}},\phin{\bar{4}},\Qb{(\wt{25})})|^2 J_2^{(2)}(\tilde{p_{16}},\tilde{p_{25}})\bigg]\nonumber\\
&&\:\:-\frac{1}{N_c^2}\bigg[2C_4^0(\qbi{3},\q{5},\qb{6},\qi{4})|\cm_4(\Q{\tilde{1}},\Qb{\tilde{2}},\qbi{\bar{3}},\qi{\bar{4}})|^2 J_2^{(2)}(\tilde{p_1},\tilde{p_2})\nonumber\\
&&\:\:\:\:\:\:\:\:\:\:\: +2C_4^0(\qi{4},\qb{6},\q{5},\qbi{3})|\cm_4(\Q{\tilde{1}},\Qb{\tilde{2}},\qbi{\bar{3}},\qi{\bar{4}})|^2 J_2^{(2)}(\tilde{p_1},\tilde{p_2})\bigg]\bigg\}.\label{eq.sub2}
\eeqa

The choice of one of the hard radiators in the $B_4^0$ antennae used to subtract the triple collinear limits is rather ambiguous. Any of the three particles that are not involved in the limit can be chosen as a radiator, since the only role it plays is to recoil and ensure overall momentum conservation. We decided to use always antennae with massless hard radiators.

\subsubsection{The $qq'\rightarrow Q\bar{Q}qq'$ process}
By squaring eq.(\ref{eq.meni}), using eq.(\ref{eq.deci}) to eliminate the interferences terms, and crossing $\qb{4}$ and $\qpb{6}$ to the initial state, the amplitude squared for the process $qq'\rightarrow Q\bar{Q}qq'$ in terms of partial amplitudes squared is obtained. It reads
\beqa
\lefteqn{|M_6^0(\Q{1},\Qb{2},\q{3},\qi{4},\qp{5},\qpi{6})|^2=g^8(N_c^2-1)}\nonumber\\
&&\bigg\{ N_c \bigg( |\cm_6(\Q{1},\qi{4};;\q{3},\qpi{6};;\qp{5},\Qb{2})|^2+|\cm_6(\Q{1},\qpi{6};;\q{3},\Qb{2};;\qp{5},\qi{4})|^2\bigg) \nonumber\\
&&+\frac{1}{N_c}\bigg( |\cm_6(\Q{1},\qi{4};;\q{3},\Qb{2};;\qp{5},\qpi{6})|^2+|\cm_6(\Q{1},\qpi{6};;\q{3},\qi{4};;\qp{5},\Qb{2})|^2\nonumber\\
&&\:\:\:\:\:\:\:\:\: +|\cm_6(\Q{1},\Qb{2};;\q{3},\qpi{6};;\qp{5},\qi{4})|^2-3|\cm_6(\Q{1},\Qb{2};;\q{3},\qi{4};;\qp{5},\qpi{6})|^2\bigg)\bigg\}.\label{eq:me2ni}
\eeqa
The singularities of this amplitude are: the $\q{3}||\qi{4}$ and $\qp{5}||\qpi{6}$ collinear limits, the $\q{3}||\qi{4}+\qp{5}||\qpi{6}$ double collinear limit, and the $\q{3}||\qi{4}||\qp{5}$ and $\q{3}||\qp{5}||\qpi{6}$ triple collinear limits. To subtract these divergencies from the partonic differential cross section we use the following subtraction term
\beqa
\lefteqn{\ds^S_{qq'\rightarrow Q\bar{Q}qq'}=\norm_{LO} \left( \frac{\alpha_s}{2\pi} \right)^2 \frac{\bar{C}(\epsilon)^2}{C(\epsilon)^2}\dphi_4(p_1,p_2,p_3,p_5;p_4,p_6)}\nonumber\\
&&\times\bigg\{ N_c \bigg[ \frac{1}{2}E_3^0(\Q{1},\q{3},\qi{4}) \bigg ( |\cm_5(\Q{(\wt{13})},\gli{\bar{4}},\qpi{6};;\qp{5},\Qb{2})|^2\nonumber\\
&&\:\:\:\:\:\:\:\:\:\:\:\:\:\:\:\:\:\:\:\:\:\:\:\:\:\:\:\:\:+|\cm_5(\Q{(\wt{13})},\qpi{6};;\qp{5},\gli{\bar{4}},\Qb{2})|^2 \bigg) J_2^{(3)}(\wt{p_{13}},p_2,p_5)\nonumber\\
&&\:\:\:\:\:\:\:\:\:\:\: +\frac{1}{2}E_3^0(\Qb{2},\qb{3},\qbi{4}) \bigg ( |\cm_5(\Q{1},\gli{\bar{4}},\qpi{6};;\qp{5},\Qb{(\wt{23})})|^2\nonumber\\
&&\:\:\:\:\:\:\:\:\:\:\:\:\:\:\:\:\:\:\:\:\:\:\:\:\:\:\:\:\:+|\cm_5(\Q{1},\qpi{6};;\qp{5},\gli{\bar{4}},\Qb{(\wt{23})})|^2 \bigg) J_2^{(3)}(p_1,\wt{p_{23}},p_5)\nonumber\\
&&\:\:\:\:\:\:\:\:\:\:\: +\frac{1}{2}E_3^0(\Q{1},\qp{5},\qpi{6}) \bigg( |\cm_5(\Q{(\wt{15})},\gli{\bar{6}},\qi{4};;\q{3},\Qb{2})|^2\nonumber\\
&&\:\:\:\:\:\:\:\:\:\:\:\:\:\:\:\:\:\:\:\:\:\:\:\:\:\:\:\:\:+ |\cm_5(\Q{(\wt{15})},\qi{4};;\q{3},\gli{\bar{6}},\Qb{2})|^2 \bigg) J_2^{(3)}(\wt{p_{15}},p_2,p_3)\nonumber\\
&&\:\:\:\:\:\:\:\:\:\:\: +\frac{1}{2}E_3^0(\Qb{2},\qp{5},\qpi{6}) \bigg( |\cm_5(\Q{1},\gli{\bar{6}},\qi{4};;\q{3},\Qb{(\wt{25})})|^2\nonumber\\
&&\:\:\:\:\:\:\:\:\:\:\:\:\:\:\:\:\:\:\:\:\:\:\:\:\:\:\:\:\:+ |\cm_5(\Q{1},\qi{4};;\q{3},\gli{\bar{6}},\Qb{(\wt{25})})|^2 \bigg) J_2^{(3)}(p_1,\wt{p_{25}},p_3)\nonumber\\
&&\:\:\:\:\:\:\:\:\:\:\: +\bigg( B_4^0(\q{3},\qpi{6},\qp{5},\qi{4})-\frac{1}{2}E_3^0(\Q{1},\qp{5},\qpi{6})A_3^0(\q{3},\gli{\bar{6}},\qi{4})\nonumber\\
&&\:\:\:\:\:\:\:\:\:\:\:\:\:\:\:\:\:\:\:\:\:\:\:\:\:\:\:\:\:\:\:\:\:\:-\frac{1}{2}E_3^0(\Qb{2},\qp{5},\qpi{6})A_3^0(\q{3},\gli{\bar{6}},\qi{4})\bigg) |\cm_4(\Q{\tilde{1}},\Qb{\tilde{2}},\qbi{\bar{6}},\qi{\bar{4}})|^2 J_2^{(2)}(\tilde{p_1},\tilde{p_2})\nonumber\\
&&\:\:\:\:\:\:\:\:\:\:\: +\bigg( B_4^0(\qp{5},\qi{4},\q{3},\qpi{6})-\frac{1}{2}E_3^0(\Q{1},\q{3},\qi{4})A_3^0(\qp{5},\gli{\bar{4}},\qpi{6})\nonumber\\
&&\:\:\:\:\:\:\:\:\:\:\:\:\:\:\:\:\:\:\:\:\:\:\:\:\:\:\:\:\:\:\:\:\:\:-\frac{1}{2}E_3^0(\Qb{2},\q{3},\qi{4})A_3^0(\qp{5},\gli{\bar{4}},\qpi{6})\bigg) |\cm_4(\Q{\tilde{1}},\Qb{\tilde{2}},\qpbi{\bar{4}},\qpi{\bar{6}})|^2 J_2^{(2)}(\tilde{p_1},\tilde{p_2})\nonumber\\
&&\:\:\:\:\:\:\:\:\:\:\: -\frac{1}{2}E_3^0(\Q{1},\q{3},\qi{4})E_3^0(\Qb{2},\qp{5},\qpi{6})\bigg( |\cm_4(\Q{(\wt{13})},\gli{\bar{4}},\gli{\bar{6}},\Qb{(\wt{25})})|^2\nonumber\\
&&\:\:\:\:\:\:\:\:\:\:\:\:\:\:\:\:\:\:\:\:\:\:\:\:\:\:\:\:\:\:\:\:\:\:\:\:\:\:\:\:\:\:\:\:+|\cm_4(\Q{(\wt{13})},\gli{\bar{6}},\gli{\bar{4}},\Qb{(\wt{25})})|^2 \bigg) J_2^{(2)}(\tilde{p_{13}},\tilde{p_{25}})\nonumber\\
&&\:\:\:\:\:\:\:\:\:\:\: -\frac{1}{2}E_3^0(\Q{1},\qp{5},\qpi{6})E_3^0(\Qb{2},\q{3},\qi{4})\bigg( |\cm_4(\Q{(\wt{15})},\gli{\bar{4}},\gli{\bar{6}},\Qb{(\wt{23})})|^2 \nonumber\\
&&\:\:\:\:\:\:\:\:\:\:\:\:\:\:\:\:\:\:\:\:\:\:\:\:\:\:\:\:\:\:\:\:\:\:\:\:\:\:\:\:\:\:\:\:+   |\cm_4(\Q{(\wt{15})},\gli{\bar{6}},\gli{\bar{4}},\Qb{(\wt{23})})|^2 \bigg)J_2^{(2)}(\tilde{p_{15}},\tilde{p_{23}})\bigg]\nonumber\\
&&\:\:+\frac{1}{N_c}\bigg[\frac{1}{2}E_3^0(\Q{1},\q{3},\qi{4}) \bigg ( |\cm_5(\Q{(\wt{13})},\gli{\bar{4}},\Qb{2};;\qp{5},\qpi{6})|^2\nonumber\\
&&\:\:\:\:\:\:\:\:\:\:\:\:\:\:\:\:\:\:\:\:\:\:\:\:\:\:\:\:\:+ |\cm_5(\Q{(\wt{13})},\Qb{2};;\qp{5},\gli{\bar{4}},\qpi{6})|^2\nonumber\\
&&\:\:\:\:\:\:\:\:\:\:\:\:\:\:\:\:\:\:\:\:\:\:\:\:\:\:\:\:\: -2 |\cm_5(\Q{(\wt{13})},\Qb{2},\qp{5},\qpi{6},\phin{\bar{4}})|^2 \bigg) J_2^{(3)}(\wt{p_{16}},p_2,p_5)\nonumber\\
&&\:\:\:\:\:\:\:\:\:\:\: +\frac{1}{2}E_3^0(\Qb{2},\q{3},\qi{4}) \bigg ( |\cm_5(\Q{1},\gli{\bar{4}},\Qb{(\wt{23})};;\qp{5},\qpi{6})|^2\nonumber\\
&&\:\:\:\:\:\:\:\:\:\:\:\:\:\:\:\:\:\:\:\:\:\:\:\:\:\:\:\:\:+ |\cm_5(\Q{1},\Qb{(\wt{23})};;\qp{5},\gli{\bar{4}},\qpi{6})|^2\nonumber\\
&&\:\:\:\:\:\:\:\:\:\:\:\:\:\:\:\:\:\:\:\:\:\:\:\:\:\:\:\:\:-2|\cm_5(\Q{1},\Qb{(\wt{23})},\qp{5},\qpi{6},\phin{\bar{4}})|^2 \bigg) J_2^{(3)}(p_1,\wt{p_{23}},p_5)\nonumber\\
&&\:\:\:\:\:\:\:\:\:\:\: +\frac{1}{2}E_3^0(\Q{1},\qp{5},\qpi{6}) \bigg( |\cm_5(\Q{(\wt{15})},\gli{\bar{6}},\Qb{2};;\q{3},\qi{4})|^2\nonumber\\
&&\:\:\:\:\:\:\:\:\:\:\:\:\:\:\:\:\:\:\:\:\:\:\:\:\:\:\:\:\:  +  |\cm_5(\Q{(\wt{15})},\Qb{2};;\q{3},\gli{\bar{6}},\qb{4})|^2\nonumber\\
&&\:\:\:\:\:\:\:\:\:\:\:\:\:\:\:\:\:\:\:\:\:\:\:\:\:\:\:\:\:-2 |\cm_5(\Q{(\wt{15})},\Qb{2},\q{3},\qi{4},\phin{\bar{6}})|^2 \bigg) J_2^{(3)}(\wt{p_{15}},p_2,p_3)\nonumber\\
&&\:\:\:\:\:\:\:\:\:\:\: +\frac{1}{2}E_3^0(\Qb{2},\qp{5},\qpi{6}) \bigg( |\cm_5(\Q{1},\gli{\bar{6}},\Qb{(\wt{25})};;\q{3},\qi{4})|^2\nonumber\\
&&\:\:\:\:\:\:\:\:\:\:\:\:\:\:\:\:\:\:\:\:\:\:\:\:\:\:\:\:\:+|\cm_5(\Q{1},\Qb{(\wt{25})};;\q{3},\gli{\bar{6}},\qi{4})|^2\nonumber\\
&&\:\:\:\:\:\:\:\:\:\:\:\:\:\:\:\:\:\:\:\:\:\:\:\:\:\:\:\:\:-2|\cm_5(\Q{1},\Qb{(\wt{25})},\q{3},\qi{4},\phin{\bar{6}})|^2 \bigg) J_2^{(3)}(p_1,\wt{p_{25}},p_3)\nonumber\\
&&\:\:\:\:\:\:\:\:\:\:\: -\bigg( B_4^0(\q{3},\qpi{6},\qp{5},\qi{4})-\frac{1}{2}E_3^0(\Q{1},\qp{5},\qpi{6})A_3^0(\q{3},\gli{\bar{6}},\qi{4})\nonumber\\
&&\:\:\:\:\:\:\:\:\:\:\:\:\:\:\:\:\:\:\:\:\:\:\:\:\:\:\:\:\:\:\:\:\:\:-\frac{1}{2}E_3^0(\Qb{2},\qp{5},\qpi{6})A_3^0(\q{3},\gli{\bar{6}},\qi{4})\bigg) |\cm_4(\Q{\tilde{1}},\Qb{\tilde{2}},\qbi{\bar{6}},\qi{\bar{4}})|^2 J_2^{(2)}(\tilde{p_1},\tilde{p_2})\nonumber\\
&&\:\:\:\:\:\:\:\:\:\:\: -\bigg( B_4^0(\qp{5},\qi{4},\q{3},\qpi{6})-\frac{1}{2}E_3^0(\Q{1},\q{3},\qi{4})A_3^0(\qp{5},\gli{\bar{4}},\qpi{6})\nonumber\\
&&\:\:\:\:\:\:\:\:\:\:\:\:\:\:\:\:\:\:\:\:\:\:\:\:\:\:\:\:\:\:\:\:\:\:-\frac{1}{2}E_3^0(\Qb{2},\q{3},\qi{4})A_3^0(\qp{5},\gli{\bar{4}},\qpi{6})\bigg) |\cm_4(\Q{\tilde{1}},\Qb{\tilde{2}},\qpbi{\bar{4}},\qpi{\bar{6}})|^2 J_2^{(2)}(\tilde{p_1},\tilde{p_2})\nonumber\\
&&\:\:\:\:\:\:\:\:\:\:\: +\frac{1}{2}E_3^0(\Q{1},\q{3},\qi{4})E_3^0(\Qb{2},\qp{5},\qpi{6})|\cm_4(\Q{(\wt{13})},\phin{\bar{4}},\phin{\bar{6}},\Qb{(\wt{25})})|^2 J_2^{(2)}(\tilde{p_{13}},\tilde{p_{25}})\nonumber\\
&&\:\:\:\:\:\:\:\:\:\:\: +\frac{1}{2}E_3^0(\Q{1},\qp{5},\qpi{6})E_3^0(\Qb{2},\q{3},\qi{4}) |\cm_4(\Q{(\wt{15})},\phin{\bar{4}},\phin{\bar{6}},\Qb{(\wt{23})})|^2 J_2^{(2)}(\tilde{p_{15}},\tilde{p_{23}})\bigg]\bigg\}.\nonumber\\ \label{eq.sub3}
\eeqa
As for the previous process only massless four-parton antennae whose integrated form is available \cite{antennannlo,Gionata,radja} have been used.

\subsubsection{The $qq\rightarrow Q\bar{Q}qq$ process: identical quark flavour contributions}
As we did for the $q\bar{q}\rightarrow Q\bar{Q}q\bar{q}$ process, we split the amplitude corresponding to the $qq\rightarrow Q\bar{Q}qq$ process into a piece that contains the same diagrams as the non-identical quark flavour case, and another piece which consists of diagrams which are specific to the identical flavour case. After squaring eq.(\ref{eq.mei}) we obtain 
\beqa
&&\hspace{-3mm}|M_6^0(\Q{1},\Qb{2},\q{3},\qi{4},\q{5},\qi{6})|^2=\frac{1}{2}\bigg(|M_6^0(\Q{1},\Qb{2},\q{3},\qi{4},\qp{5},\qpi{6})|^2+|M_6^0(\Q{1},\Qb{2},\q{3},\qpi{4},\qp{5},\qi{6})|^2\nonumber\\
&&\hspace{5mm}-2\re \left(M_6^0(\Q{1},\Qb{2},\q{3},\qi{4},\qp{5},\qpi{6}))M_6^0(\Q{1},\Qb{2},\q{3},\qpi{4},\qp{5},\qi{6})\dagger \right)\bigg).\label{eq.qqbarqqbarid}
\eeqa
The factor $1/2$ is a symmetry factor that accounts for the presence of identical particles in the final state. The interference term comes with a minus sign since the amplitudes differ by the interchange of the initial state quark labels. The identical-flavour-only amplitude squared can be written in terms of colour-ordered amplitudes as:
\beqa
\lefteqn{|M_6^0(\Q{1},\Qb{2},\q{3},\qi{4},\q{5},\qi{6})|^2_{\rm{IO}}=g^8(N_c^2-1)\frac{1}{2}}\nonumber\\
&&\times\bigg\{ N_c \bigg(|\cm_6(\Q{1},\qi{6};;\q{3},\qpi{4};;\qp{5},\Qb{2})|^2 + |\cm_6(\Q{1},\qpi{4};;\q{3},\Qb{2};;\qp{5},\qi{6})|^2  \bigg) \nonumber\\
&&\:\:\:\:\:\:\:\:\:\:\:+ 2\re(\cm_6(\Q{1},\qpi{4};;\q{3},\Qb{2};;\qp{5},\qi{6})\cm_6(\Q{1},\qi{4};;\q{3},\Qb{2};;\qp{5},\qpi{6})^{\dagger})\nonumber\\
&&\:\:\:\:\:\:\:\:\:\:\:+ 2\re(\cm_6(\Q{1},\qpi{4};;\q{3},\qi{6};;\qp{5},\Qb{2})\cm_6(\Q{1},\qi{4};;\q{3},\qpi{6};;\qp{5},\Qb{2})^{\dagger})\nonumber\\
&&\:\:\:\:\:\:\:\:\:\:\:+ 2\re(\cm_6(\Q{1},\qi{6};;\q{3},\Qb{2};;\qp{5},\qpi{4})\cm_6(\Q{1},\qpi{6};;\q{3},\Qb{2};;\qp{5},\qi{4})^{\dagger})\nonumber\\
&&\:\:\:\:\:\:\:\:\:\:\:+ 2\re(\cm_6(\Q{1},\qi{6};;\q{3},\qpi{4};;\qp{5},\Qb{2})\cm_6(\Q{1},\qpi{6};;\q{3},\qi{4};;\qp{5},\Qb{2})^{\dagger})\nonumber\\
&&\:\:\:\:\:\:\:\:\:\:\:- 2\re(\cm_6(\Q{1},\Qb{2};;\q{3},\qi{6};;\qp{5},\qpi{4})\cm_6(\Q{1},\Qb{2};;\q{3},\qi{4};;\qp{5},\qpi{6})^{\dagger})\nonumber\\
&&\:\:+\frac{1}{N_c}\bigg(|\cm_6(\Q{1},\qi{6};;\q{3},\Qb{2};;\qp{5},\qpi{4})|^2+|\cm_6(\Q{1},\qpi{4};;\q{3},\qi{6};;\qp{5},\Qb{2})|^2\nonumber\\
&&\:\:\:\:\:\:\:\:\:\:\: +|\cm_6(\Q{1},\Qb{2};;\q{3},\qpi{4};;\qp{5},\qi{6})|^2-3|\cm_6(\Q{1},\Qb{2};;\q{3},\qi{6};;\qp{5},\qpi{4})|^2\bigg)\nonumber\\
&&\:\:-\frac{1}{N_c^2}\bigg(6\re(\cm_6(\Q{1},\Qb{2};;\q{3},\qi{6};;\qp{5},\qpi{4})\cm_6(\Q{1},\Qb{2};;\q{3},\qi{4};;\qp{5},\qpi{6})^{\dagger})\nonumber\\
&&\:\:\:\:\:\:\:\:\:\:\: -2\re(\cm_6(\Q{1},\Qb{2};;\q{3},\qi{6};;\qp{5},\qpi{4})\cm_6(\Q{1},\Qb{2};;\q{3},\qpi{6};;\qp{5},\qi{4})^{\dagger})\nonumber\\
&&\:\:\:\:\:\:\:\:\:\:\: -2\re(\cm_6(\Q{1},\Qb{2};;\q{3},\qpi{4};;\qp{5},\qi{6})\cm_6(\Q{1},\Qb{2};;\q{3},\qi{4};;\qp{5},\qpi{6})^{\dagger})\bigg)\bigg\}.\label{eq:me2i}
\eeqa
Apart from the interference terms, this amplitude squared is the same as that of the non identical flavour quark case with $p_4$ and $p_6$ interchanged. The interferences are singular in the $\q{3}||\qi{4}||\q{5}$ and $\q{3}||\q{5}||\qi{6}$ triple collinear limits, and these singularities are subtracted with C-type four-parton antennae. Thus, we can write our subtraction term for the identical-flavour-only case as
\beqa
\ds^S_{qq\rightarrow Q\bar{Q}qq,{\rm{IO}}}=\frac{1}{2}\ds^S_{qq'\rightarrow Q\bar{Q}qq'}(p_4 \leftrightarrow p_6) -\frac{1}{2} \norm_{LO} \left( \frac{\alpha_s}{2\pi} \right)^2 \frac{\bar{C}(\epsilon)^2}{C(\epsilon)^2} \dphi_4(p_1,p_2,p_3,p_5;p_4,p_6)\nonumber\\
\hspace{5mm}\times\left( 1 - \frac{1}{N_c^2} \right)\bigg( 2C_4^0(\qi{4},\qi{6},\q{3},\q{5}) |\cm_4(\Q{\tilde{1}},\Qb{\tilde{2}},\qbi{\bar{6}},\qi{\bar{4}})|^2 J_2^{(2)}(\tilde{p_1},\tilde{p_2})\nonumber\\
+2C_4^0(\qi{6},\qi{4},\q{3},\q{5}) |\cm_4(\Q{\tilde{1}},\Qb{\tilde{2}},\qbi{\bar{4}},\qi{\bar{6}})|^2 J_2^{(2)}(\tilde{p_1},\tilde{p_2})\bigg).\nonumber\\ \label{eq.sub4}
\eeqa

\subsection{Infrared structure of the subtraction terms}\label{sec.cancelation}
In all partonic processes considered above, examining the terms which capture the single unresolved limits in the respective subtraction terms (which belong to the subtraction term ${\rm d}\hat{\sigma}^{(S,a)}_{NNLO}$ given in eqs.(\ref{eq.sub2aff}-\ref{eq.sub2aii})) which are made of three-parton antennae multiplied by five-parton reduced matrix elements, one can determine how the infrared poles that arise in these terms upon integration over the appropriate one-particle antenna phase spaces will have to cancel against the poles present in the real-virtual five-parton contributions. 

Looking at each process individually, we find that only in the subtraction term related to the process $ q \bar{q} \to Q \bar{Q} q'\bar{q}'$ the poles generated after integration over the antenna phase space will have to cancel against explicit poles present in the one-loop contributions at the five-parton level. These poles have to cancel against explicit poles related to the process $q \bar{q}\to Q \bar{Q} g$ at one-loop. Indeed, in this case the subtraction term at the six-parton level involves the same process (i.e. $q \bar{q }\to Q \bar{Q} g$ at tree-level) multiplied by a massive E-type three-parton antenna. This antenna captures the final-final quark-antiquark collinear features and its integrated form is related to the massive ${\bf I}^{(1)}$-type operator ${\bf I}^{(1)}_{Qg,N_{F}}$ defined in \cite{us} and quoted in eq.(\ref{eq.integratedE03ff}). These statements are valid both for the leading and subleading colour contributions.

For all other processes, the $\epsilon$ poles in the integrated ${\rm d}\sigma^{S,a}_{NNLO}$ subtraction terms will have to cancel against poles present in mass-factorisation counterterms. This can be seen from the fact that the ${\rm d}\sigma^{S,a}_{NNLO}$ pieces of those subtraction terms only contain initial-final E-type three-parton antennae. These capture the massless initial-final single collinear features of the real radiation matrix elements squared and the only $\epsilon$ pole present in their integrated form is proportional to the splitting kernel $p_{qg}^{0}$, as it can be seen from eq.(\ref{eq.integratedE03if}).

\section{Results}\label{sec.results}
For all processes considered in section \ref{sec.subterms} we have tested how well the subtraction terms approximate the double real matrix elements in all their single and double unresolved regions of phase space, so that their difference can be numerically evaluated in four dimensions. In this section we describe how this testing procedure is performed.

To verify how well the subtraction terms approximate the double real contribution, we have used RAMBO \cite{Rambo} to generate phase space points in the vicinity of the singular regions and computed the ratio
\beq
R=\frac{\ds^{RR}_{NNLO}}{\ds^S_{NNLO}}
\eeq
for each of these points. As before, $\ds^{RR}_{NNLO}$ stands for the double real radiation corrections while $\ds^S_{NNLO}$ is the corresponding subtraction term. In each unresolved limit we defined a control variable $x$ that allowed us to vary the proximity of the phase space points to the singularity. For the difference $\ds^{RR}_{NNLO}-\ds^S_{NNLO}$ to be finite and numerically integrable in four dimensions, the ratio $R$ should approach unity as we get close to any singularity. The phase space points were generated with a fixed centre-of-mass energy of $\sqrt{s}=1000$ GeV, the heavy fermions were given a mass of $174.3$ GeV, and the two hard jets were required to have $p_T > 50$ GeV.

In the following, we shall present our results for one particular partonic process, namely for the $q\bar{q}\rightarrow Q\bar{Q}q'\bar{q}'$ process which we separate into identical and non-identical quark contributions.

\subsection{The $q\bar{q}\rightarrow Q\bar{Q}q'\bar{q}'$ process}
We start by presenting the behaviour of the subtraction terms in double unresolved limits, before describing how those approximate the real matrix elements in single unresolved regions of phase spaces.

\subsubsection{Double soft limit}
For the  process $q\bar{q}\rightarrow Q\bar{Q}q'\bar{q}'$, the double soft phase space configurations are characterised by the $Q\bar{Q}$ pair taking nearly the full center of mass energy of the event $s$, leaving the massless final state $q\bar{q}$ pair with almost zero energy as depicted in fig.1(a).
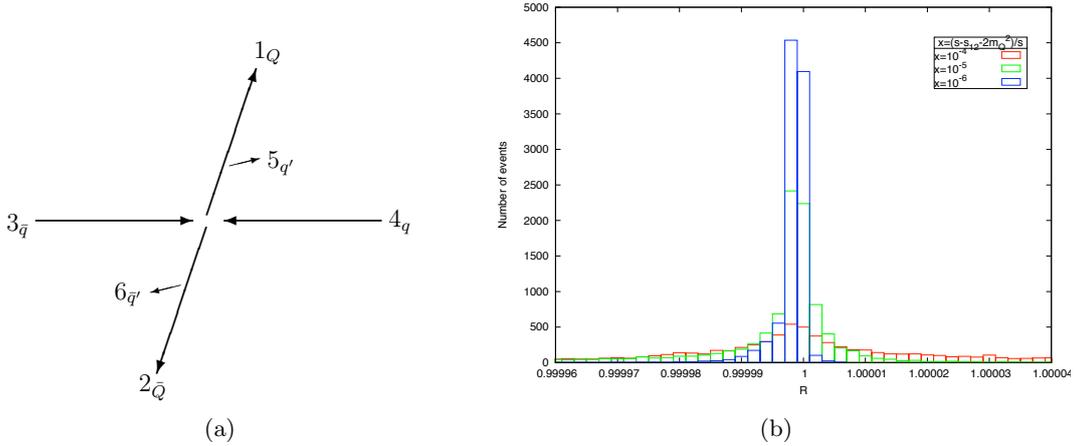
\begin{figure}[ht]
\centering
\subfigure[]{
\scalebox{0.8}{
\setlength{\unitlength}{4144sp}

\begingroup\makeatletter\ifx\SetFigFont\undefined
\gdef\SetFigFont#1#2#3#4#5{
  \reset@font\fontsize{#1}{#2pt}
  \fontfamily{#3}\fontseries{#4}\fontshape{#5}
  \selectfont}
\fi\endgroup
\begin{picture}(3296,2656)(1576,-2585)

\thicklines
{\put(3021,-1323){\vector(-1,-3){364.500}}
}
{\put(4321,-1276){\vector(-1, 0){1170}}
}
{\put(3026,-1235){\vector( 1, 3){364.500}}
}

\thinlines
{\put(2832,-1754){\vector(-4,-1){222.353}}
}
{\put(3196,-871){\vector( 4, 1){222.353}}

\put(3200,-61){\makebox(0,0)[lb]{\smash{{\SetFigFont{12}{14.4}{\rmdefault}{\mddefault}{\updefault}{$1_Q$}
}}}}
\put(2350,-2550){\makebox(0,0)[lb]{\smash{{\SetFigFont{12}{14.4}{\rmdefault}{\mddefault}{\updefault}{$2_{\bar{Q}}$}
}}}}
}
\put(1370,-1350){\makebox(0,0)[lb]{\smash{{\SetFigFont{12}{14.4}{\rmdefault}{\mddefault}{\updefault}{$3_{\bar{q}}$}
}}}}
\put(4200,-1321){\makebox(0,0)[lb]{\smash{{\SetFigFont{12}{14.4}{\rmdefault}{\mddefault}{\updefault}{$4_q$}
}}}}
\put(3300,-880){\makebox(0,0)[lb]{\smash{{\SetFigFont{12}{14.4}{\rmdefault}{\mddefault}{\updefault}{$5_{q'}$}
}}}}
\put(2170,-1861){\makebox(0,0)[lb]{\smash{{\SetFigFont{12}{14.4}{\rmdefault}{\mddefault}{\updefault}{$6_{\bar{q}'}$}
}}}}
\thicklines
{\put(1756,-1276){\vector( 1, 0){1170}}
}
\end{picture}
\label{fig.pic1}
}}
\subfigure[]{
\includegraphics[scale=0.6]{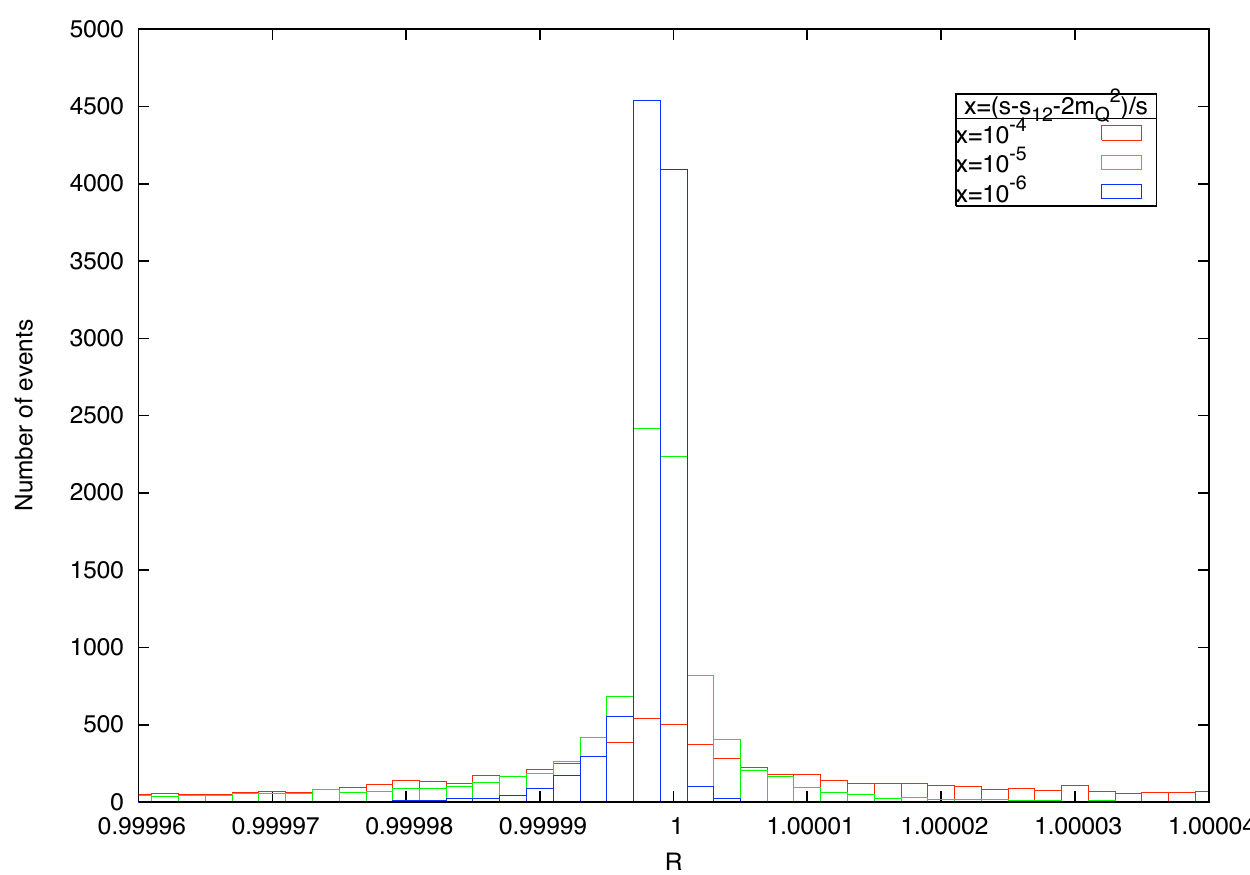}
\label{fig.ds}
}
\caption[]{\subref{fig.pic1} Ilustration of a double soft event. \subref{fig.ds} Distribution of R for 10000 double soft phase space points.}
\end{figure}
In fig.1(b) we show the ratio between the double real radiation matrix element and the subtraction term for three different values of $x=(s-s_{12}-2m_Q^2)/s$. It can be seen that as the soft $q\bar{q}$ pair takes a smaller share of the total energy, i.e. as $x$ becomes smaller, the peak of the distribution around $R=1$ is sharper. This is a sign that the approximation improves as the limit is approached.

\subsubsection{Triple collinear limits}
The other double unresolved configurations in which the amplitude for this process is singular are the triple collinear limits $\qbi{3}||\qp{5}||\qpb{6}$ and $\qi{4}||\qp{5}||\qpb{6}$. 
\begin{figure}[ht]
\centering
\subfigure[]{
\scalebox{0.8}{
\setlength{\unitlength}{4144sp}

\begingroup\makeatletter\ifx\SetFigFont\undefined
\gdef\SetFigFont#1#2#3#4#5{
  \reset@font\fontsize{#1}{#2pt}
  \fontfamily{#3}\fontseries{#4}\fontshape{#5}
  \selectfont}
\fi\endgroup
\begin{picture}(2927,2640)(1576,-2581)
\thicklines
{\put(3021,-1323){\vector(-1,-3){364.500}}
}
{\put(4321,-1276){\vector(-1, 0){1170}}
}
{\put(3026,-1235){\vector( 1, 3){364.500}}
}
{\put(1910,-1243){\vector( 4, 1){641.412}}
}
{\put(1911,-1304){\vector( 4,-1){684}}
}
{\put(1756,-1276){\vector( 1, 0){1170}}
}
\put(1340,-1340){\makebox(0,0)[lb]{\smash{{\SetFigFont{12}{14.4}{\rmdefault}{\mddefault}{\updefault}{$3_{\bar{q}}$}
}}}}
\put(4230,-1321){\makebox(0,0)[lb]{\smash{{\SetFigFont{12}{14.4}{\rmdefault}{\mddefault}{\updefault}{$4_q$}
}}}}
\put(2415,-2600){\makebox(0,0)[lb]{\smash{{\SetFigFont{12}{14.4}{\rmdefault}{\mddefault}{\updefault}{$2_{\bar{Q}}$}
}}}}
\put(3200,-61){\makebox(0,0)[lb]{\smash{{\SetFigFont{12}{14.4}{\rmdefault}{\mddefault}{\updefault}{$1_Q$}
}}}}
\put(2490,-1100){\makebox(0,0)[lb]{\smash{{\SetFigFont{12}{14.4}{\rmdefault}{\mddefault}{\updefault}{$5_{q'}$}
}}}}
\put(2490,-1550){\makebox(0,0)[lb]{\smash{{\SetFigFont{12}{14.4}{\rmdefault}{\mddefault}{\updefault}{$6_{\bar{q}'}$}
}}}}
\end{picture}
\label{fig.pic2}
}}
\subfigure[]{
\includegraphics[scale=0.6]{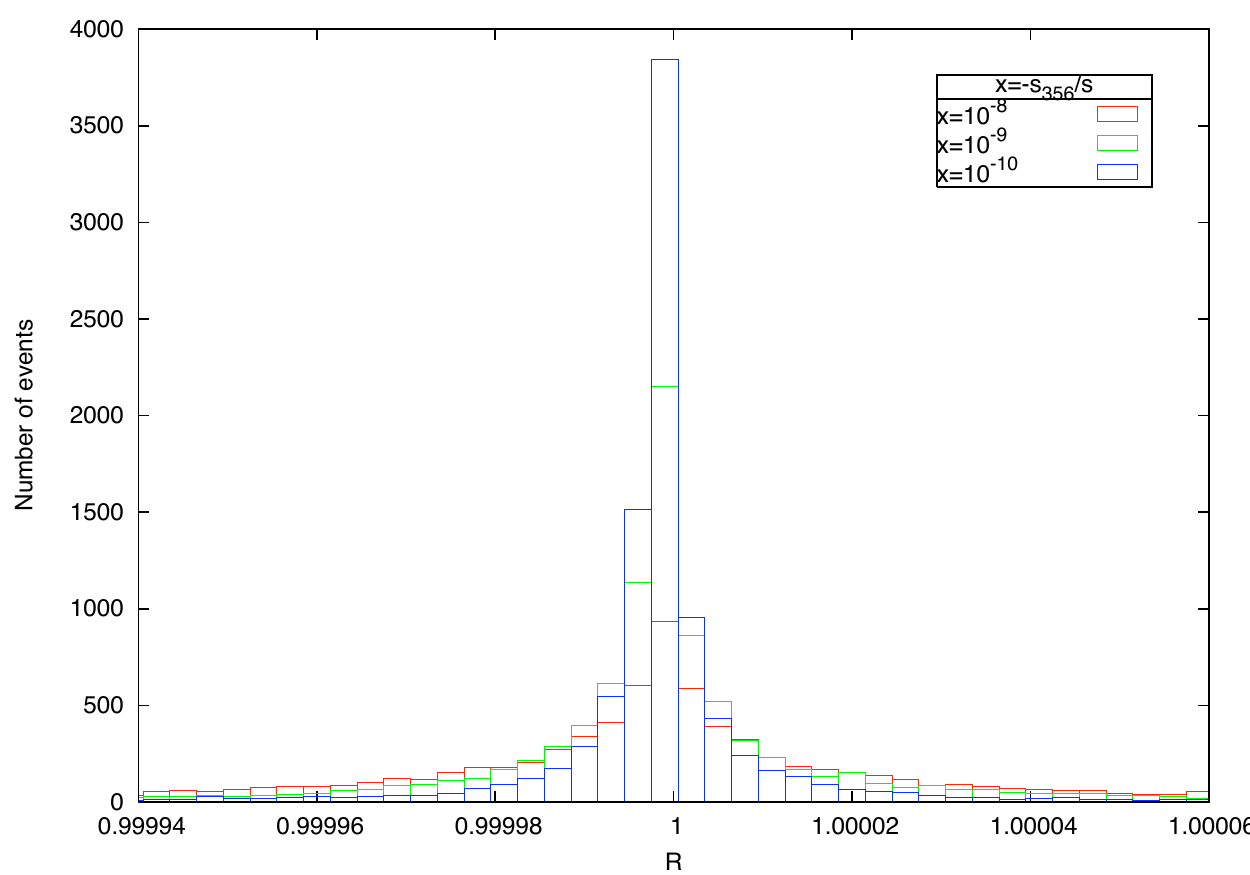}
\label{fig.tc3561}
}
\caption[]{\subref{fig.pic2} Topology of the triple collinear limit \subref{fig.tc3561} Distribution of R for 10000 double triple collinear space points.}
\end{figure}
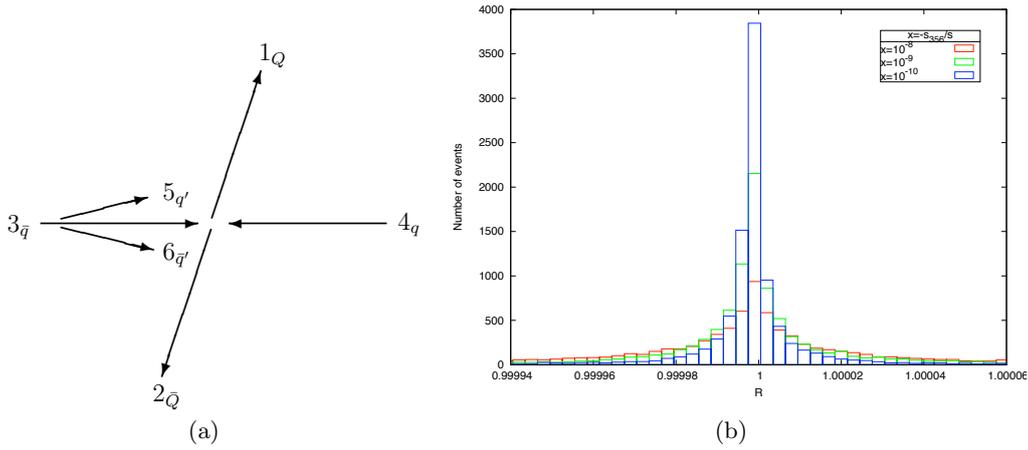
In fig.2(b) we show how, as we make $x=-s_{356}/s$ smaller, that is, as we get closer in phase space to the singularity of the real radiation matrix element, the histogram becomes more pronouncedly peaked around unity, signaling again that the approximation is good.
Similar results which are not shown are obtained for this process 
for the triple collinear limit involving parton $\qi{4}$.

\subsubsection{Single collinear limit}\label{sec.singlecoll}
The only single unresolved limit in which the matrix element for this process is divergent is the final-final collinear limit $\qp{5}||\qpb{6}$. The topology of this kinematical configuration is shown in fig.3(a), and the divergence is approached as the ratio $x=s_{56}/s$ becomes closer to zero.
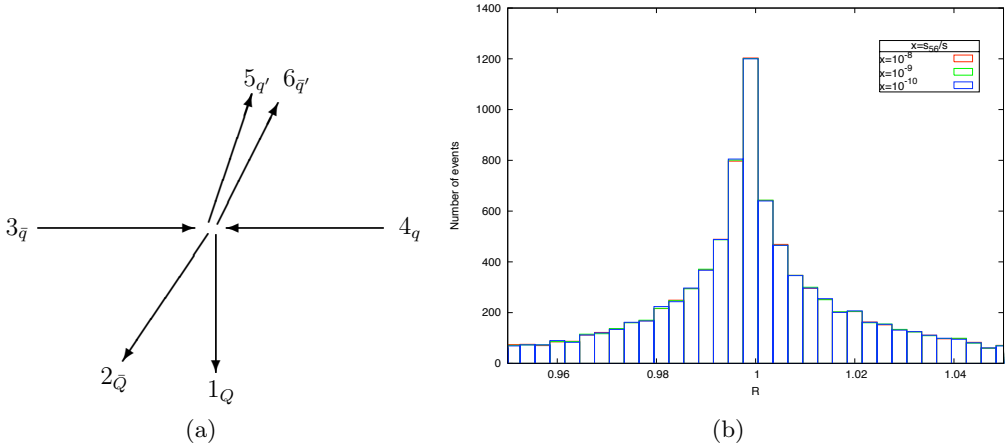
\begin{figure}[ht]
\centering
\subfigure[]{
\scalebox{0.8}{
\setlength{\unitlength}{4144sp}

\begingroup\makeatletter\ifx\SetFigFont\undefined
\gdef\SetFigFont#1#2#3#4#5{
  \reset@font\fontsize{#1}{#2pt}
  \fontfamily{#3}\fontseries{#4}\fontshape{#5}
  \selectfont}
\fi\endgroup
\begin{picture}(2927,2370)(1576,-2536)
\thicklines
{\put(3021,-1323){\vector(-2,-3){630.615}}
}
{\put(4321,-1276){\vector(-1, 0){1170}}
}
{\put(3090,-1243){\vector( 1, 2){449}}
}
{\put(3078,-1322){\vector( 0,-1){1035}}
}
{\put(3026,-1235){\vector( 1, 3){319.700}}
}
{\put(1756,-1276){\vector( 1, 0){1170}}
}
\put(1350,-1325){\makebox(0,0)[lb]{\smash{{\SetFigFont{12}{14.4}{\rmdefault}{\mddefault}{\updefault}{$3_{\bar{q}}$}
}}}}
\put(4260,-1321){\makebox(0,0)[lb]{\smash{{\SetFigFont{12}{14.4}{\rmdefault}{\mddefault}{\updefault}{$4_q$}
}}}}
\put(2840,-2536){\makebox(0,0)[lb]{\smash{{\SetFigFont{12}{14.4}{\rmdefault}{\mddefault}{\updefault}{$1_Q$}
}}}}
\put(2050,-2440){\makebox(0,0)[lb]{\smash{{\SetFigFont{12}{14.4}{\rmdefault}{\mddefault}{\updefault}{$2_{\bar{Q}}$}
}}}}
\put(3110,-220){\makebox(0,0)[lb]{\smash{{\SetFigFont{12}{14.4}{\rmdefault}{\mddefault}{\updefault}{$5_{q'}\:\:6_{\bar{q}'}$}
}}}}

\end{picture}
\label{fig.pic3}
}}
\subfigure[]{
\includegraphics[scale=0.6]{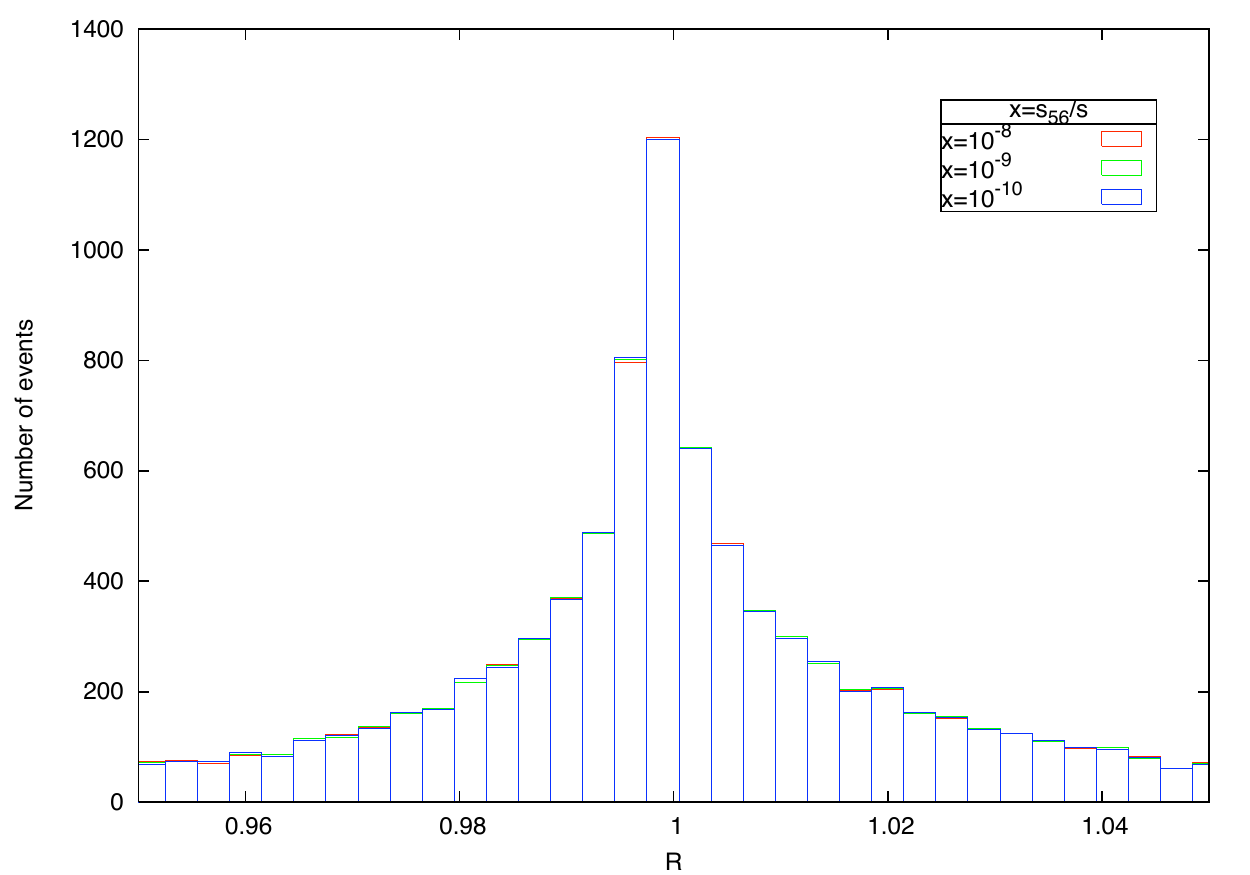}
\label{fig.coll561}
}
\caption[]{\subref{fig.pic3} Kinematics of the final-final single collinear limit. \subref{fig.coll561} Distribution of R for 10000 single collinear phase space points.}
\end{figure}

As it was discussed in section \ref{subsec.ang}, the azimuthal terms that the antenna subtraction term develops in its single collinear limits and those acquired by the real radiation matrix element do no match. This is the reason why the distributions in fig.3(b) are not very sharply peaked, and this is also the reason why the peaks do not become sharper as $x$ becomes smaller. As explained in section \ref{subsec.ang} these angular terms cancel, both in the matrix element squared and in the subtraction term, when a point in the collinear region of phase space is combined with another point that differs from the original one by a rotation of $\pi/2$ around the collinear axis.
\begin{figure}[ht]
\centering
\includegraphics[scale=0.6]{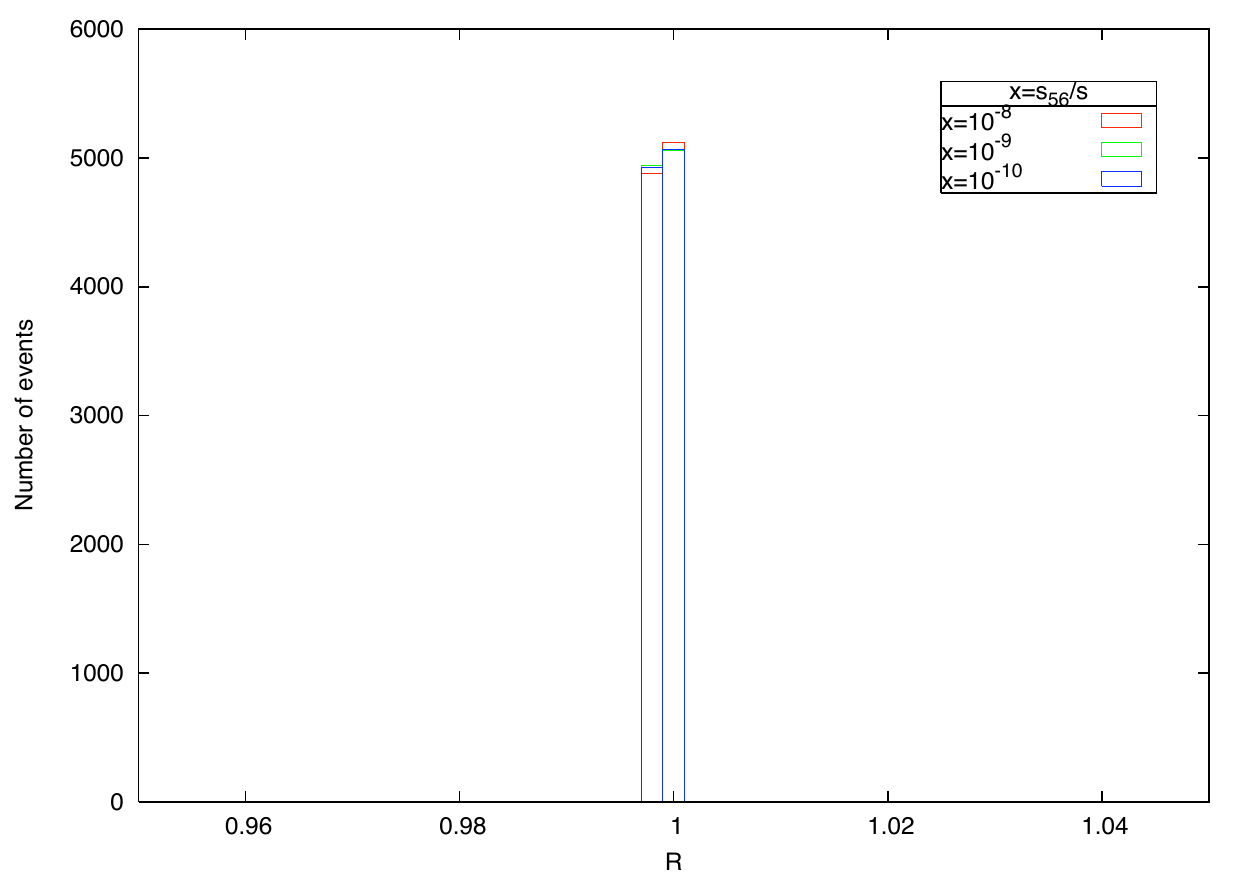}
\label{fig.coll56A1}
\caption[]{Distribution of R for 10000 single collinear phase space points after azimuthal terms have been eliminated by combining phase points related by a $\pi/2$ rotation about the collinear axis.}
\end{figure}
In fig.4 we see how the ratio of the real radiation matrix element and the subtraction term is substancially improved when the aforementioned combination of phase space points is performed.

\subsection{The $q\bar{q}\rightarrow Q\bar{Q}q\bar{q}$ process: identical quark flavour contributions}
When the incoming $q\bar{q}$ pair is of the same flavour as the outgoing one, the corresponding amplitude becomes singular in more regions of phase space, and new diagrams which are divergent in the regions where the non-identical flavour contributions were not singular are introduced. As we did in the previous section, we shall consider the identical-flavour-only piece, which contains the triple collinear limits $\qbi{3}||\q{5}||\qb{6}$ and $\qi{4}||\q{5}||\qb{6}$, the double collinear limit $\qbi{3}||\qb{6}+\qi{4}||\q{5}$ and the initial-final single collinear limits $\qbi{3}||\qb{6}$ and $\qi{4}||\q{5}$.

\subsubsection{Triple collinear limits}
The only limits shared  by the identical and non-identical quark flavour contributions are the triple collinear limits $\qbi{3}||\q{5}||\qb{6}$ and $\qi{4}||\q{5}||\qb{6}$, although the splitting functions corresponding to these limits in each case are different.
\begin{figure}[ht]\label{fig.tc3562}
\centering
\includegraphics[scale=0.6]{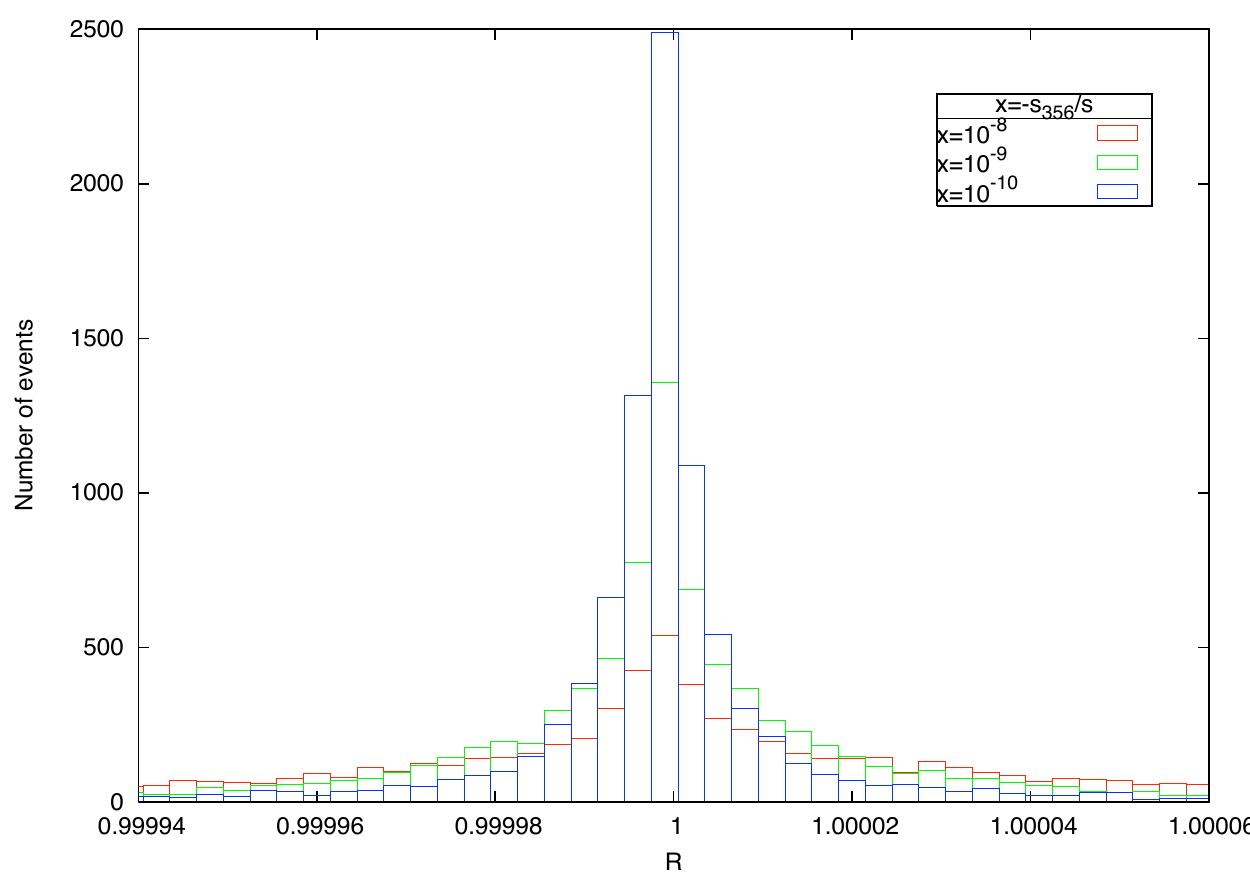}
\caption[]{Distribution of R for 10000 triple collinear phase space points.}
\end{figure}

As it can be seen in fig.5, the ratio between the real radiation corrections and the subtraction term converges to unity as the limit is approached. Again, similar results which are not shown are obtained for the limit involving the incoming quark $\qi{4}$.\\

\subsubsection{Double collinear limit}\label{sec.dc}
Another important feature of the identical-flavour-only contributions is that, containing amplitudes in which both incoming massless particles go through to the final state, double collinear limits are possible as depicted in fig.6(a).
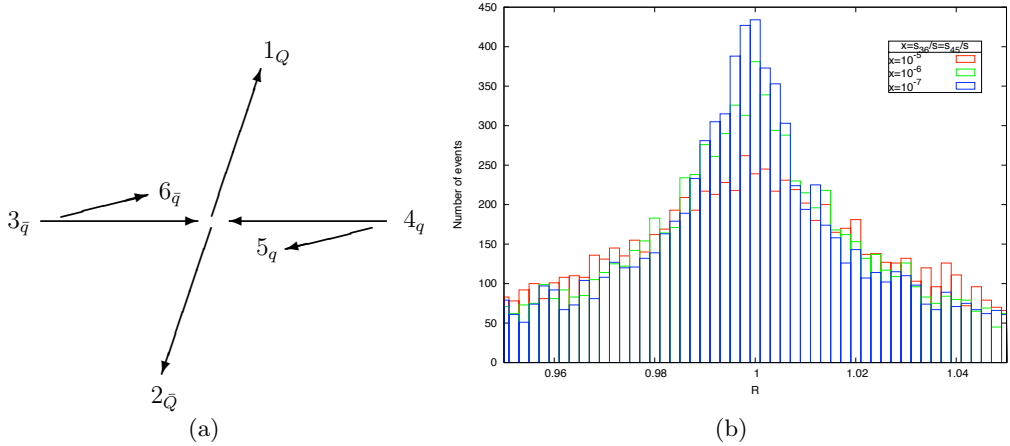
\begin{figure}[ht]
\centering
\subfigure[]{
\scalebox{0.8}{
\setlength{\unitlength}{4144sp}

\begingroup\makeatletter\ifx\SetFigFont\undefined
\gdef\SetFigFont#1#2#3#4#5{
  \reset@font\fontsize{#1}{#2pt}
  \fontfamily{#3}\fontseries{#4}\fontshape{#5}
  \selectfont}
\fi\endgroup
\begin{picture}(2927,2640)(1576,-2581)

\thicklines
{\put(3021,-1323){\vector(-1,-3){364.500}}
}
{\put(1756,-1276){\vector( 1, 0){1170}}
}
{\put(4321,-1276){\vector(-1, 0){1170}}
}
{\put(3026,-1235){\vector( 1, 3){364.500}}
}
{\put(1910,-1243){\vector( 4, 1){641.412}}
}
{\put(4212,-1327){\vector(-4,-1){641.412}}
}
\put(1340,-1330){\makebox(0,0)[lb]{\smash{{\SetFigFont{12}{14.4}{\rmdefault}{\mddefault}{\updefault}{$3_{\bar{q}}$}
}}}}
\put(4270,-1321){\makebox(0,0)[lb]{\smash{{\SetFigFont{12}{14.4}{\rmdefault}{\mddefault}{\updefault}{$4_q$}
}}}}
\put(2460,-1110){\makebox(0,0)[lb]{\smash{{\SetFigFont{12}{14.4}{\rmdefault}{\mddefault}{\updefault}{$6_{\bar{q}}$}
}}}}
\put(2400,-2630){\makebox(0,0)[lb]{\smash{{\SetFigFont{12}{14.4}{\rmdefault}{\mddefault}{\updefault}{$2_{\bar{Q}}$}
}}}}
\put(3230,-61){\makebox(0,0)[lb]{\smash{{\SetFigFont{12}{14.4}{\rmdefault}{\mddefault}{\updefault}{$1_Q$}
}}}}
\put(3180,-1530){\makebox(0,0)[lb]{\smash{{\SetFigFont{12}{14.4}{\rmdefault}{\mddefault}{\updefault}{$5_q$}
}}}}

\end{picture}
\label{fig.pic4}
}}
\subfigure[]{
\includegraphics[scale=0.6]{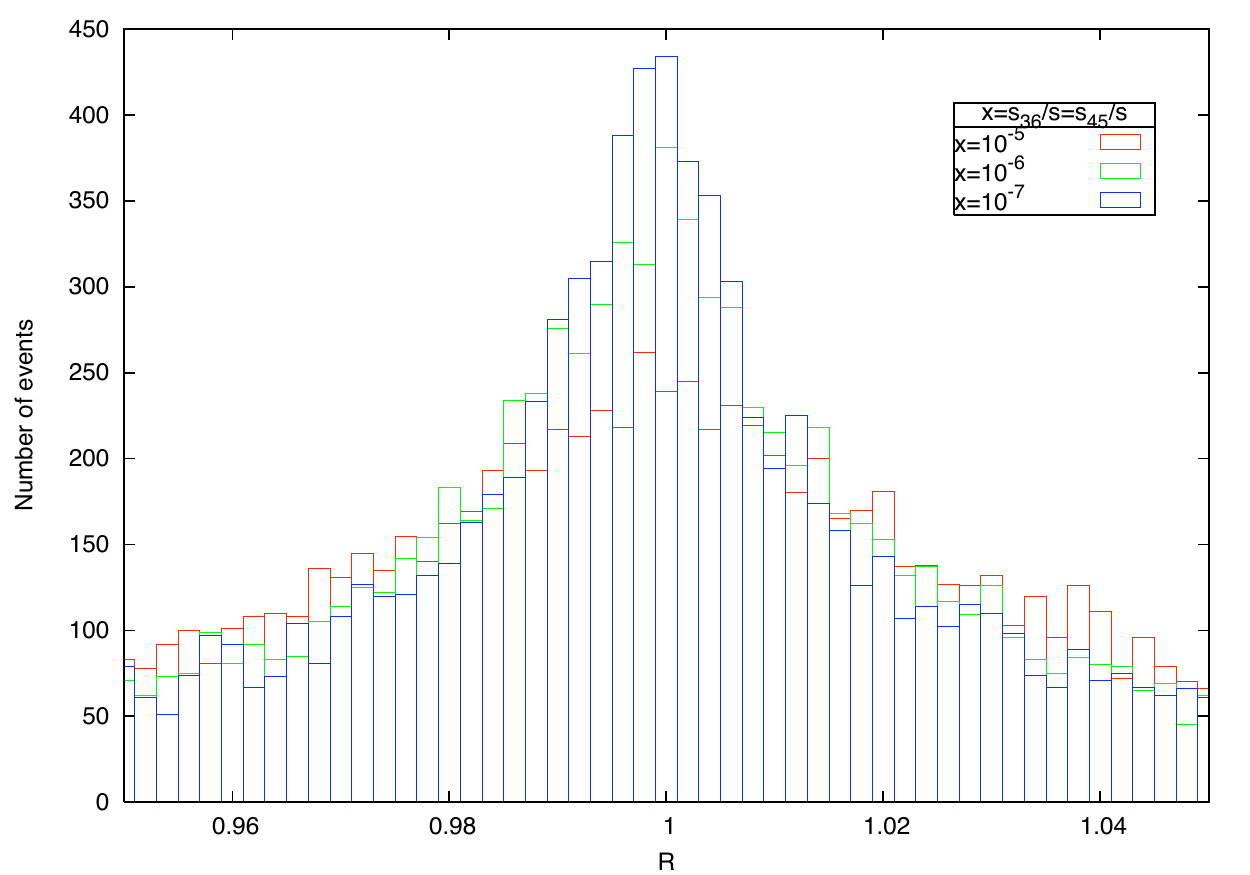}
\label{fig.dc2}
}
\caption[]{\subref{fig.pic4} Topology of a double collinear limit with two initial final collinearities. \subref{fig.dc2} Distribution of R for 10000 double collinear phase space points.}
\end{figure}
It can be seen in fig.6(b) that, in the double collinear limit, the presence of azimuthal terms also spoils the convergence of the subtraction term to the real radiation corrections. These terms can be averaged out by combining four points in phase space related by two $\pi/2$ rotations about each of the two collinear directions respectively. 
\begin{figure}[ht]
\centering
\includegraphics[scale=0.6]{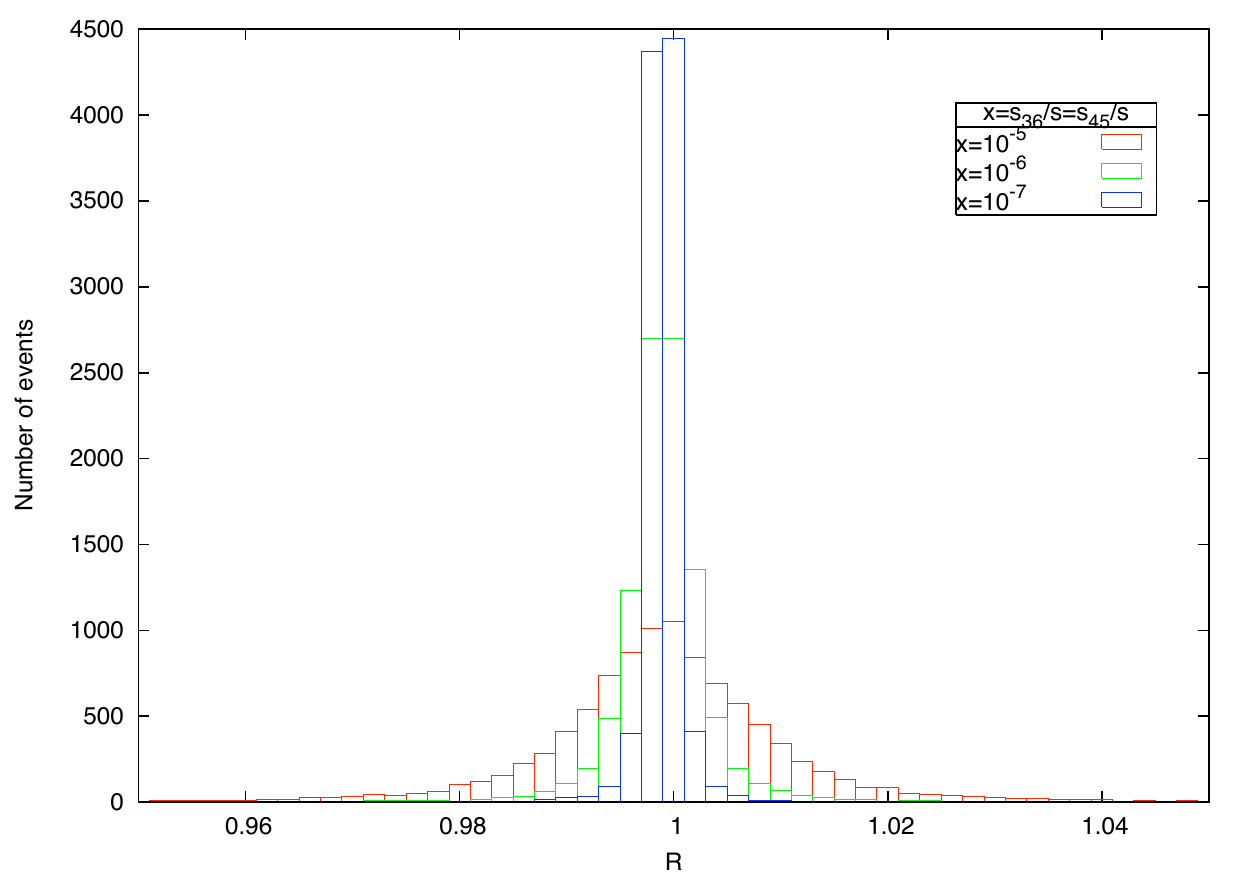}
\label{fig.dca2}
\caption[]{Distribution of R for 10000 double collinear phase space points with azimuthal terms averaged out.}
\end{figure}
The improvement of the convergence when these angular averaging is carried out can be clearly seen in fig.7.

\subsubsection{Single collinear limits}
As opposed to the non-identical flavour case, in which there was only one collinear limit involving the massless final state $q\bar{q}$ pair, the identical flavour contributions contain two initial-final collinear limits: $\qbi{3}||\qb{6}$ and $\qi{4}||\q{5}$. The topology of these unresolved limits is illustrated in fig.8(a).
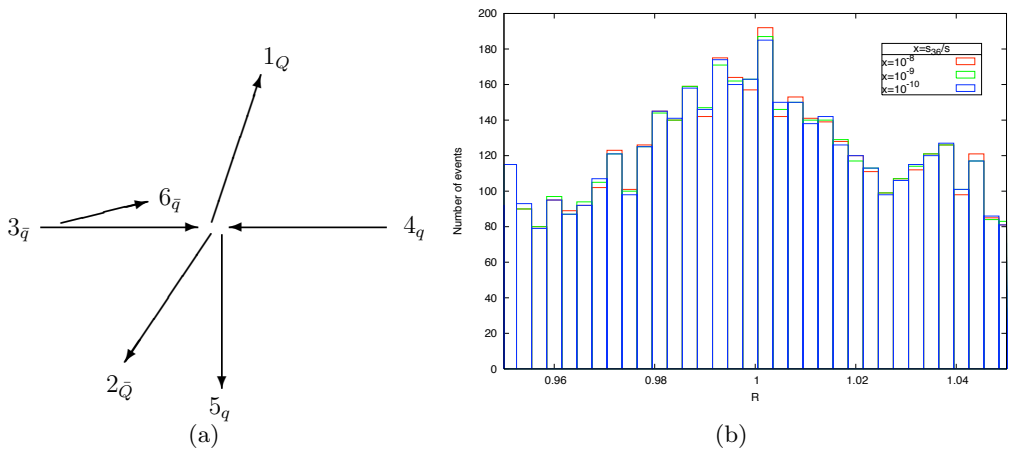
\begin{figure}[ht]
\centering
\subfigure[]{
\scalebox{0.8}{
\setlength{\unitlength}{4144sp}

\begingroup\makeatletter\ifx\SetFigFont\undefined
\gdef\SetFigFont#1#2#3#4#5{
  \reset@font\fontsize{#1}{#2pt}
  \fontfamily{#3}\fontseries{#4}\fontshape{#5}
  \selectfont}
\fi\endgroup
\begin{picture}(2927,2640)(1576,-2581)

\thicklines
{\put(3021,-1323){\vector(-2,-3){640}}
}
{\put(1756,-1276){\vector( 1, 0){1170}}
}
{\put(4321,-1276){\vector(-1, 0){1170}}
}
{\put(3026,-1235){\vector( 1, 3){364.500}}
}
{\put(1910,-1243){\vector( 4, 1){641.412}}
}
{\put(3100,-1327){\vector(0,-1){1152.6}}
}
\put(1340,-1330){\makebox(0,0)[lb]{\smash{{\SetFigFont{12}{14.4}{\rmdefault}{\mddefault}{\updefault}{$3_{\bar{q}}$}
}}}}
\put(4270,-1321){\makebox(0,0)[lb]{\smash{{\SetFigFont{12}{14.4}{\rmdefault}{\mddefault}{\updefault}{$4_q$}
}}}}
\put(2460,-1110){\makebox(0,0)[lb]{\smash{{\SetFigFont{12}{14.4}{\rmdefault}{\mddefault}{\updefault}{$6_{\bar{q}}$}
}}}}
\put(2070,-2500){\makebox(0,0)[lb]{\smash{{\SetFigFont{12}{14.4}{\rmdefault}{\mddefault}{\updefault}{$2_{\bar{Q}}$}
}}}}
\put(3230,-61){\makebox(0,0)[lb]{\smash{{\SetFigFont{12}{14.4}{\rmdefault}{\mddefault}{\updefault}{$1_Q$}
}}}}
\put(2830,-2670){\makebox(0,0)[lb]{\smash{{\SetFigFont{12}{14.4}{\rmdefault}{\mddefault}{\updefault}{$5_q$}
}}}}

\end{picture}
\label{fig.pic5}
}}
\subfigure[]{
\includegraphics[scale=0.6]{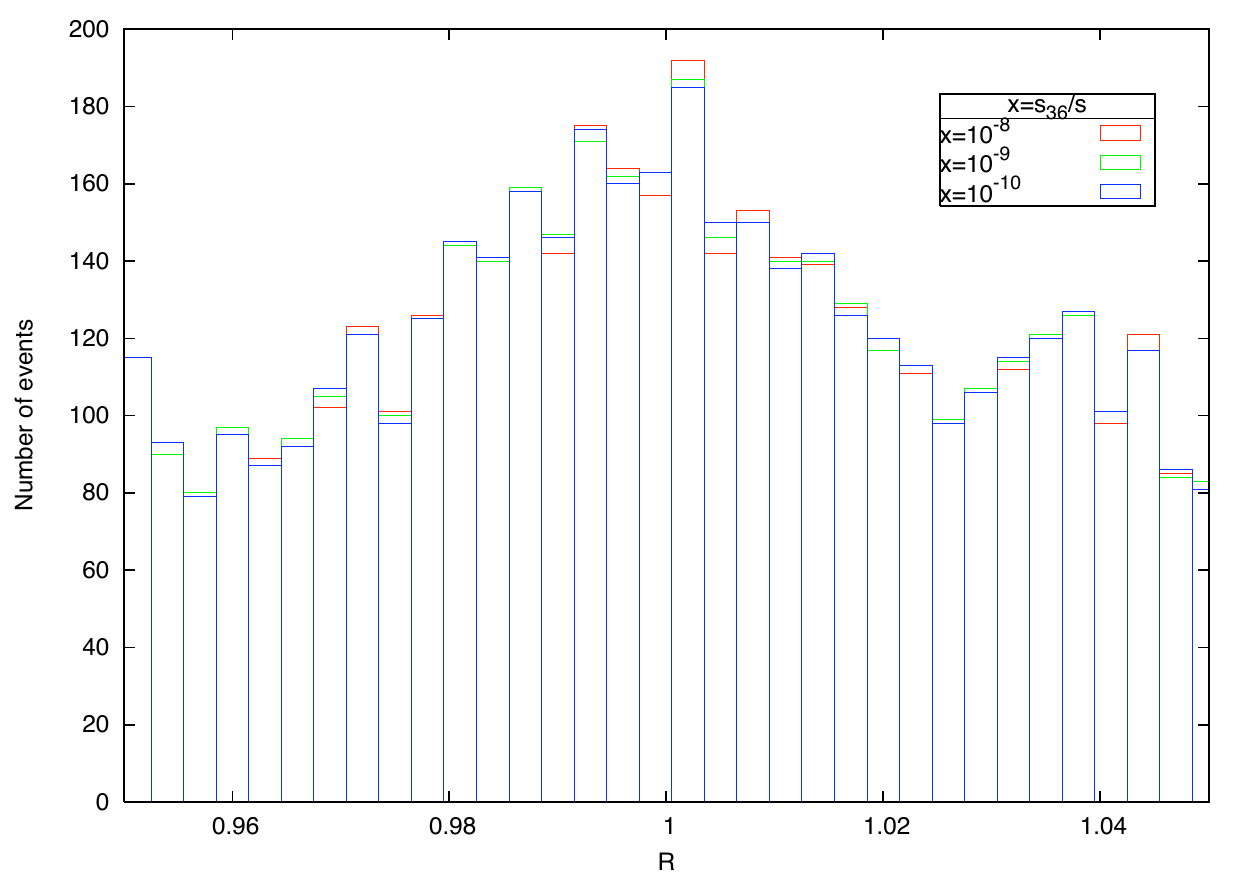}
\label{fig.coll362}
}
\caption[]{\subref{fig.pic5} Topology of an initial-final single collinear event. \subref{fig.coll362} Distribution of R for 10000 phase space points in the single collinear region.}
\end{figure}
In fig.8(b) we show how, in the presence of angular terms, the distributions for the ratio $R$ are not peaked around unity. As shown in fig.9, the convergence of the subtraction term is again improved when we combine two events that differ by a rotation of $\pi/2$ of the collinear pair about the collinear axis. This combination of phase space points cancels out the azimuthal correlations, as it did in the single collinear limit discussed in section (\ref{sec.singlecoll}). However, it should be mentioned that for initial-final limits this procedure is not as straightforward as it is in the final-final case, since the rotation of the collinear pair takes the incoming particle involved in the collinear limit out of the beam axis \footnote{To compensate for this, the rotated phase space trajectory has to be boosted to a frame in which the incoming particles are back-to-back, and then another rotation is needed in order to bring the incoming momenta back to the beam axis.}.

\begin{figure}[ht]
\centering
\includegraphics[scale=0.6]{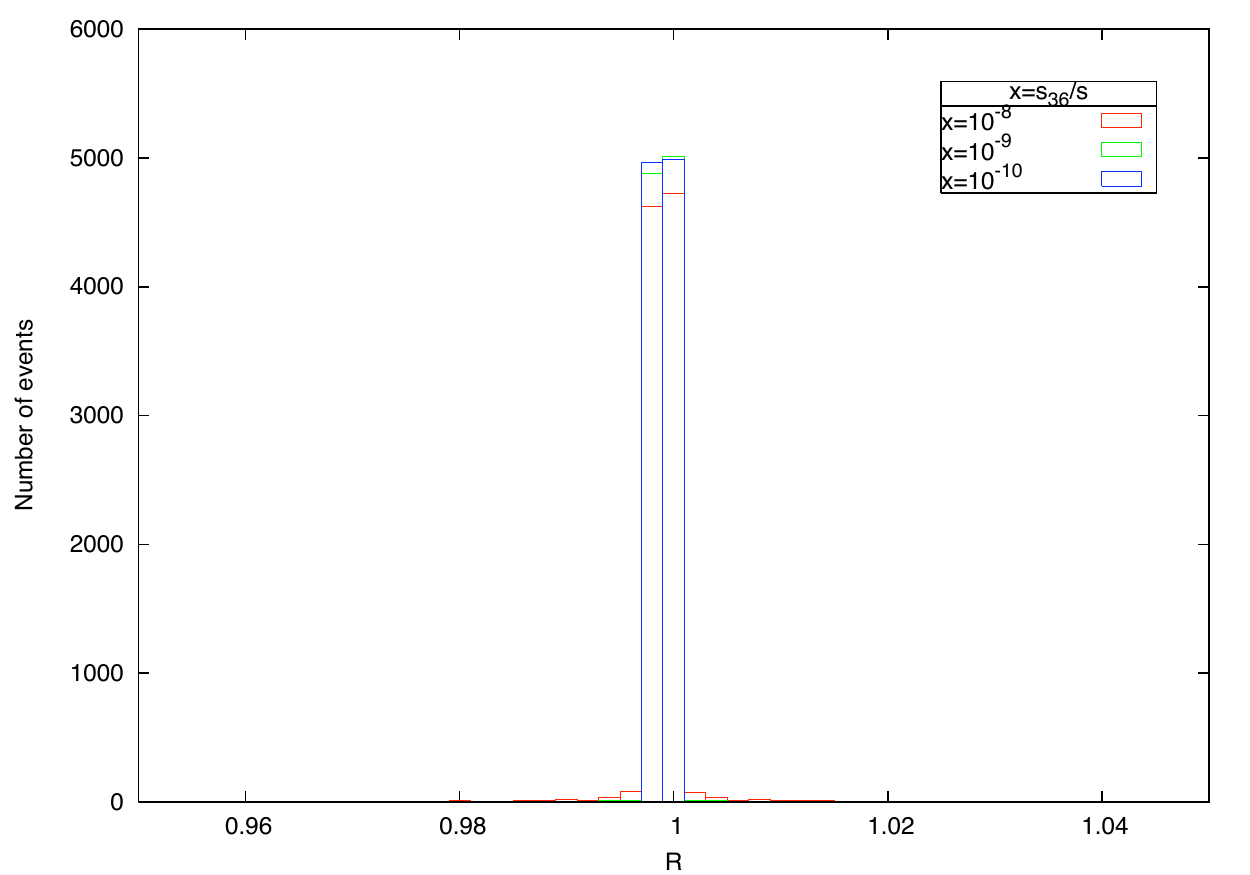}
\label{fig.coll36a2}
\caption[]{Distribution of R for 10000 single collinear phase space points after azimuthal terms have been eliminated by combining phase points related by a $\pi/2$ rotation about the collinear axis.}
\end{figure}
Similar results were obtained for the $\qi{4}||\q{5}$ collinear limit.\\

The same checks have been performed on the subtraction terms (\ref{eq.sub3}) and (\ref{eq.sub4}) corresponding to the processes $qq'\rightarrow Q\bar{Q}qq'$ and $qq\rightarrow Q\bar{Q}qq$ respectively. The convergence of these subtraction terms to their corresponding squared matrix elements has been tested in the same way as described above, and the numerical results obtained are very similar to the ones shown in the histograms of this section and related to the processes $q\bar{q}\rightarrow Q\bar{Q}q'\bar{q}'$ and $q\bar{q}\rightarrow Q\bar{Q}q\bar{q}$.

\section{Conclusions}\label{sec.conclusions}
We presented the extension of the antenna formalism required for the calculation of hadronic processes involving a pair of massive final states in association with jets at the NNLO level. We applied this subtraction framework to the computation of the double real radiation corrections to the hadronic $t \bar{t}$ differential cross section arising from partonic processes with fermions only. In section \ref{sec.formalism}, we outlined the general structure of the subtraction terms which capture all single and double unresolved features of the real matrix elements squared, while  focussing on the differences caused by the presence of massive final states compared to the NNLO antenna formalism developed for the computation of massless QCD reactions at hadron colliders~\cite{joao,RVnew}. The genuine unknown NNLO tree-level massive four-parton antenna functions required to capture the double unresolved singular behaviour of the real matrix elements when the two unresolved partons are colour-connnected, are derived and their limiting behaviour is presented in sections \ref{sec.antennae} and \ref{sec.limits} respectively. The subtraction terms given in leading and subleading colour contributions are constructed in section~\ref{sec.subterms} for all partonic processes involved. The subtraction of double soft singularities from a fermion-antifermion pair in subleading-colour contributions requires a special treatment which is also explained in that section. Finally, in section \ref{sec.results} we checked the validity of our subtraction terms numerically by verifying that they approximate the real matrix elements in all the single and double unresolved configurations in a point-by point manner. The numerical results showed that the combination of antenna functions multiplied by reduced matrix elements present in the subtraction terms correctly describes the infrared singularity structure of the real matrix elements  in all limits including also single collinear limits. This is achieved provided the azimuthal terms associated with these collinear limits are correctly handled.

The double real subtraction terms related to the all fermion processes presented here provide a substantial step towards the calculation of the NNLO corrections to top quark pair production at the LHC. Future steps include in particular the computation of the remaining double real subtraction terms related to partonic channels involving gluons and  the computation of mixed real-virtual contributions for all partonic channels involved.

\section{Acknowledgements}
We would like to thank Joao Pires for many useful and stimulating discussions. This research was supported by the Swiss National Science Foundation (SNF) under contract PP0022-118864 and in part by the European Commission through the 'LHCPhenoNet' Initial Training Network PITN-GA-2010-264564', which are hereby acknowledged.

\appendix
\section{Three-parton antennae}\label{sec.3partonantennae}
In this appendix we list the three-parton antenna functions used in the subtraction terms of section \ref{sec.subterms} together with their single unresolved limits. The massive antennae that we give here can be found in \cite{mathias,us}, while massless ones are given in \cite{antennannlo,daleo}. All the single unresolved factors can be found in section \ref{subsec.factors}.

Two types of antennae are needed: E-type antenna functions, which are used to subtract single and double collinear limits in all the different pieces of the subtraction terms, and A-type antenna functions, which are needed in the $\ds^{S,b}_{NNLO}$ pieces since they are the reduced matrix elements to which $B_4^0$ antennae collapse in their spurious single collinear limits.

\subsection{A-type antennae}
The only final-final A-type antenna that we use in our subtraction terms is massive. It has been computed and integrated in \cite{mathias}, and its unintegrated form is
\beqa
A^{0}_{3}(\Q{1},\gl{3},\Qb{2})& =& \frac{1}{ \left(E_{cm}^2 + 2 m_{Q}^2\right)} \left(\frac{2 s_{12}^2}{s_{13} s_{23}}+\frac{2s_{12}}{s_{13}}+\frac{2s_{12}}{s_{23}}+\frac{s_{23}}{s_{13}}+\frac{s_{13}}{s_{23}}\right.\nonumber\\
&&\left. + m_{Q}^2 \left(\frac{8 s_{12}}{s_{13} s_{23}}-\frac{2s_{12}}{s_{13}^2}-\frac{2s_{12}}{s_{23}^2}-\frac{2s_{23}}{s_{13}^2}-\frac{2}{s_{13}}-\frac{2}{s_{23}}-\frac{2 s_{13}}{s_{23}^2}\right)\right.\nonumber\\
&&\left. + m_{Q}^4 \left(-\frac{8}{s_{23}^2}-\frac{8}{s_{13}^2}\right)\right)+\order{\epsilon},\label{eq:A03mff}
\eeqa
with $E_{cm}^2=(p_1+p_2+p_3)^2$. This antenna has only a single soft limit:
\beq
A_3^0(\Q{1},\gl{3},\Qb{2})\stackrel{^{p_3 \rightarrow0}}{\longrightarrow} \ssoft{1}{3}{2}(m_Q,m_Q).
\eeq

The flavour violating initial-final antenna reads
 \beqa
A_3^0(\Q{1},\gl{3},\qi{2})&=&\frac{1}{\left( Q^2-m_Q^2 \right)}\left(\frac{2s_{12}^2}{s_{13}s_{23}}+\frac{2s_{12}}{s_{13}}-\frac{2s_{12}}{s_{23}}+\frac{s_{13}}{s_{23}}+\frac{s_{23}}{s_{13}}\right.\nonumber\\
&& \left. -m_Q^2\left( \frac{2s_{12}}{s_{13}^2}+\frac{2s_{23}}{s_{13}^2}-\frac{2}{s_{13}}\right)\right)+\order{\epsilon},
\label{eq.Afl}
\eeqa
where $Q^2=-(p_1-p_2+p_3)^2$. The integrated form of eq.(\ref{eq.Afl}) can be found in \cite{us}. This antenna has a soft gluon limit as well as an initial-final collinear limit:
\beqa
&&A_3^0(\Q{1},\gl{3},\qi{2})\stackrel{^{p_3 \rightarrow0}}{\longrightarrow}\soft{1}{3}{2}(m_Q,0),\\
&&A_3^0(\Q{1},\gl{3},\qi{2})\stackrel{^{\hat{2}_q||3_g}}{\longrightarrow}\frac{1}{s_{23}}P_{qg\leftarrow Q}(z).
\eeqa

Two different initial-initial A-type antenna functions are used in our subtraction terms. They are given by
\beqa
&&A_3^0(\qbi{1},\gl{3},\qi{2})=\frac{1}{Q^2}\left(-\frac{2s_{12}^2}{s_{13}s_{23}}+\frac{2s_{12}}{s_{13}}+\frac{2s_{12}}{s_{23}}-\frac{s_{13}}{s_{23}}-\frac{s_{23}}{s_{13}}\right)+\order{\epsilon},\label{eq.Aii}\\
&&A_3^0(\q{1},\gli{3},\qi{2})=\frac{1}{Q^2}\left(-\frac{2s_{12}^2}{s_{13}s_{23}}+\frac{2s_{12}}{s_{13}}-\frac{2s_{12}}{s_{23}}-\frac{s_{13}}{s_{23}}-\frac{s_{23}}{s_{13}}\right)+\order{\epsilon},\label{eq.Aii2}
\eeqa
with $Q^2=-(p_3-p_1-p_2)^2$ in eq.(\ref{eq.Aii}) and $Q^2=-(p_1-p_2-p_3)^2$ in eq.(\ref{eq.Aii2}). These antennae have been integrated in \cite{daleo}. Their infrared limits are
\beqa
&&A_3^0(\qbi{1},\gl{3},\qi{2})\stackrel{^{p_3 \rightarrow0}}{\longrightarrow}S_{132},\\
&&A_3^0(\qbi{1},\gl{3},\qi{2})\stackrel{^{\hat{1}_{\bar{q}}||3_g}}{\longrightarrow}\frac{1}{s_{13}}P_{qg\leftarrow Q}(z),\\
&&A_3^0(\qbi{1},\gl{3},\qi{2})\stackrel{^{\hat{2}_q||3_g}}{\longrightarrow}\frac{1}{s_{23}}P_{qg\leftarrow Q}(z),
\eeqa
and
\beq
A_3^0(\q{1},\gli{3},\qi{2})\stackrel{^{1_{q}||\hat{3}_g}}{\longrightarrow}\frac{1}{s_{13}}P_{q\bar{q}\leftarrow G}(z)
\eeq

\subsection{E-type antennae}
For the subtraction of final-final collinear limits we use
\beq\label{eq.Effm}
 E^{0}_{3}(\Q{1},\q{3},\qb{4}) = \frac{1}{\left(E_{cm}^2-m_{Q}^2 \right)^2} \left( s_{13} + s_{14} + \frac{s_{13}^2}{s_{34}} +\frac{s_{14}^2}{s_{34}} - 2 E_{cm} m_{Q}\right) + \order{\epsilon},
\eeq
where $E_{cm}^2=(p_1+p_3+p_4)^2$. The integrated form of this antenna can be found in \cite{mathias}, and its only infrared limit is
\beq
 E^{0}_{3}(\Q{1},\q{3},\qb{4})\stackrel{^{\q{3}||\qb{4}}}{\longrightarrow}\frac{1}{s_{34}}P_{q\bar{q}\rightarrow G}(z)
\eeq

On the other hand, to subtract initial-final collinear singularities we use the following initial-final antenna
\beq\label{eq.Eifm}
E_3^0(\Q{1},\q{3},\qi{4})=-\frac{1}{\left(Q^2+m_Q^2\right)^2}\left(-s_{14}+s_{13}-\frac{s_{13}^2}{s_{34}}-\frac{s_{14}^2}{s_{34}}
-2m_{Q}m_{\chi}\right)+\order{\epsilon},
\eeq
with $Q^2=-(p_1+p_3-p_4)^2$, and $m_{\chi}=\sqrt{-Q^2}$ is the mass of the incoming neutralino denoted by $\chi$ and present in the process 
$\chi q \to Q q$ defining this initial-final ${E}_3^0$ antenna~\cite{GehrmannDeRidder:2005hi}. The only infrared limit that this antenna has is 
\beq
E_3^0(\Q{1},\q{3},\qi{4})\stackrel{^{3_q||\hat{4}_q}}{\longrightarrow}\frac{1}{s_{34}} P_{gq\leftarrow Q}(z).
\eeq

\subsection{Integrated forms of massive $E_3^0$ antennae}\label{sec.appendixintegrated}
To supplement the discussion given in section \ref{sec.cancelation} on the infrared structure associated  with subtraction terms capturing  single unresolved limits, we list the integrated forms of the two three-parton antenna functions that we use in those pieces of the subtraction terms.

After integration over the corresponding $1\rightarrow 3$ antenna phase space, the massive final-final $E_3^0$ antenna function reads \cite{mathias}  
\beqa
\lefteqn{{\cal E}^{0}_{3}(\Q{1}, \qp{3},\qpb{4})=}\nonumber\\
&&\hspace{15mm}-4 {\bf I}^{(1)}_{Qg,N_{F}}\left(\epsilon,s_{Q q'\bar{q}'},m_{Q},0,\frac{m_{Q}^2}{m_{Q}^2+s_{Qq'\bar{q}'}}\right)\nonumber\\
&&\hspace{15mm}-\frac{1}{6(1-\mu^2)^3}\bigg[ 6+ 3\mu-14\mu^2+14\mu^4-3\mu^5 -14\mu^2+14\mu^4-3\mu^5-6\mu^6\nonumber\\
&&\hspace{15mm}-2\mu^3(-3-3\mu+\mu^3)\ln(\mu^2)\bigg]+{\cal O}(\epsilon)\label{eq.integratedE03ff}
\eeqa
where the infrared singularity operator is given by \cite{cdt2,us}
\beq\label{eq:Ione}
{\bf I}^{(1)}_{Q g,N_{F}} \left(\epsilon,s_{Q g},m_{Q},0,\frac{m_{Q}^2}{s_{Qg}+m_{Q}^2}\right) =  
\frac{e^{\epsilon \gamma_E}}{2\Gamma(1-\epsilon)}\, \left[\frac{s_{Q g}+m_{Q}^2}{s_{Qg}^2}\right]^{\epsilon}\left(\frac{1}{6\epsilon} \right).
\eeq

The integrated form of the massive initial-final $E_3^0$ antenna is given by \cite{us}
\beqa
&&{\cal E}_3^0(4_q;3_q,1_{Q})=\frac{e^{\epsilon\gamma_E}}{\Gamma(1-\epsilon)}
\left[ Q^2+m_{Q}^2 \right]^{-\epsilon}
\times \Bigg \{ -\frac{1}{2\epsilon}\;p_{qg}^{(0)}(x) \nonumber \\
&&\hspace{10mm} +\frac{1}{2x}(2-2x+x^2) \left[-2+2\ln(1-x)-\ln(x) -\ln(1-x_0x) \right]\nonumber\\
&&\hspace{10mm}+\frac{Q^2(-1+3x-2x^2)+m_Q^2(1+x)+2m_Qm_{\chi}(1-x)}{2(Q^2(1-x)+m_Q^2)}\Bigg \}+{\cal O}(\epsilon) \label{eq.integratedE03if}
\eeqa
where the splitting kernel $p^{(0)}_{qg}$ reads
\beq
p^{(0)}_{qg}(x)=1-2x+2x^2.
\eeq

\section{Phase space mappings}
In this appendix we give the various phase space mappings used in the construction of our subtraction terms. These mappings must (a) conserve four momentum and (b) maintain the on-shellness of the particles involved. Three configurations need to be considered: final-final, initial-final and initial-initial, and depending if the mapping is related to a single or a double unresolved configuration $3\to 2$ 
or $4\to 2$ mapping procedures are needed. 

For processes involving massive particles the mapping procedures remain unaltered. 
As mentioned in section \ref{sec.formalism}, provided all variables and momenta are expressed in terms of invariants of the form $s_{ij}=2p_i\cdot p_j$, massless phase mappings can be used for processes with massive particles. For final-final configurations we use the mappings given in \cite{3jet}, whereas initial-final and initial-initial mappings are taken from \cite{daleo}.

\subsection{Single unresolved mappings} 
Subtraction terms built with three-parton antennae, namely those present at NLO and those related to 
the $\ds^{S,a}_{NNLO}$  subtraction terms, need the definition of $\{3\rightarrow 2\}$ phase space mappings in all three configurations.

\subsubsection{Final-final configurations}
In this case, the mapping can be symbolically understood as 
$(i,j,k)\rightarrow(\wt{ij},\wt{jk})$, and it is defined by
\beqa
&&\wt{p}_{ij}^\mu=x p_i^\mu+r p_j^\mu+z p_k^\mu\nonumber\\
&&\wt{p}_{jk}^\mu=(1-x) p_i^\mu+(1-r) p_j^\mu+(1-z) p_k^\mu
\eeqa
where
\beqa
&& x=\frac{1}{2(s_{ij}+s_{ik})}\left[ (1+\rho)s_{ijk}-2 r s_{jk}\right]\nonumber\\
&& z=\frac{1}{2(s_{jk}+s_{ik})}\left[ (1-\rho)s_{ijk}-2 r s_{ij}\right]\nonumber\\
&&\rho^2=1+\frac{4r(1-r)s_{ij}s_{jk}}{s_{ijk}s_{ik}}\nonumber\\
&&r=\frac{s_{jk}}{s_{ij}+s_{jk}}.
\eeqa
This mapping preserves momentum conservation 
\beq
\wt{p}_{ij}+\wt{p}_{jk}=p_i+p_j+p_k
\eeq
and satisfies the following properties
\beqa
&&\wt{p}_{ij}\rightarrow p_i\hspace{20mm}\wt{p}_{jk}\rightarrow p_k\hspace{20mm}\text{when $j\rightarrow 0$},\nonumber\\
&&\wt{p}_{ij}\rightarrow p_i+p_j\hspace{12mm}\wt{p}_{jk}\rightarrow p_k\hspace{20mm}\text{when $i||j$},\nonumber\\
&&\wt{p}_{ij}\rightarrow p_i\hspace{20mm}\wt{p}_{jk}\rightarrow p_j+p_k\hspace{12mm}\text{when $j||k$},
\eeqa
which guarantee the correct subtraction of infrared singularities.

\subsubsection{Initial-final configurations}
For these cases the mappings are of the form $(\hat{i},j,k)\rightarrow(\hat{\bar{i}},\wt{jk})$, where the bar on the initial state momentum indicates that it is only rescaled. The remapped momenta are given in terms of the original ones by    
\beqa
&&\bar{p}_i^\mu=x_i p_i^\mu\nonumber\\
&&\wt{p}_{jk}^\mu=p_j^\mu+p_k^\mu-(1-x_i) p_i^\mu,
\eeqa
where
\beq
x_i=\frac{s_{ij}+s_{ik}-s_{jk}}{s_{ij}+s_{ik}}.
\eeq
This mapping satisfies the following properties, which are needed for the correct subtraction of infrared singularities:
\beqa
&&x_i p_i \rightarrow p_i\hspace{20mm}\wt{p}_{jk}\rightarrow p_k\hspace{20mm}\text{when $j\rightarrow 0$},\nonumber\\
&&x_i p_i \rightarrow p_i\hspace{20mm}\wt{p}_{jk}\rightarrow p_j+p_k\hspace{12mm}\text{when $j||k$},\nonumber\\
&&x_i p_i \rightarrow (1-z_i)p_i \hspace{7.8mm}\wt{p}_{jk}\rightarrow p_k\hspace{20mm}\text{when $p_j=z_i p_i$}.
\eeqa

\subsubsection{Initial-initial configurations}
The mapping used for these configurations is of the form $\{\hat{i},j,\hat{k},...,m,n,...\}\rightarrow \{\hat{\bar{i}},\hat{\bar{k}},...,\tilde{m},\tilde{n},...  \}$. It is explicitely given by
\beqa
&&\bar{p}_i^\mu=x_i p_i^\mu\nonumber\\
&&\bar{p}_j^\mu=x_k p_k^\mu\nonumber\\
&&\tilde{p}_n^\mu=p_n^\mu-\frac{2p_n\cdot (q+\tilde{q})}{(q+\tilde{q})^2}(q^\mu+\tilde{q}^\mu)+\frac{2p_n\cdot q}{q^2}\tilde{q}^\mu
\eeqa
where
\beq
q^\mu=p_i^\mu+p_k^\mu-p_j^\mu, \hspace{25mm} \tilde{q}^\mu=\tilde{p}_i^\mu+\tilde{p}_k^\mu.
\eeq
The particular feature about initial-initial mappings is that all final state momenta, even those that are not actually part of the antenna, need to be remapped in order to preserve momentum conservation. This is due to the fact that $\tilde{q}$ lies along the beam axis while $q$ is in general not along the beam axis. The rescaling variables $x_i$ and $x_k$ are given by
\beqa
x_i = \sqrt{\frac{s_{ik}-s_{jk}}{s_{ik}-s_{ij}}}\sqrt{\frac{s_{ik}-s_{ij}-s_{jk}}{s_{ik}}},\nonumber\\
x_k = \sqrt{\frac{s_{ik}-s_{ij}}{s_{ik}-s_{jk}}}\sqrt{\frac{s_{ik}-s_{jk}-s_{ij}}{s_{ik}}},
\eeqa
and the properties satisfied by this mapping which ensure the correct subtraction of infrared singularities are
\beqa
&&x_i p_i \rightarrow p_i\hspace{20mm}x_k p_k \rightarrow p_k\hspace{20mm}\text{when $j\rightarrow 0$},\nonumber\\
&&x_i p_i \rightarrow (1-z_i)p_i \hspace{7.25mm}x_k p_k \rightarrow p_k\hspace{20mm}\text{when $p_j=z_i p_i$},\nonumber\\
&&x_i p_i \rightarrow p_i \hspace{20mm}x_k p_k \rightarrow (1-z_k)p_k\hspace{7mm}\text{when $p_j=z_k p_k$}.
\eeqa

\subsection{Double unresolved mappings}
The use of four-parton antenna functions in the $\ds^{S,b}_{NNLO}$ parts of the subtraction terms requires the definition of adequate $\{4\rightarrow 2\}$ mappings which smoothly interpolate the kinematics of the different unresolved configurations.

\subsubsection{Final-final configurations}
In this case the mapping is $(i,j,k,l)\rightarrow (\wt{ijk},\wt{jkl})$, and the remapped momenta are given by
\beqa
&&\wt{p}_{ijk}^\mu=x p_i^\mu + r_1 p_j^\mu + r_2 p_k^\mu + z p_l^\mu\nonumber\\
&&\wt{p}_{jkl}^\mu=(1-x) p_i^\mu + (1-r_1) p_j^\mu + (1-r_2) p_k^\mu + (1-z) p_l^\mu\label{eq.mapffd}
\eeqa
with
\beqa
&&r_1=\frac{s_{jk}+s_{jl}}{s_{ij}+s_{jk}+s_{kl}}\nonumber\\
&&r_2=\frac{s_{kl}}{s_{ik}+s_{jk}+s_{kl}}\nonumber\\
&&x=\frac{1}{2(s_{ij}+s_{ik}+s_{il})}\bigg[ (1+\rho)s_{ijkl}-r_1(s_{jk}+2s_{jl})-r_2(s_{jk}+2s_{kl})\nonumber\\
&&\hspace{20mm} +(r_1-r_2)\frac{s_{ij}s_{kl}-s_{ik}s_{kl}}{s_{il}}\bigg]\nonumber\\
&&z=\frac{1}{2(s_{il}+s_{jl}+s_{kl})}\bigg[ (1-\rho)s_{ijkl}-r_1(s_{jk}+2s_{ij})-r_2(s_{jk}+2s_{ik})\nonumber\\
&&\hspace{20mm} -(r_1-r_2)\frac{s_{ij}s_{kl}+s_{ik}s_{jl}}{s_{il}}\bigg]\nonumber\\
&&\rho=\bigg[ 1+\frac{(r_1-r_2)^2}{s_{il}^2s_{ijkl}^2}\lambda(s_{ij},s_{kl},s_{il},s_{jk},s_{ik},s_{jl})\nonumber\\
&&\hspace{7mm}+\frac{1}{s_{il}s_{ijkl}}\bigg\{2\left( r_1(1-r_2)+r_2(1-r_1)\right)\left( s_{ij}s_{kl}+s_{ik}s_{jl}-s_{jk}s_{il} \right)\nonumber\\
&&\hspace{22mm}+4r_1(1-r_1)s_{ij}s_{jl}+4r_2(1-r_2)s_{ik}s_{kl}\bigg\}\bigg]^{\frac{1}{2}}.
\eeqa
As usually, the function $\lambda(u,v,w)$ is defined as
\beq
\lambda(u,v,w)=u^2+v^2+w^2-2(uv+uw+vw).
\eeq  
The behaviour of these mappings in the different double unresolved limits is
\beqa
&&\wt{p}_{ijk}\rightarrow p_i \hspace{30mm} \wt{p}_{jkl}\rightarrow p_l \hspace{30mm}\text{when $j,k\rightarrow 0$}\nonumber\\
&&\wt{p}_{ijk}\rightarrow p_i + p_j \hspace{21.5mm} \wt{p}_{jkl}\rightarrow p_l \hspace{30mm}\text{when $k\rightarrow 0$ and $i||j$}\nonumber\\
&&\wt{p}_{ijk}\rightarrow p_i+p_j+p_k \hspace{13mm} \wt{p}_{jkl}\rightarrow p_l \hspace{30mm}\text{when $i||j||k$}\nonumber\\
&&\wt{p}_{ijk}\rightarrow p_i \hspace{30mm} \wt{p}_{jkl}\rightarrow p_j+p_k+p_l \hspace{13mm}\text{when $j||k||l$}\nonumber\\
&&\wt{p}_{ijk}\rightarrow p_i \hspace{30mm} \wt{p}_{jkl}\rightarrow p_l+p_k \hspace{21.5mm}\text{when $j\rightarrow 0$ and $k||l$}\nonumber\\
&&\wt{p}_{ijk}\rightarrow p_i + p_j \hspace{21.5mm} \wt{p}_{jkl}\rightarrow p_l+p_k \hspace{21.5mm}\text{when $i||j$ and $k||l$}.
\eeqa
This behaviour ensures that infrared singularities are properly subtracted. Furthermore, it can be shown that in single unresolved limits, the momentum mappings defined in eq.(\ref{eq.mapffd}) collapse into an NLO $3 \to 2$ mapping. This allows the subtraction of single unresolved limits in four-parton antenna functions with products of three-parton antennae.

\subsubsection{Initial-final configurations}
For initial-final subtraction terms the mappings are $(\hat{i},j,k,l)\rightarrow (\hat{\bar{i}},\wt{jkl})$, and the remapped momenta are given by
\beqa
&&\bar{p}_i^\mu=x_i p_i^\mu\nonumber\\
&&\wt{p}_{ijk}^\mu=p_j^\mu+p_k^\mu+p_l^\mu-(1-x_i)p_i^\mu\label{eq.mapifd}
\eeqa
with the rescaling variable
\beq
x_i=\frac{s_{ij}+s_{ik}+s_{il}-s_{jk}-s_{jl}-s_{kl}}{s_{ij}+s_{ik}+s_{il}}.
\eeq
The properties satisfied by these mappings in double unresolved limits are
\beqa
&& x_i p_i \rightarrow p_i \hspace{30mm} \wt{p}_{jkl}\rightarrow p_l \hspace{30mm}\text{when $j,k\rightarrow 0$}\nonumber\\
&& x_i p_i \rightarrow p_i \hspace{30mm} \wt{p}_{jkl}\rightarrow p_k+p_l \hspace{21.5mm}\text{when $j\rightarrow 0$ and $k||l$}\nonumber\\
&& x_i p_i \rightarrow (1-z_i) p_i \hspace{17mm} \wt{p}_{jkl}\rightarrow p_l \hspace{30mm}\text{when $k\rightarrow 0$ and $p_j=z_i p_i$}\nonumber\\
&& x_i p_i \rightarrow p_i \hspace{30mm} \wt{p_{jkl}}\rightarrow p_j+p_k+p_l \hspace{13mm}\text{when $j||k||l$}\nonumber\\
&& x_i p_i \rightarrow (1-z_i) p_i \hspace{17mm} \wt{p}_{jkl}\rightarrow p_l \hspace{30mm}\text{when $p_j+p_k = z_i p_i$}\nonumber\\
&& x_i p_i \rightarrow (1-z_i) p_i \hspace{17mm} \wt{p}_{jkl}\rightarrow p_k+p_l \hspace{21.5mm}\text{when $p_j=z_i p_i$ and $k||l$}.\nonumber\\
\eeqa
As before, it can be shown that in single unresolved limits eq.(\ref{eq.mapifd}) collapses into an NLO mapping.

\subsubsection{Initial-initial configurations}
Like in the single unresolved case, phase space mappings for these present configurations consist of a rescaling of the initial state hard radiators momenta, and a boost of all final state momenta that preserves overall momentum conservation. Concretely the mappings are of the form $(\hat{i},j,k,\hat{l},...,m,...)\rightarrow (\hat{\bar{i}},\hat{\bar{l}},...,\tilde{m},...)$, and the remapped momenta are
\beqa
&&\bar{p}_i^\mu=x_i p_i^\mu\nonumber\\
&&\bar{p}_l^\mu=x_l p_l^\mu\nonumber\\
&&\tilde{p}_m^\mu=p_m^\mu-\frac{2p_m \cdot (q+\tilde{q})}{(q+\tilde{q})^2}(q^\mu+\tilde{q}^\mu)+\frac{2p_m \cdot q}{q^2}\tilde{q}^\mu\label{eq.mapiid}
\eeqa
where $q^\mu=p_i^\mu+p_j^\mu+p_k^\mu+p_l^\mu$ and $\tilde{q}^\mu=\tilde{p}_i^\mu+\tilde{p}_l^\mu$. The rescaling variables are given by
\beqa
&&x_i=\sqrt{\frac{s_{il}-s_{jl}-s_{kl}}{s_{il}-s_{ij}-s_{ik}}}\sqrt{\frac{s_{ij}-s_{ik}-s_{il}+s_{jk}-s_{jl}-s_{kl}}{s_{il}}}\nonumber\\
&&x_l=\sqrt{\frac{s_{il}-s_{ij}-s_{ik}}{s_{il}+s_{jl}+s_{kl}}}\sqrt{\frac{s_{ij}-s_{ik}-s_{il}+s_{jk}-s_{jl}-s_{kl}}{s_{il}}}.
\eeqa
In the double unresolved limits the mappings in eq.(\ref{eq.mapiid}) give
\beqa
&& x_i p_i \rightarrow p_i \hspace{30mm} x_l p_l \rightarrow p_l \hspace{30mm}\text{when $j,k\rightarrow 0$}\nonumber\\
&& x_i p_i \rightarrow p_i \hspace{30mm} x_l p_l \rightarrow (1-z_l)p_l \hspace{17mm}\text{when $j\rightarrow 0$ and $p_k=z_l p_l$}\nonumber\\
&& x_i p_i \rightarrow (1-z_i)p_i \hspace{17mm} x_l p_l \rightarrow p_l \hspace{30mm}\text{when $k\rightarrow 0$ and $p_j=z_i p_i$}\nonumber\\
&& x_i p_i \rightarrow (1-z_i)p_i \hspace{17mm} x_l p_l \rightarrow p_l \hspace{30mm}\text{when $p_j+p_k=z_i p_i$}\nonumber\\
&& x_i p_i \rightarrow p_i \hspace{30mm} x_l p_l \rightarrow (1-z_l)p_l \hspace{17mm}\text{when  $p_j+p_k=z_l p_l$}\nonumber\\
&& x_i p_i \rightarrow (1-z_i)p_i \hspace{17mm} x_l p_l \rightarrow (1-z_l)p_l \hspace{17mm}\text{when $p_j=z_i p_i$ and $p_k=z_l p_l$},\nonumber \\
\eeqa
while in single unresolved limits they collapse into NLO mappings.

\bibliographystyle{JHEP-2}

\providecommand{\href}[2]{#2}\begingroup\raggedright
\endgroup

\end{document}